\documentclass[fleqn,usenatbib]{mnras}

\usepackage{newtxtext,newtxmath}

\usepackage[T1]{fontenc}

\DeclareRobustCommand{\VAN}[3]{#2}
\let\VANthebibliography\thebibliography
\def\thebibliography{\DeclareRobustCommand{\VAN}[3]{##3}\VANthebibliography}

\usepackage{graphicx}	
\usepackage{amsmath}	

\newcommand{\oiii}{[O {\sc iii}]}
\newcommand{\nii}{[N {\sc ii}]}
\newcommand{\sii}{[S {\sc ii}]}
\newcommand{\oi}{[O {\sc i}]}
\newcommand{\halpha}{H$\mathrm{\alpha}$}
\newcommand{\hbeta}{H$\mathrm{\beta}$}
\newcommand{\kms}{\,km\,s$^{-1}$} 
\newcommand{\divingTD}{DIVING$^\mathrm{3D}$\,}
\newcommand{\degree}{$^{\circ}$}
\newcommand{\hii}{H {\sc ii} }
\newcommand{\pms}{$\pm$}

\title[The DIVING$^\mathrm{3D}$ Survey III]{The DIVING$^\mathrm{3D}$ Survey - Deep IFS View of Nuclei of Galaxies - III. Analysis of the nuclear region of the early-type galaxies of the sample}

\author[Ricci et al.]{
T.V. Ricci,$^{1}$\thanks{E-mail: tiago.ricci@uffs.edu.br (TVR)}
J.E. Steiner,$^{2}$\thanks{In Memorian}
R.B. Menezes,$^{3}$
K. Slodkowski Clerici,$^{4}$ and
M.D. da Silva$^{1}$
\\
$^{1}$Universidade Federal da Fronteira Sul, Campus Cerro Largo, RS 97900-000, Brazil\\
$^{2}$Instituto de Astronomia, Geof\'isica e Ci\^encias Atmosf\'ericas, Universidade de S\~ao Paulo, Brazil\\
$^{3}$Instituto Mau\'a de Tecnologia, Pra\c{c}a Mau\'a 1, 09580-900, S\~ao Caetano do Sul, SP, Brazil\\
$^{4}$Universidade Federal do Rio Grande do Sul, Departamento de Astronomia, 91501-970, Porto Alegre, RS, Brazil\\
}

\date{Accepted XXX. Received YYY; in original form ZZZ}

\pubyear{2023}

\begin{document}
\label{firstpage}
\pagerange{\pageref{firstpage}--\pageref{lastpage}}
\maketitle

\begin{abstract}
We analysed the nuclear region of all 56 early-type galaxies from the DIVING$^\mathrm{3D}$ Project, which is a statistically complete sample of objects that contains all 170 galaxies of the Southern Hemisphere with B < 12.0 mag and galactic latitude |b| < 15\degree. Observations were performed with the Integral Field Unit of the Gemini Multi-Object Spectrograph. Emission lines were detected in the nucleus of 86\pms5\% of the objects. Diagnostic diagrams were used to classify 52\pms7\% of the objects as LINERs or Seyferts, while the other 34\pms6\% galaxies without H$\beta$ or \oiii\, lines in their spectra were classified as weak emission line objects. Transition Objects are not seen in the sample, possibly because the seeing-limited data cubes of the objects allow one to isolate the nuclei of the galaxies from their circumnuclear regions, avoiding contamination from H II regions.  A broad line region is seen in 29\pms6\% of the galaxies. Of the 48 galaxies with emission-line nuclei, 41 have signs of AGNs. Some objects have also indications of shocks in their nuclei. Lenticular galaxies are more likely to have emission lines than ellipticals. Also, more luminous objects have higher \nii /H$\alpha$ ratios, which may be associated with the mass-metalicity relation of galaxies. A direct comparison of our results with the Palomar Survey indicates that the detection rates of emission lines and also of type 1 AGNs are higher in the DIVING$^\mathrm{3D}$ objects. This is a consequence of using a more modern instrument with a better spatial resolution than the Palomar Survey observations. 
\end{abstract}




\section{Introduction \label{sec:introduction} }  

Central regions of galaxies with well-defined bulges are special places, since they host supermassive black holes (SMBH) in their nuclei. Originally proposed to explain the high luminosity of quasars and the emission from the centre of some galaxies (see e.g. \citealt{1969Natur.223..690L}), SMBHs were observationally confirmed using dynamical models applied in both the stellar and gas components located in their neighbourhood \citep{1995ARA&A..33..581K,1998AJ....115.2285M,2005SSRv..116..523F,2013ARA&A..51..511K}, and, more recently, through the observations made by the Event Horizon Telescope of M87 and Sgr A$^*$ \citep{2019ApJ...875L...1E, 2022ApJ...930L..12A}. Relations between the SMBH mass ($\mathrm{M_{SMBH}}$) and global parameters of galaxies, such as stellar velocity dispersion $\sigma_*$ \citep{2000ApJ...539L...9F,2000ApJ...539L..13G,2009ApJ...698..198G, 2016ApJ...818...47S} and bulge luminosity \citep{1998AJ....115.2285M, 2004ApJ...604L..89H} suggests a co-evolution between these objects and their hosts (see \citealt{2013ARA&A..51..511K} for a review).

Quasars are among the most luminous types of Active Galactic Nuclei (AGNs), which are objects associated with the capture of matter by a SMBH in the form of an accretion disc. In the family of low luminosity AGNs (LLAGNs , see \citealt{2008ARA&A..46..475H} for a review), there are the Seyfert objects, where high ionization emission lines dominate the nuclear spectrum of the galaxy, and the low ionization nuclear emission-line regions (LINERs), with a predominance of emission lines from low ionization species. However, LINER-like spectra may also be produced by other ionization sources. Originally proposed by \citet{1980A&A....87..152H}, LINERs were associated with shock waves in the nuclear region of bright galaxies. Later, \citet{1983ApJ...264..105F} and \citet{1983ApJ...269L..37H} suggested that LINERs may be ionized by AGNs with low ionization parameter values. Populations of hot low-mass evolved stars (HOLMES) are also a possible power source for LINERs \citep{1994A&A...292...13B,2008MNRAS.391L..29S}. To confirm the presence of LLAGNs in LINERs, it is necessary to search for signatures of nuclear activity, such as the presence of broad components in permitted lines \citep{1997ApJS..112..391H,2008ARA&A..46..475H,2018MNRAS.480.1106C,2020A&A...635A..50H}, equivalent widths of the \halpha\, line [EW(\halpha)] $>$ 3 \AA\, \citep{2011MNRAS.413.1687C}, the detection of radio or X-ray cores \citep{1989MNRAS.240..591S,1994MNRAS.269..928S,2008ARA&A..46..475H,2009A&A...506.1107G,2017ApJ...835..223S, 2020ApJ...900..124B} or by the observation of coronal lines, such as the emission from the Ne V ion in mid-infrared (MIR) spectra \citep{2021ApJ...906...35S}. Given the high fraction of LINERs in galaxies from the local Universe \citep{1997ApJ...487..568H, 2003MNRAS.346.1055K, 2007MNRAS.382.1415S}, a proper association between these objects with LLAGNs is necessary for robust statistical analysis of nuclear activity in nearby objects. 

The nebular emission in the central region of elliptical and lenticular galaxies, also denominated early-type galaxies (ETGs), are usually classified as LINERs \citep{2008ARA&A..46..475H}. An important work using a statistically complete sample of ETGs focusing on the optical emission from their central regions was performed by \citet{1986AJ.....91.1062P}. In a survey of 203 ETGs from the Southern Hemisphere, these authors detected emission lines in about 55\% of the objects. A spatially extended nebular component was seen in 12\% of the galaxies. Later, \citet{1997ApJ...487..568H} found that nearly half of all ETGs from a sample containing all 486 galaxies from the Northern Hemisphere brighter than 12.5 magnitudes in the B band (Palomar Survey, \citealt{1995ApJS...98..477H}) have ionized gas, mostly classified as LINERs. In addition, 14\% of the ETGs of this sample show a broad component in the H$\alpha$ line \citep{1997ApJS..112..391H}. \citet{2008ARA&A..46..475H} argued that most of these LINERs are photoionized by LLAGNs. This conclusion is supported by the detection of both radio \citep{1989MNRAS.240..591S,1991AJ....101..148W,1994MNRAS.269..928S,2016MNRAS.458.2221N} and X-ray cores \citep{2017ApJ...835..223S, 2020ApJ...900..124B} in samples of ETGs. \citet{2010A&A...519A..40A} and \citet{2011A&A...528A..10P}, using both optical and MIR spectra, also associated the nebular emission of most objects of a sample of 65 ETGs with LLAGNs or shocks. However, other papers have revealed the importance of alternative sources of ionization in ETGs. \citet{1989ApJ...346..653K} argued that X-ray and UV emissions are not high enough to explain the nebular emission in a sample of 26 ETGs. \citet{2012ApJ...747...61Y}, using spectral data from the Sloan Digital Sky Survey (SDSS) and the Palomar Survey, detected an extended gas emission in a sample of red galaxies. They indicated the necessity of a source that follows the stellar density profiles of the objects, such as populations of post-asymptotic giant branch (pAGB) stars. Integral field spectrograph (IFS) data have also indicated the need for extended ionization sources in ETGs, with LLAGNs contributing only with the nebular emission in scales of 10 to 100 pc \citep{2010MNRAS.402.2187S,2012A&A...540A..11K,2013A&A...558A..43S, 2013A&A...555L...1P, 2015MNRAS.451.3728R,2016A&A...588A..68G, 2016MNRAS.461.3111B, 2017ApJ...837...40P}. In this sense, IFS data with high spatial resolution may provide important information about LLAGNs in the local Universe, once it is possible to resolve the central region of the galaxies in scales of 10 to 100 pc, allowing the detection of signatures of weak nuclear activity (\citealt{2014MNRAS.440.2419R,2014MNRAS.440.2442R}; \citealt{2022MNRAS.513.5935M}, hereinafter paper II) and phenomena such as compact outflows \citep{2017MNRAS.467.2612W} or changes in the direction of the gas kinematics caused by the precession of accretion discs \citep{2019MNRAS.485.5590R}.

In this work, we analysed the nuclear region of a statistically complete sample of all 56 ETGs from the \divingTD project \citep[hereinafter paper I]{2022MNRAS.510.5780S}, which contains all the 170 galaxies from the Southern Hemisphere with B $\leq$ 12.0 mag and galactic latitude |b| $>$ 15$^\mathrm{o}$. These ETGs were observed with the Gemini Multi-Object Spectrograph (GMOS) in the integral field unit (IFU) mode, which provides data cubes with seeing limited spatial resolution. Paper II presents an analysis of the nuclear region of all 57 galaxies of the \divingTD project brighter than 11.2 mag (the mini-\divingTD sample). The main goal of this third paper of the project is to characterize the unresolved nebular emission from the nuclei of the ETGs of the sample, and to search for possible signatures of nuclear activity in these objects. In a preliminary study of a sample of 10 ETGs using data taken with the GMOS-IFU, \citet{2014MNRAS.440.2419R, 2014MNRAS.440.2442R, 2015MNRAS.451.3728R} detected nebular emission in all objects, all with LINER characteristics. A broad H$\alpha$ component was detected in six objects. Now, we expand this study to a statistically complete sample of ETGs, which will allow us to obtain robust information about the nuclear activity in nearby elliptical and lenticular galaxies.

The paper is organized as follows: Section \ref{sec:observations} presents the sample properties and describes the observations. Section \ref{sec:spectral_synthesis} describes the spectral synthesis procedure to subtract the stellar component from the data cubes of the galaxies. Section \ref{sec:el_properties} shows the emission line properties from the nuclear region of the galaxies. The discussion in Section \ref{sec:discussion} is divided into three parts: Section \ref{sec:comparison_palomar} compares our results with those obtained by the Palomar Survey \citep{1997ApJS..112..315H, 1997ApJS..112..391H, 1997ApJ...487..568H, 2003ApJ...583..159H}; Section \ref{sec:ionization_sources} discusses the possible ionization sources of these ETGs and Section \ref{sec:properties_E_S0} compares the nebular emission from elliptical galaxies with those from the lenticular objects. Finally, Section \ref{sec:conclusion} presents the main conclusions of our study.

\section{Sample properties and observations} \label{sec:observations} 

\begin{figure*}
    \centering
    \includegraphics[scale=0.5]{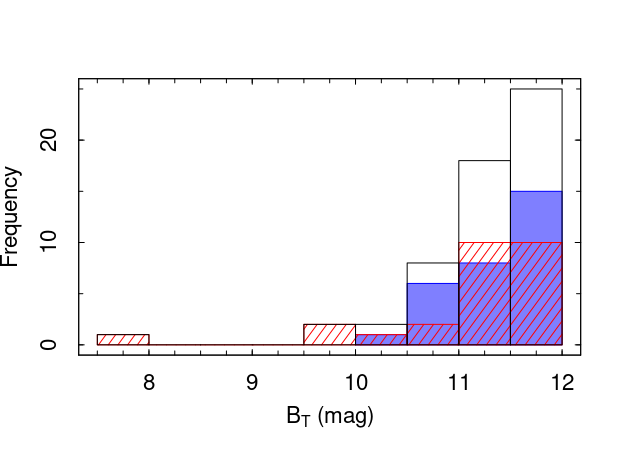}
    \includegraphics[scale=0.5]{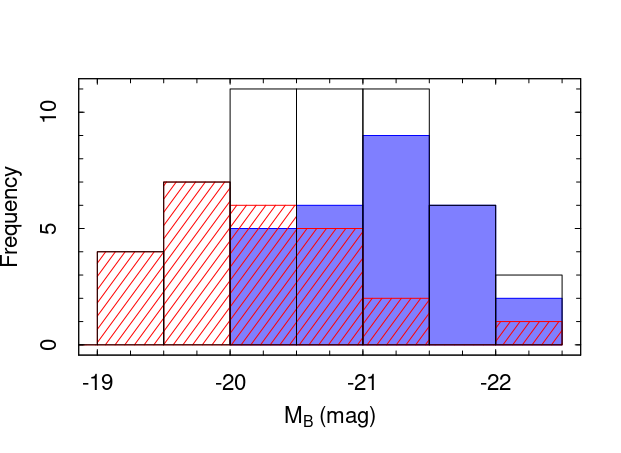}
    \includegraphics[scale=0.5]{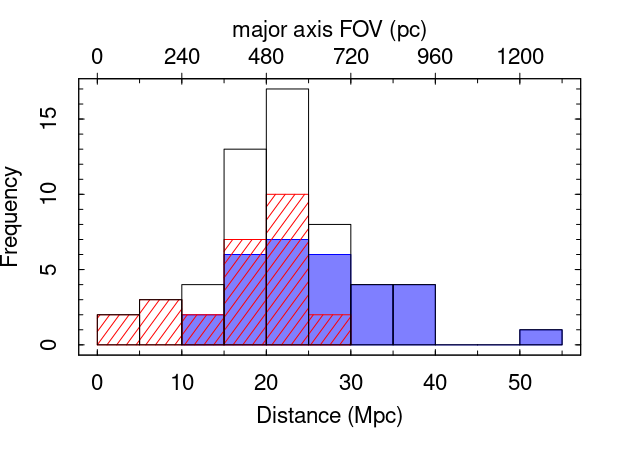}
    \includegraphics[scale=0.5]{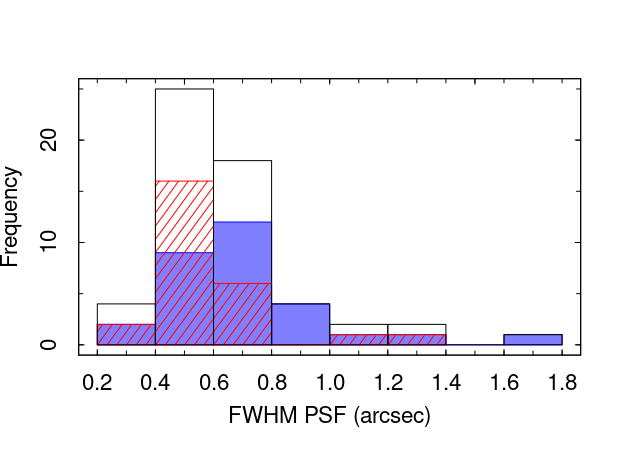}
    \caption{Basic properties of the ETGs from the \divingTD project. In all panels, the hashed red bars correspond to the lenticular galaxies, the solid blue bars are associated with the ellipticals and the white bars are related to the whole sample. Top left: distribution of the apparent B magnitudes. Top right: distribution of the absolute B magnitudes, with a median value of -20.7 mag. Bottom left: distribution of the distances, with a median value of 21.2 Mpc. Bottom right: distribution of the FWHM of the PSFs of the data cubes. }
    \label{fig:hist_sample_parameters}
\end{figure*}

In paper I, we presented all information concerning the main \divingTD sample. There is a total of 56 ETGs that belong to the \divingTD project, 30 ellipticals and 26 lenticulars. Table \ref{tab:sample_galaxies} provides some basic information about the objects, such as distances, heliocentric velocities, effective radii, spatial scales and both the apparent and absolute B-band magnitudes. Fig. \ref{fig:hist_sample_parameters} presents the distributions of the distances and of both the apparent and absolute magnitudes of the galaxies. To test if the sample objects are uniformly distributed in space, we calculated the mean value of the $V/V_{max}$ ratios, where $V$ is the volume of a sphere centred on the observer and with a radius that corresponds to the distance of the object and $V_{max}$ is the maximum volume that the object may have and still be included in the sample. If $\langle V/V_{MAX} \rangle = 0.5$, then the distribution of the galaxies in the sample is uniform in space \citep{sch68}. We found $\langle V/V_{MAX} \rangle = 0.479\pm0.039$ for the ETGs from the \divingTD project. This means that this ETG sample is representative of the local Universe. 

\begin{table*}
    \centering
    \caption{Overview of the sample of ETGs from the \divingTD project. The morphologies were taken from RC3 \citep{1991trcb.book.....D}. The heliocentric velocities (V$\mathrm{_{hel}}$) are from the NASA Extragalactic Database (NED). The apparent B magnitudes are from RSA \citep{1981RSA...C...0000S}. The absolute B magnitudes (M$\mathrm{_B}$) and the effective radii (R$\mathrm{_e}$) are from the Carnegie Irvine Galaxy Survey \citep{2011ApJS..197...21H}, except fot NGC 4105, where M$\mathrm{_B}$ is from Hyperleda \citep{2014A&A...570A..13M} and R$\mathrm{_e}$ is from RC3. The references for the distances are shown in paper I. The PSFs of the data cubes are well fitted by a Gaussian function, with a FWHM that varies in the data cubes with $\lambda^{-0.3}$ (see paper II). Here, we present the FWHM of the PSF of the deconvolved data cubes for $\lambda$ = 5500 \AA. The signal-to-noise ratio (S/N) values are for the spectra of the spaxels located at the position of the stellar photometric centres of the galaxies. In the last column, we show the observation programmes from the Gemini telescopes of each galaxy.  \label{tab:sample_galaxies}}
    \begin{tabular}{ccccccccccc}
    \hline
Galaxy & Morphology & V$\mathrm{_{hel}}$ & d & B & M$\mathrm{_B}$ & R$\mathrm{_e}$ & Spatial scale & PSF at 5500\AA & S/N & Programme \\
       &            & (km s$^{-1}$) & (Mpc) & (mag) & (mag) & (arcmin) & (pc/arcsec) & (arcsec) &  &\\
    \hline
NGC 584&E4                                      &1802&20&11.20&-21.80&0.55&97&0.78&119&GS-2013B-Q-20\\
NGC 596&cD pec?                                 &1903&22&11.88&-20.19&0.48&107&0.45&84&GS-2016B-Q-25\\
NGC 720&E5                                      &1745&31&11.15&-21.10&0.85&150&0.73&113&GS-2013B-Q-20\\
NGC 936&SB0\^{}+(rs)                               &1434&20&11.19&-20.82&0.75&97&0.53&115&GS-2014B-Q-30\\
NGC 1052&E4                                      &1488&19&11.53&-20.26&0.50&92&0.79&114&GS-2013B-Q-20\\
NGC 1201&SA0\^{}0(r)?                               &1686&20&11.56&-19.99&0.45&97&0.69&112&GS-2016B-Q-25\\
NGC 1316&SAB0\^{}0(s) pec                           &1802&21&9.60&-22.48&1.53&102&0.77&231&GS-2013B-Q-20\\
NGC 1326&(R)SB0\^{}+(r)                             &1360&19&11.34&-20.06&0.57&92&0.44&68&GS-2016B-Q-25\\
NGC 1332&S0\^{}-?(s) edge-on                        &1619&23&11.29&-20.28&0.65&112&0.52&148&GS-2013B-Q-20\\
NGC 1344&E5                                      &1169&21&11.28&-20.23&0.56&102&0.39&65&GS-2016B-Q-25\\
NGC 1380&SA0                                     &1877&21&11.10&-20.72&0.64&102&0.55&86&GS-2008B-Q-21\\
NGC 1387&SAB0\^{}-(s)                               &1302&19&11.83&-19.81&0.44&92&0.45&77&GS-2015B-Q-25\\
NGC 1395&E2                                      &1717&24&11.18&-21.36&1.10&116&0.88&182&GS-2013B-Q-20\\
NGC 1399&E1 pec                                  &1425&21&10.79&-21.44&0.91&102&0.86&184&GS-2008B-Q-21\\
NGC 1404&E1                                      &1947&19&11.06&-20.78&0.41&92&0.72&193&GS-2008B-Q-21\\
NGC 1407&E0                                      &1779&25&10.93&-21.80&1.26&121&0.76&91&GS-2013B-Q-20\\
NGC 1411&SA0\^{}-(r)?                               &983&19&11.70&-19.47&0.30&92&0.67&99&GS-2015B-Q-25\\
NGC 1427&cD                                      &1388&26&11.94&-20.18&0.56&126&0.51&95&GS-2016B-Q-25\\
NGC 1527&SAB0\^{}-(r)?                              &1212&19&11.70&-19.53&0.50&92&0.49&94&GS-2015B-Q-25\\
NGC 1533&SB0\^{}-                                   &790&24&11.71&-19.78&0.52&116&0.39&86&GS-2015B-Q-25\\
NGC 1537&SAB0\^{}- pec?                             &1429&23&11.62&-19.96&0.50&112&0.44&140&GS-2015B-Q-25\\
NGC 1543&(R)SB0\^{}0(s)                             &1176&18&11.49&-20.72&0.99&87&0.59&86&GS-2015B-Q-25\\
NGC 1549&E0-1                                    &1256&16&10.76&-20.53&0.75&78&0.58&200&GS-2013B-Q-20\\
NGC 1553&SA0\^{}0(r)                                &1080&10&10.42&-20.93&0.91&48&0.80&54&GS-2014B-Q-30\\
NGC 1574&SA0\^{}-(s)?                               &1041&20&11.19&-20.54&0.52&97&1.08&194&GS-2013B-Q-20\\
NGC 1700&E4                                      &3889&33&11.96&-21.08&0.39&160&0.40&142&GS-2013B-Q-20\\
NGC 1947&S0\^{}- pec                                &1100&20&11.86&-19.36&1.05&97&0.60&35&GS-2015B-Q-25\\
NGC 2217&(R)SB0\^{}+(rs)                            &1619&26&11.59&-20.35&0.97&126&0.65&109&GS-2013B-Q-20\\
NGC 2784&SA0\^{}0(s)?                               &691&10&11.21&-19.38&0.67&48&0.59&146&GS-2013A-Q-52\\
NGC 2974&E4                                      &1887&22&11.78&-20.69&0.58&107 &1.72&157&GS-2013A-Q-52\\
NGC 3115&S0\^{}- edge-on                            &663&10&9.98&-20.30&1.20&48&0.70&250&GS-2013A-Q-52\\
NGC 3557&E3                                      &3079&27&11.46&-21.77&0.75&131&0.76&62&GS-2013A-Q-52\\
NGC 3585&E6                                      &1434&13&10.93&-20.95&1.31&63&0.73&158&GS-2013A-Q-52\\
NGC 3904&E2-3?                                   &1576&28&11.95&-20.56&0.46&136&0.77&134&GS-2013A-Q-52\\
NGC 3923&E4-5                                    &1739&16&10.91&-21.51&1.44&78&0.72&111&GS-2013A-Q-52\\
NGC 3962&E1                                      &1815&36&11.66&-21.15&0.82&175&0.76&110&GS-2013A-Q-52\\
NGC 4105&E3                                      &1858&27&11.88&-20.83*&0.57*&131&1.02&136&GS-2013A-Q-52\\
NGC 4546&SB0\^{}-(s)?                               &1050&14&11.30&-19.95&0.43&68  &0.55&112&GS-2008A-Q-51\\
NGC 4696&cD1 pec                                 &2969&40&11.59&-22.16&1.64&194&0.59&41&GS-2013A-Q-52\\
NGC 4697&E6                                      &1241&12&10.11&-20.28&1.14&58&0.46&83&GN-2014A-Q-3\\
NGC 4753&I0                                      &1163&24&10.85&-21.07&1.16&116&0.47&84&GS-2015A-Q-3\\
NGC 4958&SB0(r)? edge-on                         &1455&21&11.48&-20.26&0.45&102&0.49&36&GS-2015A-Q-35\\
NGC 4984&(R)SAB0\^{}+(rs)                           &1279&21&11.71&-20.15&0.61&102&0.58&44&GS-2016A-Q-3\\
NGC 5018&E3?                                     &2816&38&11.71&-21.69&0.39&184&0.57&135&GS-2013A-Q-52\\
NGC 5044&E0                                      &2782&32&11.92&-21.30&0.86&155&0.84&41&GS-2013A-Q-52\\
NGC 5061&E0                                      &2061&15&11.35&-21.28&0.65&73&0.41&153&GS-2015A-Q-3\\
NGC 5102&SA0\^{}-                                   &468&4.0&10.64&-16.97&1.11&19&0.47&116&GS-2015A-Q-3\\
NGC 5128&S0 pec                                  &547&3.5&7.89&-19.94&2.71&17&0.30&25&GS-2015A-Q-3\\
NGC 6684&(R')SB0\^{}0(s)                            &883&14&11.34&-19.40&0.58&68&0.51&86&GS-2015A-Q-3\\
NGC 6868&E2                                      &2854&34&11.83&-21.28&0.85&165&0.56&71&GS-2013A-Q-52\\
NGC 7049&SA0\^{}0(s)                                &2285&30&11.64&-21.41&0.59&145&1.31&149&GS-2013A-Q-52\\
NGC 7144&E0                                      &1932&24&11.79&-20.31&0.60&116&0.53&92&GS-2015B-Q-25\\
NGC 7507&E0                                      &1546&25&11.43&-21.36&0.58&121&0.65&173&GS-2013B-Q-20\\
IC 1459&E3-4                                    &1802&29&10.96&-21.63&0.93&141&0.70&151&GS-2008B-Q-21\\
IC 4296&E                                       &3737&51&11.58&-22.41&1.14&247&1.21&105&GS-2013A-Q-52\\
IC 5328&E4                                      &3137&40&11.95&-20.64&0.56&194&0.84&67&GS-2016A-Q-3\\
 \hline
    \end{tabular}

\end{table*}

All ETGs from the \divingTD project were observed using GMOS-IFU \citep{2002PASP..114..892A, 2004PASP..116..425H}, installed on both Gemini North and South telescopes. Only one object (NGC 4697) was observed with the Gemini-North telescope; the other galaxies were observed with the Gemini-South telescope. The observation programmes of all galaxies are shown in Table \ref{tab:sample_galaxies}. The one-slit mode was used in all observations, which means that 500 fibres with a diameter of 0.2 arcsec each are dedicated to the science object, while the other 250 fibres of the instrument are responsible for observing the sky background, separated by an angular distance of 1 arcmin from the main object. These fibres are arranged linearly on the nominal location of the spectrograph slit (pseudo-slit).  All spectra that emerge from the spectrograph are stored by a set of three CCDs. This mode allows the creation of data cubes with a field of view (FOV) of 3.5$\times$5.0 arcsec$^2$ and angular resolutions that are dominated by the seeings of the observations. For the sample galaxies, the major axes of the FOVs cover projected distances $\lesssim$ 1 kpc, as shown in Fig. \ref{fig:hist_sample_parameters}. Exposure times and the number of exposures taken for each object are shown in Paper I. In 54 objects, the B600-G5323 grating was used, which provides a spectral resolution with a full width at half-maximum (FWHM) of 1.8 arcsec, measured with the O {\sc i}$\lambda$5577 skyline, and a wavelength range of 4228--7120 \AA. Two objects, NGC 4546 and NGC 4697, were observed with the R831-G5322 and R831-G5302 gratings, respectively, covering a wavelength range of 4736--6806 \AA\ with a spectral resolution of 1.3 \AA\ (FWHM), also measured with the O {\sc i}$\lambda$5577 skyline.  

All basic reduction steps for the observations, such as bias subtraction, flat-field corrections, wavelength and flux calibrations, were performed using the Gemini {\sc iraf} package. Cosmic rays were removed using the L.A. Cosmic routine \citep{2001PASP..113.1420V}. After such procedures, we built data cubes with a spatial scale of 0.05 arcsec/pixel. In addition, we also applied the complementary data treatments for GMOS data cubes described in \citet{2019MNRAS.483.3700M}. It includes correction of the differential atmospheric refraction effect using the equations proposed by \citet{1982PASP...94..715F} and \citet{1998Metro..35..133B}, Butterworth filtering \citep{2008gonzaleswoods} to remove high-frequency spatial noise, instrumental fingerprint removal (associated with low-frequency noises in both spectral and spatial dimensions) using the Principal Component Analysis (PCA) Tomography technique \citep{2009MNRAS.395...64S}, and Richardson-Lucy deconvolution \citep{1972JOSA...62...55R,1974AJ.....79..745L} to improve the angular resolution of the data cubes. A complete description of the reduction processes, including the application of these complementary techniques, is shown in paper I and also in \citet{2014MNRAS.438.2597M,2015MNRAS.450..369M}. 

The point spread functions of data cubes from GMOS-IFU that were deconvolved using the Richardson-Lucy technique are well described by a Gaussian function \citep{2019MNRAS.483.3700M}.  In a first attempt, we used the spatially unresolved emission from the broad line regions (BLR) in type 1 AGNs to measure the FWHM of the PSF of the data cubes (see Section \ref{sec:blr} and Appendix C, available online, for more details). If the galaxy does not have a BLR, we then compared images extracted from the data cubes with archival observations from the Hubble Space Telescope (HST) by following the same procedures as in paper II. In the lack of an available HST image, we assumed that the FWHM of the PSF of a given data cube is $\sim$ 75\% of the FWHM of the seeing of the observations, estimated with the acquisition image obtained with the GMOS
imager in filter r (SDSS system). This factor was obtained by deconvolving the data cubes of the standard stars that were obtained with the observation programmes and is compatible with the measurements performed using the type 1 AGNs or by the comparison of the data cubes with the HST images. We present the FWHM of the PSFs from the data cubes in Table \ref{tab:sample_galaxies}. Fig. \ref{fig:hist_sample_parameters} shows the distribution of these values for the sample galaxies. 

\section{Spectral synthesis} \label{sec:spectral_synthesis}

Since we are interested in characterising the emission lines from the nuclei of the sample galaxies and searching for possible signatures of LLAGNs, it is necessary to subtract the starlight from the data cubes. To do this, we used the penalized pixel fitting ({\sc ppxf}) software \citep{2004PASP..116..138C, 2017MNRAS.466..798C} to perform spectral synthesis in all spectra of the data cubes of all ETGs from the \divingTD sample. We applied the procedure using the template spectra of Simple Stellar Populations (SSPs) that are described in \citet{2015MNRAS.449.1177V}, with ages older than 30 Myr, metallicities between 0.2 and 2.4 $\mathrm{Z_\odot}$ and an initial mass function of \citet{2003PASP..115..763C}. In all cases, we fitted Gauss-Hermite functions with four moments to describe the absorption line profiles. Possible mismatches between the stellar continuum and the SSPs were corrected using 10$^\mathrm{th}$ degree order multiplicative Legendre polynomials. We then used the synthesis results to remove the stellar component from the data cubes and to produce gas cubes for all sample galaxies, i.e. data cubes containing only emission lines from ionized nebulae and also the Na I$\lambda\lambda$5890, 5896 absorption lines from the neutral interstellar medium (ISM) of the objects. In Fig. A1 of Appendix A, available online, we show the spectrum of the spaxel located at the peak of the stellar continuum from each object, together with their fitted SSPs and the residual spectra. In Table \ref{tab:sample_galaxies}, we present the signal-to-noise (S/N) ratio of these spectra shown in Fig. A1. 

\section{Nuclear emission line properties} \label{sec:el_properties}

To study the nuclei of the sample ETGs with an increased S/N, we summed all spectra from their gas cubes that were contained within apertures positioned at the stellar photometric centres of the objects, with diameters equal to the FWHM of the PSF of the respective data cube. This procedure resulted in an average increase of the S/N of the nuclear spectra of the galaxies by 24\%. Then, we corrected the fluxes of these representative spectra by assuming that the nuclear emission of the galaxies emerges from a point-like source, which is the case for LLAGNs, considering that the PSFs are well described by a Gaussian function. This procedure helps to avoid possible contamination from the circumnuclear regions of the objects, although it does not eliminate emissions from structures along our line of sight. An annulus around the nucleus of each galaxy indicates that the light from circumnuclear regions of the sample is $\sim$ 10\% of the central emission. However, assuming that most of the circumnuclear emission in ETGs comes from rotating components \citep{2006MNRAS.366.1151S, 2015MNRAS.451.3728R, 2017ApJ...837...40P}, then the contribution from off-nuclear regions along our line of sight is only significant if this structure is close to an edge-on orientation. Thus, this 10\% contribution should be taken as maximum contamination of the circumnuclear emission along the line of sight. 

The profiles of the \nii $\lambda\lambda$6548, 6583, \halpha\, and \sii $\lambda$6716, 6731 lines were fitted simultaneously by using a sum of Gaussian functions. To do this, we used the Levenberg–Marquardt algorithm implemented in the {\sc R} software environment by \citet[package \sc{minpack.lm}]{nlslm2016}. The number of Gaussian functions per emission line was set in the following way: first, we tried to fit only one Gaussian function per emission line, assuming that all profiles have the same radial velocity and velocity dispersion. The theoretical ratio \nii$\lambda$6583/\nii$\lambda$6548 = 3.06 \citep{2006agna.book.....O} was maintained fixed for the fitting procedure. If necessary, we added a second set of Gaussian functions for these emission lines, assuming again that these new components have the same radial velocity and velocity dispersion. In these cases, the theoretical \nii$\lambda$6583/\nii$\lambda$6548 ratios were maintained fixed for each set of Gaussian functions. In addition, a third Gaussian function was required in the \halpha\, line of some objects to fit a possible broad component. The statistical errors of the parameters of each set of Gaussian functions (amplitude, peak position and width) were taken from the square roots of the diagonal elements in the variance-covariance matrix, also provided by the {\sc minpack.lm} package. The fitted profiles in the \nii $\lambda\lambda$6548, 6583, \halpha\, and \sii $\lambda$6716, 6731 lines are shown in Fig. B1 of Appendix B, available online. The fitting procedure for the emission lines of NGC 4958 was performed differently by \citet{2019MNRAS.486.1138R}, since this galaxy presents broad \halpha\, components that are typical of a relativistic accretion disc. We used their results concerning the nuclear line emission of NGC 4958 in the present work, but, for completeness, we also show the fitted profiles of this object in Fig. B1. 

The presence of other important emission lines, such as \hbeta, \oiii$\lambda\lambda$4959, 5007 and \oi$\lambda\lambda$6300, 6363 was certified through a visual inspection. In NGC 1052, NGC 2217 and IC 1459, a broad \hbeta\, is also seen in their spectra. In these cases, we fitted the \hbeta\ line with the same set of Gaussian functions used to describe the \halpha\, line profile, i.e. we assumed that all Gaussian components of the \hbeta\, line have the same kinematics as the Gaussian components found for the \halpha\, line. The fitted profiles for the \hbeta\ lines of NGC 1052, NGC 2217 and IC 1459 are shown in Fig. B2 of Appendix B, together with the fitted \hbeta\, profile of NGC 4958 taken from \citet{2019MNRAS.486.1138R}. In addition,  we fitted the profiles of the \oi$\lambda\lambda$6300, 6363 lines of NGC 2217, NGC 5044 and IC 1459, since these galaxies also present visible broad components in this doublet. For NGC 2217 and NGC 5044, we fitted a sum of two Gaussian functions per \oi\, line, while for IC 1459 a sum of three Gaussian functions per \oi\, line was necessary for the profile of the doublet. For this procedure, we fixed the theoretical ratio \oi$\lambda$6300/\oi$\lambda$6363 = 3.05 \citep{2006agna.book.....O} in each Gaussian function used to describe the profiles of the doublet. We assumed that the radial velocity and velocity dispersion of the Gaussian components of the \oi\, lines are independent of the fitted profiles found for the \nii $\lambda\lambda$6548, 6583, \halpha\, and \sii $\lambda$6716, 6731 lines. The fitted \oi $\lambda\lambda$6300, 6363 profiles of NGC 2217, NGC 5044 and IC 1459 are shown in Fig. B3 of Appendix B. 

In galaxies with no obvious sign of emission lines in the nuclear spectrum, we proceeded in the following way: usually, these objects have only the \nii $\lambda\lambda$6548, 6583 + \halpha\, lines that may be described as a sum of three Gaussian functions, one per emission line. So, we tried to fit the profiles of these lines using the same procedure described above. In most cases, the algorithm finds a reasonable solution. To check if the emission lines are statistically relevant, we resampled the observed nuclear spectrum of each galaxy 1000 times using the Bootstrap method. For each resampled spectrum, we fitted again the line profiles, and then we calculated their fluxes by integrating the Gaussian functions. The standard deviation of the 1000 measurements of the fluxes provides an estimate of the error of this parameter for these three lines. An emission line is said to be detected if the S/N ratio of its calculated flux is higher than three. By using this methodology, we conclude that 48 galaxies have emission lines in their nuclei. The limiting case where the emission lines are confirmed is NGC 596. We did not detect any nebular emission for NGC 1201, NGC 1344, NGC 1407, NGC 1427, NGC 1527, NGC 1549, NGC 3904 and NGC 3923. For NGC 1543, NGC 7144 and NGC 7507, we detected only the \halpha\, line, while for NGC 1399 only the \nii$\lambda$6584 line is statistically relevant. 

Since the \divingTD sample is statistically complete, we may determine the fraction of galactic nuclei that contains emission lines in the local Universe. We propose that (86\pms5)\% of nearby ETGs have emission lines in their nuclear spectra, while (14\pms5)\% have pure absorption line nuclei (AbN). The errors were calculated using a binomial distribution, i.e. they correspond to the uncertainties of the fractions of emission line nuclei and pure absorption line nuclei in the local Universe that may be obtained from a sample of 56 objects.

\subsection{Narrow components \label{sec:nlr}}

\begin{figure}
    \includegraphics[scale=0.5]{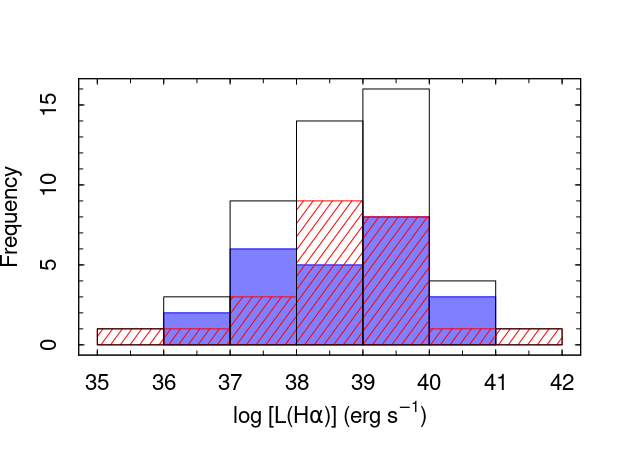}
    \includegraphics[scale=0.5]{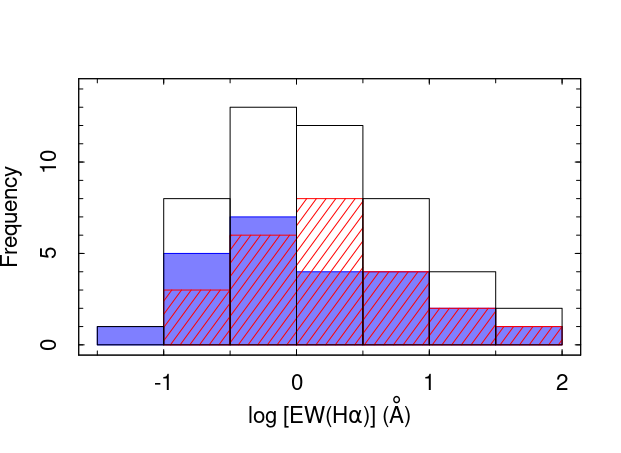}
    \includegraphics[scale=0.5]{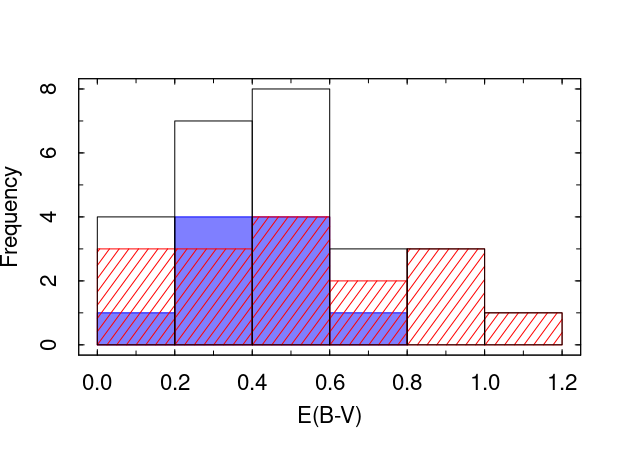}
    \caption{Histograms of some emission line parameters. In all panels, the hashed red bars correspond to the lenticular galaxies, the solid blue bars are associated with the ellipticals and the white bars are related to the whole sample. Top panel: \halpha\, luminosity. Central panel: equivalent width of \halpha. Bottom panel: Colour excess.}
    \label{fig:histograms_comparison}
\end{figure}

We calculated the flux of the narrow components of the \nii $\lambda\lambda$6548, 6583, \halpha\, and \sii $\lambda$6716, 6731 lines by integrating their fitted profiles. For the \hbeta, \oiii $\lambda$5007 and \oi $\lambda$6300 lines, we calculated their fluxes by performing a direct numerical integration for most of the sample galaxies. In the cases where we had to fit the profiles of the \hbeta\, and \oi\, lines, the fluxes were calculated by integrating the fitted Gaussian functions related only to the narrow components. For some objects, the \oiii $\lambda$5007 line fell in the gap between the first and the second CCDs of the GMOS-IFU instrument. In these situations, the flux of the \oiii $\lambda$5007 line was calculated by integrating the \oiii $\lambda$4959 line and then multiplying the result by the theoretical ratio \oiii$\lambda$5007/\oiii$\lambda$4959 =  2.917 \citep{2006agna.book.....O}. In Table \ref{tab:nlf_flux_infos}, we present the observed flux (i.e. not corrected for extinction effects) of the narrow component of the \halpha\, line (f(\halpha)$_n$), the EW(\halpha) and the colour excess [E(B-V)], calculated using the observed \halpha/\hbeta\, ratio, assuming an intrinsic \halpha/\hbeta=3.1, taken from photoionization models for LINER objects \citep{1983ApJ...264..105F, 1983ApJ...269L..37H}, and the extinction law of \citet{1989ApJ...345..245C} ($R_V$ = 3.1). The \nii$\lambda$6583/\halpha, (\sii$\lambda$6716+\sii$\lambda$6731)/\halpha, \oi$\lambda$6300/\halpha, \halpha/\hbeta\, and \oiii$\lambda$5007/\hbeta\ line ratios, corrected for the extinction effects (where possible), are also shown in Table \ref{tab:nlf_flux_infos}. Histograms with the distributions of EW(\halpha), E(B-V) and L(\halpha) are shown in Fig. \ref{fig:histograms_comparison}.

\begin{table*}
    \centering
    \caption{Flux measurements for the narrow components of the nuclear emission lines. The \halpha\, flux is in units of 10$^{-15}$ erg s$^{-1}$ cm$^{-2}$. The equivalent width of the \halpha\, line is in units of \AA. The line ratios are corrected for extinction effects in the galaxies where the colour excess was calculated. \label{tab:nlf_flux_infos}}
    \begin{tabular}{ccccccccc}
    \hline
    Galaxy & f(\halpha)$_n$ & (\halpha/\hbeta)$_n$ & E(B-V) & \nii/\halpha & \sii/\halpha & \oi/\halpha & \oiii/\hbeta & EW(\halpha)$_n$  \\
    \hline
NGC 584&3.7\pms0.6&5.2\pms1.0&0.50\pms0.19&2.2\pms0.4&1.01\pms0.23&0.35\pms0.07&3.9\pms0.6&0.81\pms0.13 \\
NGC 596&0.26\pms0.05&--&--&1.5\pms0.4&--&--&--&0.12\pms0.03 \\
NGC 720&0.94\pms0.09&--&--&1.07\pms0.14&0.57\pms0.12&--&--&0.44\pms0.04 \\
NGC 936&14.4\pms0.7&3.00\pms0.21&0.00\pms0.07&1.39\pms0.10&0.89\pms0.07&0.309\pms0.018&1.56\pms0.12&3.55\pms0.18 \\
NGC 1052&213\pms3&4.93\pms0.15&0.45\pms0.03&1.002\pms0.021&1.177\pms0.024&0.85\pms0.03&3.01\pms0.17&43.9\pms0.7 \\
NGC 1316&2.62\pms0.14&7.9\pms0.5&0.91\pms0.06&2.03\pms0.13&1.17\pms0.09&0.164\pms0.011&3.12\pms0.10&1.77\pms 0.10\\
NGC 1326&4.6\pms0.4&9.1\pms1.6&1.05\pms0.17&1.67\pms0.18&1.56\pms0.15&0.36\pms0.03&1.8\pms0.3&2.26\pms0.18 \\
NGC 1332&10.5\pms0.6&4.8\pms0.4&0.43\pms0.08&1.09\pms0.08&0.45\pms0.05&--&1.38\pms0.11&2.07\pms0.12 \\
NGC 1380&3.7\pms0.3&6.2\pms0.8&0.67\pms0.12&1.88\pms0.22&0.83\pms0.10&0.23\pms0.03&1.88\pms0.19&2.77\pms0.25 \\
NGC 1387&3.2\pms0.3&3.7\pms0.5&0.18\pms0.12&1.19\pms0.16&0.43\pms0.09&--&--&1.58\pms0.15\\
NGC 1395&0.205\pms0.018&--&--&2.04\pms0.21&1.02\pms0.12&--&--&0.54\pms0.05\\
NGC 1399&0.09\pms0.05&--&--&4.4\pms2.7&--&--&--&0.032\pms0.019\\
NGC 1404&0.97\pms0.16&--&--&0.99\pms0.24&--&--&--&0.25\pms0.04\\
NGC 1411&1.62\pms0.08&2.5\pms0.4&0.00\pms0.14&2.65\pms0.15&1.52\pms0.10&1.05\pms0.08&3.8\pms0.7&0.69\pms0.04\\
NGC 1533&8.9\pms0.3&3.37\pms0.19&0.08\pms0.05&1.52\pms0.07&0.68\pms0.04&0.432\pms0.018&1.56\pms0.11&6.10\pms0.23\\
NGC 1537&3.19\pms0.16&--&--&1.26\pms0.08&0.37\pms0.06&--&--&0.87\pms0.04\\
NGC 1543&0.45\pms0.07&3.6\pms1.4&0.1\pms0.4&0.44\pms0.13&--&--&--&0.31\pms0.05\\
NGC 1553&2.54\pms0.13&5.2\pms0.8&0.50\pms0.14&3.31\pms0.19&1.47\pms0.09&0.90\pms0.09&6.6\pms1.1&1.62\pms0.08\\
NGC 1574&11.2\pms0.3&5.9\pms0.5&0.62\pms0.08&1.62\pms0.05&0.45\pms0.03&0.17\pms0.03&2.58\pms0.21&0.919\pms0.023\\
NGC 1700&1.20\pms0.16&--&--&3.1\pms0.5&--&--&--&0.27\pms0.04\\
NGC 1947&1.66\pms0.03&4.0\pms0.3&0.25\pms0.07&0.96\pms0.03&1.34\pms0.04&0.192\pms0.018&1.61\pms0.18&2.87\pms0.06\\
NGC 2217&46.8\pms0.5&4.22\pms0.11&0.30\pms0.03&2.47\pms0.03&1.492\pms0.018&0.602\pms0.015&5.69\pms0.22&10.67\pms0.11\\
NGC 2784&6.0\pms0.7&7.9\pms2.6&0.91\pms0.32&1.01\pms0.16&--&--&2.7\pms0.9&0.62\pms0.07\\
NGC 2974&24\pms3&3.0\pms0.6&0.00\pms0.21&2.5\pms0.4&1.53\pms0.25&0.46\pms0.06&2.8\pms0.5&2.5\pms0.3\\
NGC 3115&0.167\pms0.020&2.2\pms0.5&0.00\pms0.21&0.45\pms0.12&0.70\pms0.12&--&--&0.29\pms0.03\\
NGC 3557&4.4\pms0.3&5.3\pms0.6&0.52\pms0.11&2.66\pms0.26&0.84\pms0.10&0.19\pms0.03&2.42\pms0.21&2.65\pms0.20\\
NGC 3585&4.78\pms0.25&--&--&1.56\pms0.11&0.51\pms0.06&--&--&0.67\pms0.03\\
NGC 3962&14.4\pms0.9&4.5\pms0.4&0.36\pms0.08&1.97\pms0.15&1.69\pms0.12&0.227\pms0.016&1.86\pms0.12&3.68\pms0.22\\
NGC 4105&5.6\pms0.3&3.5\pms0.3&0.11\pms0.09&1.43\pms0.11&1.16\pms0.09&--&2.85\pms0.24&1.05\pms0.06\\
NGC 4546&54\pms4&4.0\pms0.3&0.26\pms0.08&2.07\pms0.19&0.53\pms0.07&0.61\pms0.05&3.32\pms0.10&6.6\pms0.5\\
NGC 4696&4.96\pms0.22&2.98\pms0.19&0.00\pms0.06&2.11\pms0.11&1.43\pms0.09&0.247\pms0.018&1.51\pms0.09&8.5\pms0.4\\
NGC 4697&0.0764\pms0.0052&--&--&1.80\pms0.15&0.32\pms0.08&--&--&0.64\pms0.04\\
NGC 4753&1.72\pms0.09&--&--&3.23\pms0.18&0.94\pms0.08&--&--&0.56\pms0.03\\
NGC 4958&7.3\pms0.3&5.0\pms0.4&0.47\pms0.07&1.48\pms0.08&0.85\pms0.04&0.81\pms0.03&2.58\pms0.20&16.5\pms0.7\\
NGC 4984&289\pms43&9\pms1&0.99\pms0.14&0.76\pms0.16&0.22\pms0.07&0.122\pms0.018&7.86\pms0.16&47\pms7\\
NGC 5018&5.05\pms0.17&--&--&2.24\pms0.09&0.78\pms0.05&--&--&0.744\pms0.024\\
NGC 5044&10.8\pms0.6&3.89\pms0.29&0.22\pms0.07&2.54\pms0.15&2.26\pms0.13&0.50\pms0.07&1.51\pms0.13&21.4\pms1.2\\
NGC 5061&0.88\pms0.12&--&--&4.2\pms0.6&--&--&--&0.157\pms0.021\\
NGC 5102&11.9\pms1.3&--&--&1.6\pms0.3&--&--&--&0.57\pms0.06\\
NGC 5128&0.070\pms0.010&5.7\pms1.4&0.59\pms0.25&1.18\pms0.23&0.87\pms0.17&0.19\pms0.03&4.8\pms1.0&4.4\pms0.6\\
NGC 6684&0.53\pms0.08&--&--&2.2\pms0.4&1.6\pms0.3&--&--&0.152\pms0.024\\
NGC 6868&9.6\pms0.4&4.9\pms0.3&0.45\pms0.06&2.79\pms0.15&1.66\pms0.09&0.359\pms0.018&2.57\pms0.12&7.1\pms0.3\\
NGC 7049&8.6\pms0.7&2.66\pms0.27&0.00\pms0.10&1.98\pms0.21&1.39\pms0.16&0.075\pms0.013&1.38\pms0.09&1.29\pms0.11\\
NGC 7144&0.33\pms0.09&--&--&--&--&--&--&0.18\pms0.05\\
NGC 7507&2.7\pms0.6&6\pms3&0.6\pms0.5&0.22\pms0.20&--&--&--&0.37\pms0.08\\
IC 1459&49.2\pms1.9&4.1\pms0.4&0.27\pms0.08&2.60\pms0.13&1.35\pms0.06&0.38\pms0.06&2.42\pms0.19&16.4\pms0.6\\
IC 4296&12.7\pms0.9&3.8\pms0.3&0.21\pms0.07&3.51\pms0.28&1.81\pms0.14&0.32\pms0.03&2.32\pms0.09&4.1\pms0.3\\
IC 5328&5.4\pms0.6&2.8\pms0.4&0.00\pms0.14&2.5\pms0.4&1.32\pms0.19&0.18\pms0.02&1.18\pms0.12&1.29\pms0.14\\
\hline

    \end{tabular}

\end{table*}

\begin{table*}
    \centering
    \caption{Emission line parameters. Columns (1)–(4): gas kinematics of sets 1 and 2 of the NLR. Column (5): total luminosity (NLR + BLR, when present) of \halpha. For the galaxies where the E(B - V) values were available, then both components were corrected for reddening effects. The uncertainties related to the luminosity of the \halpha\, line took into account the errors in the fluxes, in E(B - V) and in the distances of the galaxies. Column (6): electronic density of the NLR. Column (7): ionized gas mass of the NLR. Note that some objects have only superior or inferior values for n$\mathrm{_e}$ and M$\mathrm{_{ion}}$.}
    \begin{tabular}{cccccccc}
    \hline
    Galaxy&FWHM$\mathrm{_{(n1)}}$&FWHM$\mathrm{_{(n2)}}$&V$\mathrm{_{r(n1)}}$&V$\mathrm{_{r(n2)}}$& log L(\halpha) & n$\mathrm{_e}$ & M$\mathrm{_{ion}}$ \\
    &(km s$^{-1}$)&(km s$^{-1}$)&(km s$^{-1}$)&(km s$^{-1}$)&(erg s$^{-1}$) &(10$^2$ cm$^{-3})$ &(10$^3$ M$_\odot$) \\
    &(1)&(2)&(3)&(4)&(5)&(6)&(7) \\
    \hline
NGC 584&147\pms18&431\pms39&93\pms4&56\pms10&39.31\pms0.17&< 8.7 & > 1  \\
NGC 596&279\pms35&--&22\pms15&--&37.17\pms0.11&--&-- \\
NGC 720&508\pms29&--&-70\pms16&--&38.03\pms0.25&> 14 &< 0.17  \\
NGC 936&276\pms7&972\pms34&8\pms2&32\pms14&38.83\pms0.18&9.4$_{-3.3}^{+3.4}$&1.6$_{-0.7}^{+1.3}$ \\
NGC 1052&355\pms7&702\pms5&-23\pms2&-189\pms4&40.72\pms0.05&19$_{-2}^{+2}$&33$_{-6}^{+5}$ \\
NGC 1316&199\pms4&925\pms37&-43\pms1&-55\pms14&39.10\pms0.07&2.5$_{-1.0}^{+2.4}$&11$_{-5}^{+8}$ \\
NGC 1326&102\pms8&273\pms8&-54\pms2&-34\pms2&39.40\pms0.19&3.7$_{-2.1}^{+2.3}$&15$_{-8}^{+24}$ \\
NGC 1332&163\pms18&843\pms28&22\pms6&-2\pms11&39.28\pms0.12&5.0$_{-3.2}^{+6.4}$&8.4$_{-4.6}^{+17.4}$ \\
NGC 1380&310\pms11&689\pms34&28\pms2&1\pms8&38.98\pms0.14&4.3$_{-2.7}^{+5.0}$&4.9$_{-3.0}^{+7.0}$ \\
NGC 1387&148\pms6&486\pms63&-34\pms2&-41\pms19&38.33\pms0.15&> 9 &< 0.48  \\
NGC 1395&562\pms21&--&-64\pms9&--&37.16\pms0.07&--&-- \\
NGC 1399&208\pms38&--&40\pms14&--&36.68\pms0.26&--&-- \\
NGC 1404&515\pms56&--&-18\pms25&--&37.62\pms0.24&--&-- \\
NGC 1411&304\pms6&--&-21\pms2&--&38.57\pms0.06&5.3$_{-1.5}^{+2.0}$&0.29$_{-0.12}^{+0.14}$ \\
NGC 1533&222\pms4&897\pms24&-4\pms1&-69\pms9&38.88\pms0.34&17$_{-4}^{+8}$&0.99$_{-0.79}^{+0.71}$ \\
NGC 1537&593\pms16&--&18\pms9&--&38.31\pms0.06&> 9 &< 0.56  \\
NGC 1543&181\pms25&--&4\pms10&--&37.41\pms0.49&--&-- \\
NGC 1553&392\pms6&--&125\pms2&--&38.65\pms0.25&9.7$_{-1.6}^{+2.8}$&0.21$_{-0.15}^{+0.17}$ \\
NGC 1574&226\pms4&--&1\pms1&--&39.67\pms0.09&3.4$_{-1.5}^{+2.6}$&16$_{-7}^{+13}$ \\
NGC 1700&565\pms26&--&13\pms13&--&38.20\pms0.20&--&-- \\
NGC 1947&227\pms3&--&39\pms1&--&38.15\pms0.18&< 10 &> 2.4  \\
NGC 2217&443\pms2&--&-11\pms1&--&40.39\pms0.17&10.5$_{-0.5}^{+0.4}$&16$_{-7}^{+6}$ \\
NGC 2784&593\pms37&--&76\pms14&--&39.20\pms0.25&--&-- \\
NGC 2974&323\pms17&668\pms45&-33\pms4&7\pms10&39.60\pms0.18&4.8$_{-3.1}^{+8.7}$&6.4$_{-4.9}^{+8.3}$ \\
NGC 3115&562\pms45&--&71\pms19&--&36.44\pms0.18&6.3$_{-5.0}^{+16.8}$&0.0065$_{-0.0051}^{+0.0206}$ \\
NGC 3557&286\pms10&618\pms28&29\pms3&-92\pms13&39.57\pms0.17&4.7$_{-2.2}^{+6.3}$&6.4$_{-4.4}^{+7.2}$ \\
NGC 3585&613\pms18&--&-30\pms10&--&37.99\pms0.23&8.1$_{-4.9}^{+10.4}$&0.27$_{-0.21}^{+0.43}$ \\
NGC 3962&283\pms5&732\pms45&-59\pms1&-80\pms13&39.74\pms0.10&3.1$_{-1.4}^{+1.6}$&39$_{-16}^{+34}$ \\
NGC 4105&497\pms15&--&-25\pms5&--&39.11\pms0.06&< 3.2 &> 4  \\
NGC 4546&318\pms9&805\pms37&-47\pms2&0\pms13&39.37\pms0.10&14$_{-7}^{+24}$&3.6$_{-2.3}^{+3.3}$ \\
NGC 4696&309\pms14&958\pms19&-80\pms4&195\pms13&38.97\pms0.08&3.5$_{-1.3}^{+1.5}$&6.0$_{-2.3}^{+4.0}$ \\
NGC 4697&327\pms12&--&-12\pms5&--&36.15\pms0.06&--&-- \\
NGC 4753&271\pms5&--&52\pms2&--&38.08\pms0.06&6.2$_{-2.7}^{+6.2}$&0.42$_{-0.23}^{+0.31}$ \\
NGC 4958&175\pms3&627\pms24&-9\pms1&32\pms7&39.10\pms0.18&10$_{-2}^{+2}$&2.7$_{-1.5}^{+1.3}$ \\
NGC 4984&94\pms8&226\pms19&4\pms3&-47\pms16&41.25\pms0.36&> 2 &< 1884  \\
NGC 5018&366\pms6&--&-21\pms2&--&38.93\pms0.10&5.1$_{-2.2}^{+3.3}$&3.7$_{-1.6}^{+2.8}$ \\
NGC 5044&190\pms20&1273\pms9&-66\pms7&12\pms5&39.84\pms0.08&4.3$_{-0.8}^{+0.7}$&12$_{-3}^{+4}$ \\
NGC 5061&472\pms18&--&-21\pms8&--&37.39\pms0.22&--&-- \\
NGC 5102&308\pms26&--&-54\pms12&--&37.36\pms0.07&--&-- \\
NGC 5128&115\pms7&404\pms74&-20\pms2&-48\pms25&35.65\pms0.28&6.0$_{-4.4}^{+21.5}$&0.0016$_{-0.0015}^{+0.0027}$ \\
NGC 6684&297\pms21&--&14\pms9&--&37.07\pms0.09&--&-- \\
NGC 6868&310\pms4&1021\pms30&78\pms1&18\pms10&39.60\pms0.21&4.7$_{-1.2}^{+1.5}$&19$_{-10}^{+13}$ \\
NGC 7049&180\pms6&522\pms36&-2\pms2&-85\pms16&39.56\pms0.10&< 4.3 &> 4  \\
NGC 7144&249\pms56&--&41\pms21&--&37.37\pms0.13&--&-- \\
NGC 7507&557\pms89&--&56\pms32&--&38.94\pms0.50&--&-- \\
IC 1459&343\pms11&798\pms13&-84\pms2&-62\pms3&40.31\pms0.08&8.4$_{-1.6}^{+1.6}$&25$_{-6}^{+8}$ \\
IC 4296&202\pms8&655\pms13&-3\pms2&29\pms3&40.27\pms0.09&5.3$_{-1.1}^{+1.8}$&27$_{-10}^{+11}$ \\
IC 5328&278\pms13&277\pms72&-37\pms12&-245\pms72&39.53\pms0.19&< 8.9 &> 2.2  \\
\hline
    \end{tabular}
    \label{tab:nlr_kin_lum_infos}
\end{table*}

The kinematic information of both sets of Gaussian functions that were used to fit the narrow components of the \nii $\lambda\lambda$6548, 6583, \halpha\, and \sii $\lambda$6716, 6731 lines were calculated using the peak position (radial velocity) and the FWHM of the fitted profiles. In addition, we corrected the observed FWHM values for the instrumental broadening effect by making $\mathrm{FWHM_{intrinsic}^2 = FWHM_{observed}^2 - FWHM_{instrumental}^2}$, where $\mathrm{FWHM_{instrumental}}$ corresponds to the spectral resolution of the data cubes. The results are shown in Table \ref{tab:nlr_kin_lum_infos}. All radial velocities are presented with respect to the heliocentric velocity of the sample galaxies, shown in Table \ref{tab:sample_galaxies}. Although these results may be associated with gaseous discs or non-keplerian motions, such as inflows or outflows, a proper interpretation of the kinematics of the nebular structure of the sample ETG will be done in a future analysis of the whole FOV of the observations.

For the galaxies where we fitted the profiles of the \oi $\lambda\lambda$6300, 6363 doublet, the kinematic results regarding these lines are shown in Table \ref{tab:fitted_oi_infos}. One may see that the narrow components of this doublet have the same kinematics as the narrow components of the \nii $\lambda\lambda$6548, 6583, \halpha\, and \sii $\lambda$6716, 6731 lines in NGC 2217, NGC 5044 and IC 1459. This indicates that a fraction of the \oi $\lambda\lambda$6300, 6363 doublet is emerging from the narrow line regions (NLR) of these objects.

\begin{table*}
    \centering
    \caption{Parameters of the fitted components of the \oi$\lambda\lambda$6300, 6363 doublet for the objects where a broad feature is seen in this feature. ``n$_1$'' and ``n$_2$'' correspond to the two sets used to describe the narrower component and ``b'' is related to the broad component of this doublet.}
    \begin{tabular}{cccccccc}
    \hline
    Galaxy & FWHM(\oi)$\mathrm{_{(n1)}}$ & FWHM(\oi)$\mathrm{_{(n2)}}$ & FWHM(\oi)$\mathrm{_{(b)}}$ & V$\mathrm{_{r}}$(\oi)$\mathrm{_{(n1)}}$ & V$\mathrm{_{r}}$(\oi)$\mathrm{_{(n2)}}$ & V$\mathrm{_{r}}$(\oi)$\mathrm{_{(b)}}$ & f(\oi)$\mathrm{_{(b)}}$ \\ 
    & (km s$^{-1}$) & (km s$^{-1}$) & (km s$^{-1}$) & (km s$^{-1}$) & (km s$^{-1}$) & (km s$^{-1}$) & (10$^{-15}$ erg s$^{-1}$ cm$^{-2}$) \\
    \hline
    NGC 2217 & 444\pms8 & -- & 1982\pms78  & -13\pms3 & -- & -54\pms29  & 24\pms2  \\
    NGC 5044 & -- & 1094\pms92 & 3088\pms187 & -- & -28\pms25 & 192\pms58  & 5.2\pms0.6 \\ 
    IC 1459 & 277\pms79 & 784\pms76 & 2720\pms294 & -77\pms18 & -58\pms14 & 197\pms105 & 14\pms3\\
    \hline
    \end{tabular}
    \label{tab:fitted_oi_infos}
\end{table*}

The electron density n$\mathrm{_e}$ was calculated with the \sii$\lambda$6716/\sii$\lambda$6731 ratio using the {\sc pyneb} code \citep{2015A&A...573A..42L}, assuming an electron temperature $\mathrm{T_e}$ = 10000 K. The results are presented in Table \ref{tab:nlr_kin_lum_infos}. In most cases, n$\mathrm{_e}$ $<$ 1000 cm$^{-3}$. It is worth mentioning that in AGNs, much of the \sii\, lines are formed in the partially ionized zone of the nebula. According to \citet{2020MNRAS.498.4150D}, objects that the \sii\, method provided values of n$\mathrm{_e}$ $<$ 1000 cm$^{-3}$ may have their electron density measurements underestimated by a factor of 2. 

Finally, the ionized gas mass M$\mathrm{_{ion}}$ of the NLR was estimated as

\begin{equation}
	\mathrm{M_{ion}=\frac{L(H\alpha)m_H}{\epsilon n_e}=\frac{2.2\times10^7L_{40}(H\alpha)}{n_e} M_\odot,}
	\label{eqionizedmass}
\end{equation}
where $\mathrm{L_{40}(H\alpha)}$ is the \halpha\, luminosity from the NLR in units of 10$^{40}$ erg s$^{-1}$ and $\mathrm{\epsilon}$ = 3.84$\times$10$^{-25}$ erg s$^{-1}$ cm$^3$ is the H$\alpha$ line emissivity (\citealt{2006agna.book.....O}, case B). The results are presented in Table \ref{tab:nlr_kin_lum_infos}. Again, because of the possible underestimation of the n$\mathrm{_e}$ measurements, the M$\mathrm{_{ion}}$ results may be higher than their real values by a factor of 2. We note that the mass of ionized gas has a range of 10$^2$ -- 10$^4$ M$_\odot$, which is in accordance with previous works in ETGs \citep{1986AJ.....91.1062P, 1996A&AS..120..463M}.

\subsection{Broad components \label{sec:blr}}

In 23 galaxies, we added a broad component to account for the line fluxes in the \halpha+\nii\, region. One way to check if this component comes from a BLR is to extract an image of the red wing of this feature from the gas cube. If the result is an unresolved source in the centre of the bulge, we may confirm that this object contains a visible BLR. This is the case for 16 objects. Thus, we propose that (29\pms6)\% of the ETGs in the local Universe have type-1 AGNs. For the other seven galaxies, the broad component is probably related to a bad star subtraction, since type K and M giant stars may generate a bump near \halpha\, that may mimic a broad emission line component \citep{1997ApJS..112..391H}. In Fig. C1 in Appendix C, available online, we present the radial profiles of the images from the red wing of the broad components of these 16 galaxies, together with the radial profile of the images from the stellar continuum, taken from the main data cubes. We note that the BLR profiles are narrower than the stellar profiles. Since the BLR is a spatially unresolved source, we may use these radial profiles to estimate the PSF of the data cubes of these 16 objects. The resulting FWHM for these PSFs are shown in Table \ref{tab:sample_galaxies} and the fitted Gaussian functions to the radial profiles are presented in Fig. C1. 

We found that in these 16 cases where a BLR is present, the FWHM of the broad component is higher than 2100 \kms, as shown in Table \ref{tab:blr_infos}. In addition, the radial velocity, the observed \halpha\, flux for the broad component [f(\halpha)$_b$] and the \halpha/\hbeta\, ratio in the four galaxies where a broad component is seen in \hbeta\,  are also shown in Table \ref{tab:blr_infos}. The \halpha\, luminosities for the broad components [L(\halpha)$_b$] that are shown in Table \ref{tab:blr_infos} were corrected for the reddening effects with the E(B-V) values obtained for the narrow components, even for the galaxies where \halpha/\hbeta\, is available, since this ratio may be affected by collisional effects in the BLR \citep{2006agna.book.....O}. This is the reason why the \halpha/\hbeta\, results for the BLR are much higher than the values of this ratio observed in the NLR. The total \halpha\, luminosities [L(\halpha), BLR + NLR], corrected for extinction effects (where possible), are shown in Table \ref{tab:nlr_kin_lum_infos}. Finally, the ratios between the \halpha\, luminosities from the BLR and from the NLR [(\halpha)$_b$/(\halpha)$_n$] are shown in Table \ref{tab:blr_infos}.

\begin{table*}
    \centering
    \caption{Flux and kinematic measurements taken from the broad component. The \halpha\, luminosities were corrected for the reddening effects with the E(B - V) values presented in Table \ref{tab:nlf_flux_infos}. One should have in mind that this correction may be underestimated, since we considered the colour excesses measured for the NLR.}    \begin{tabular}{ccccccc}
    \hline
Galaxy & f(\halpha)$_b$ & log L(\halpha)$_b$ & (\halpha)$_b$/(\halpha)$_n$ & FWHM$_b$(\halpha) & V$_{rb}$(\halpha) & (\halpha/\hbeta)$_b$  \\     
& (10$^{-15}$ erg s$^{-1}$ cm$^{-2}$) & (erg s$^{-1}$) & &(km s$^{-1}$)&(km s$^{-1}$)&  \\
    \hline
NGC 584&8.8\pms1.7&39.16\pms0.23&2.4\pms1.7 &3119\pms297&516\pms96&-- \\
NGC 1052&188\pms9&40.40\pms0.07& 0.88\pms0.19 &2550\pms40&-122\pms13&11\pms1\\
NGC 1411&6.8\pms0.4&38.47\pms0.06&4.2\pms1.6 &2125\pms69&259\pms30&--\\
NGC 1553&9.8\pms0.7&38.55\pms0.31&3.8\pms3.9 &2597\pms90&656\pms37&--\\
NGC 1574&10\pms1&39.35\pms0.13& 0.91\pms0.37&2489\pms222&1203\pms112&--\\
NGC 2217&106\pms2&40.24\pms0.23&2.3\pms1.7 &2688\pms29&215\pms10&12\pms1\\
NGC 2974&44\pms9&39.41\pms0.24&1.8\pms1.4 &2680\pms176&473\pms48&--\\
NGC 3115&0.076\pms0.014&35.93\pms0.24&0.45\pms0.36 &3194\pms214&758\pms91&--\\
NGC 3557&7.7\pms1.3&39.37\pms0.23&1.7\pms1.3 &3088\pms210&486\pms66&--\\
NGC 4105&5.9\pms1.7&38.82\pms0.06&1.1\pms0.3 &2950\pms391&567\pms148&--\\
NGC 4958&22\pms1&39.58\pms0.18&3.0\pms1.7 &2990\pms60&226\pms14&11\pms1\\
NGC 5044&22\pms2&39.66\pms0.11&2.0\pms0.7 &2972\pms89&436\pms18&--\\
NGC 7049&25\pms2&39.43\pms0.12&2.9\pms1.2 &2463\pms88&388\pms35&--\\
IC 1459&57\pms5&40.03\pms0.11&1.2\pms0.4 &3497\pms138&488\pms36&25\pms100\\
IC 4296&23\pms3&40.08\pms0.12&1.8\pms0.7 &2535\pms114&560\pms42&--\\
IC 5328&13\pms1&39.38\pms0.25&2.4\pms1.9 &3577\pms190&1101\pms104&--\\
\hline
    \end{tabular}
    
    \label{tab:blr_infos}
\end{table*}

For the galaxies where we detected broad components in the \oi $\lambda\lambda$6300, 6363 doublet, only in NGC 5044 the FWHM of this component is similar to the broad component found for \halpha, although the radial velocity of the \halpha\, line is higher than the one detected for the  \oi\, lines (see Table \ref{tab:fitted_oi_infos}). For both NGC 2217 and IC 1459, the broad \oi\, components are narrower and have lower radial velocity values than the broad features detected in \halpha. One possible interpretation for these components in the \oi $\lambda\lambda$6300, 6363 doublet is related to nuclear outflows, since shocks may enhance the \oi\, emission \citep{2006agna.book.....O,2008ApJS..178...20A}. Indeed, all three objects present pieces of evidence of outflowing gas, as stated by IFS data \citep{2015MNRAS.451.3728R, 2017MNRAS.470.1703D, 2017ApJS..232...11T}. We may also speculate that this component is emerging from the outermost region of the BLR, in a position where $\mathrm{n_e} \sim$ 10$^6$ cm$^{-3}$ (critical density of the \oi $\lambda\lambda$6300, 6363 doublet, \citealt{2006agna.book.....O}). However, a proper study of this component is beyond the scope of this paper.  

\subsection{Classification of the nuclear spectra \label{sec:diagnostic_diagrams}}

\begin{figure*}
    \centering
    \includegraphics[scale=0.4]{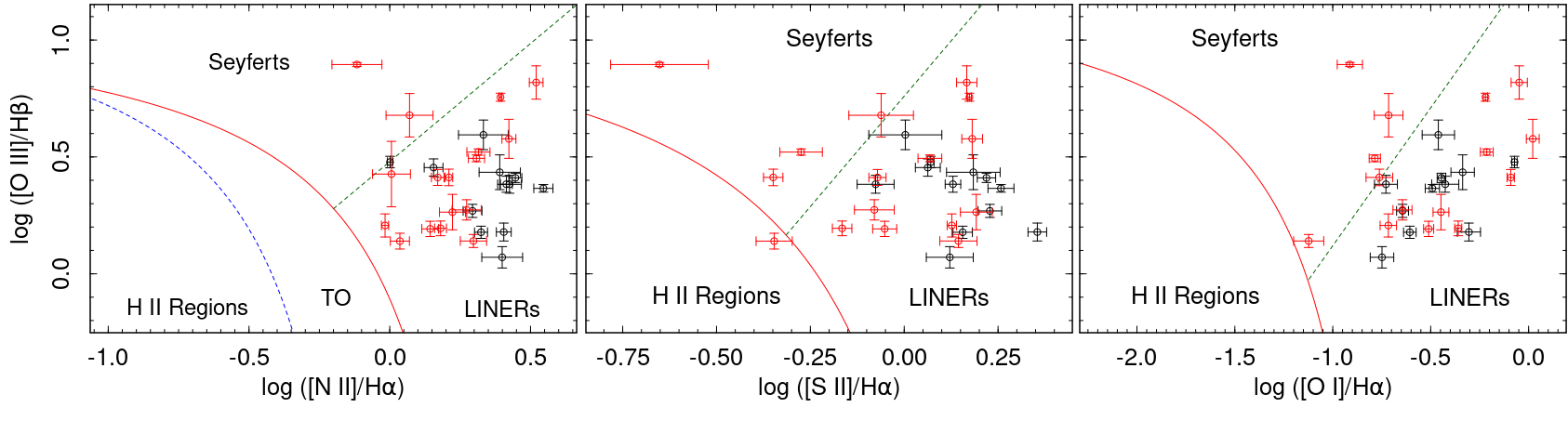}
    \caption{Diagnostic diagrams. The red circles are related to lenticular galaxies, while the black circles correspond to ellipticals. The blue dashed line is for the empirical division between H II regions and AGNs \citep{2003MNRAS.346.1055K}, the red solid line is for the maximum starburst line proposed by \citet{2001ApJ...556..121K} and the green dashed line is for the LINER–Seyfert division suggested by \citet{2010MNRAS.403.1036C} for the diagram with the \nii/\halpha\, ratio and \citet{2006MNRAS.372..961K} for the other diagrams. TO are Transition Objects.}
    \label{fig:diagnostic_diagrams_d3d}
\end{figure*}

To classify the nuclear emission from the ETG sample, we analysed the emission line ratios using BPT diagnostic diagrams \citep{1981PASP...93....5B,1987ApJS...63..295V}. We used the classification scheme proposed by \citet{2006MNRAS.372..961K}, which divides the objects as Seyferts, LINERs, \hii\, regions and Transition Objects (TO). Fig. \ref{fig:diagnostic_diagrams_d3d} shows three plots comparing \oiii/\hbeta\, with \nii/\halpha, \sii/\halpha\, and \oi/\halpha. Elliptical and lenticular galaxies are presented with different colours. Of the 48 nuclei with emission lines, only the 29 objects with measured \oiii/\hbeta\, ratio were classified. In Table \ref{tab:nuc_classification}, we show three classifications for these 29 objects. The first comes from the BPT with the \nii/\halpha\, ratio, the second one is from the BPT with the \sii/\halpha\, ratio and the third is from the BPT with the \oi/\halpha\, ratio.

\begin{table*}
    \centering
    \caption{Classification of the nuclear spectra of the galaxies. We also inform if the objects have a broad component (taken from Table \ref{tab:blr_infos}), an X-ray core (taken from \citealt{2017ApJ...835..223S} and \citealt{2020ApJ...900..124B}), or the coronal [Ne {\sc v}] emission in the MIR range (taken from \citealt{2013MNRAS.432..374R}). For the nuclear type: L = LINER; S = Seyfert; H = \hii\, region; AbN = Pure absorption line nucleus; WELO = weak emission line object.}
    \begin{tabular}{ccccc} 
    \hline
    Galaxy & Nuclear type & Broad component & X-ray core & MIR [Ne {\sc v}] emission \\
    \hline
NGC 584& L/L/L&yes&yes&no\\
NGC 596&WELO&no&no&no\\
NGC 720&WELO&no&yes&no\\
NGC 936& L/L/L&no&yes&no\\
NGC 1052& L/L/L&yes&yes&no\\
NGC 1201&AbN&--&no&no\\
NGC 1316&L/L/S&no&yes&no\\
NGC 1326& L/L/L&no&no&no\\
NGC 1332& L/H/--&no&yes&no\\
NGC 1344&AbN&--&no&no\\
NGC 1380& L/L/L&no&yes&no\\
NGC 1387&WELO&no&yes&no\\
NGC 1395&WELO&no&yes&no\\
NGC 1399&WELO&no&yes&no\\
NGC 1404&WELO&no&yes&no\\
NGC 1407&AbN&--&yes&no\\
NGC 1411&L/L/L&yes&no&no\\
NGC 1427&AbN&--&yes&no\\
NGC 1527&AbN&--&no&no\\
NGC 1533&L/L/L&no&no&no\\
NGC 1537&WELO&no&no&no\\
NGC 1543&WELO&no&no&no\\
NGC 1549&AbN&--&no&no\\
NGC 1553&L/L/L&yes&yes&yes\\
NGC 1574&L/S/S&yes&no&no\\
NGC 1700&WELO&no&yes&no\\
NGC 1947&L/L/L&no&no&no\\
NGC 2217&L/L/L&yes&yes&no\\
NGC 2784&L/--/--&no&no&no\\
NGC 2974&L/L/L&yes&no&yes\\
NGC 3115&WELO&yes&yes&no\\
NGC 3557&L/L/L&yes&no&yes\\
NGC 3585&WELO&no&yes&no\\
NGC 3904&AbN&--&no&no\\
NGC 3923&AbN&--&yes&no\\
NGC 3962&L/L/L&no&no&yes\\
NGC 4105&L/L/--&yes&no&no\\
NGC 4546&L/S/L&no&no&no\\
NGC 4696&L/L/L&no&yes&no\\
NGC 4697&WELO&no&yes&no\\
NGC 4753&WELO&no&yes&no\\
NGC 4958&L/L/L&yes&no&no\\
NGC 4984&S/S/S&no&no&no\\
NGC 5018&WELO&no&yes&no\\
NGC 5044&L/L/L&yes&yes&yes\\
NGC 5061&WELO&no&no&no\\
NGC 5102&WELO&no&yes&no\\
NGC 5128&S/S/S&no&yes&yes\\
NGC 6684&WELO&no&no&no\\
NGC 6868&L/L/L&no&yes&yes\\
NGC 7049&L/L/S&yes&yes&no\\
NGC 7144&WELO&no&no&no\\
NGC 7507&WELO&no&yes&no\\
IC 1459&L/L/L&yes&yes&yes\\
IC 4296&L/L/L&yes&yes&yes\\
IC 5328&L/L/L&yes&no&no\\    
\hline
    \end{tabular}
    \label{tab:nuc_classification}
\end{table*}

We found that (52\pms7)\% of the sample ETGs are classified as LINERs or Seyfert galaxies. No Transitions Objects are seen among the objects. Only one galaxy (NGC 1332) has a nucleus that may be classified as an \hii\, region using the \oiii/\hbeta\, $\times$ \sii/\halpha\, diagram. However, within the errors of the line ratios, this object may have a LINER or even a Seyfert nucleus. One may also note that one object is a bona-fide Seyfert galaxy (NGC 4984). 

For the other 19 galactic nuclei without measured \oiii/\hbeta\, ratios, we note that their EW(\halpha) $<$ 1.6 \AA. According to \citet{2010MNRAS.403.1036C}, galaxies with such low values of EW(\halpha) are usually retired galaxies, i.e. they have LINER-like spectra, but their ionization sources are associated with HOLMES instead of LLAGNs. In this paper, we will designate those objects as ``weak emission line objects'' (WELO) in Table \ref{tab:nuc_classification} to separate them from the other galactic nuclei that were classified using the BPT diagrams, since a fraction of these galaxies may host a LLAGN, as will be discussed in Section \ref{sec:ionization_sources}. WELOs correspond to 34\pms6\% of all sample ETGs. 

\section{Discussion} \label{sec:discussion}

\subsection{Comparison with the results from the Palomar Survey} \label{sec:comparison_palomar}

The sample selection of the \divingTD project is quite similar to the Palomar Survey. Thus, a comparison of the results obtained with both studies is straightforward. To do so, we first selected all ETGs brighter than 12.0 mag (B band) and |b| $>$ 15\degree from the Palomar Survey that belongs to the Northern Hemisphere\footnote{A few galaxies from the Palomar Survey are from the Southern Hemisphere, (e.g. NGC 1052)}. This selection resulted in 78 ETGs. We obtained all original spectra from the Palomar Survey, available in the NASA Extragalactic Database (NED). There are two spectra per object, one with a spectral coverage of $\sim$ 4230--5110 \AA\ and a spectral resolution of 4 \AA, and the other covering a spectral range of $\sim$6210--6860 \AA\ with a resolution of $\sim$ 2.5 \AA\ (see \citealt{1995ApJS...98..477H} for more details on the observations of the galaxies from the Palomar Survey). We then applied the same methodologies we used in the analysis of the nuclear spectra from the \divingTD sample, described in Sections \ref{sec:spectral_synthesis} and \ref{sec:el_properties}. The reason for proceeding this way is that the subtraction of the stellar component performed by \citet{1997ApJS..112..315H} was done by using a combination of pure absorption-line galaxies. There are two issues here: one is to discard any differences in the comparison with the \divingTD sample that may appear due to distinct stellar subtraction procedures. The other is that one of the galaxies that \citet{1997ApJS..112..315H} used to build the starlight templates is NGC 3115, which contains emission lines in its spectrum, as shown in Fig. B1, on Appendix B, and in \citet{2014ApJ...796L..13M}. 

\begin{figure}
    \includegraphics[scale=0.4]{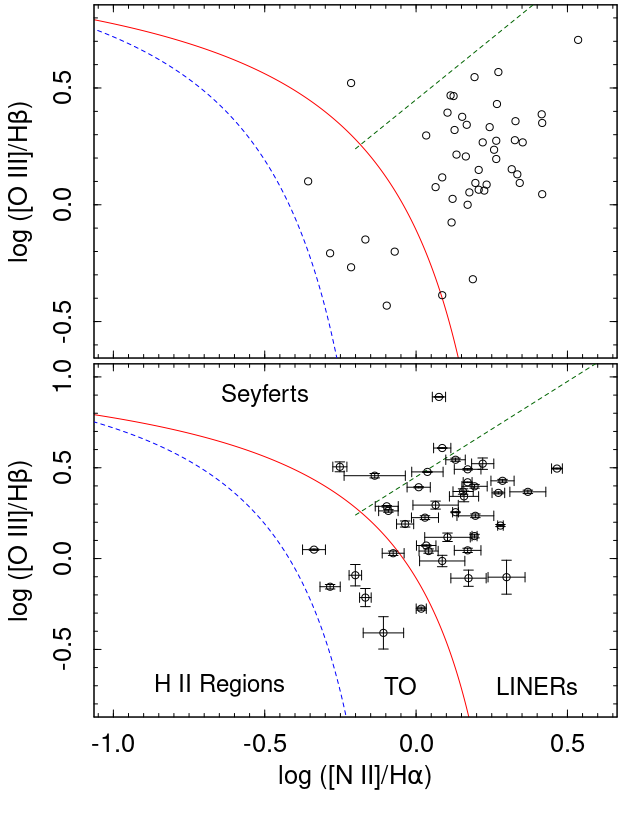}
    \caption{Diagnostic diagrams for the ETGs from the Palomar Survey. Top panel: Line ratio measurements taken from \citet{1997ApJS..112..315H}. Bottom panel: measurements obtained in this work (see the main text for more details). }
    \label{fig:palomar_bpt}
\end{figure}

The spectral synthesis procedures were performed twice for each galaxy from the Palomar Survey, one for the blue spectrum and the other for the red spectrum of each galaxy. The reason is that the spectral resolution is different in both wavelength ranges. It is worth mentioning that for these spectra, the use of multiplicative Legendre polynomials was not necessary for the SSP fitting procedure because of the narrower spectral coverage. We then fitted the line profiles by following the same procedures discussed in Section \ref{sec:el_properties}. We found that 47 ETGs from the Palomar Survey have emission lines in their nuclei, which is the same result reported by \citet{1997ApJ...487..568H}. In 40 objects, we were also able to measure the \oiii/\hbeta\, ratios.  The BPT diagram comparing both \nii/\halpha\, and \oiii/\hbeta\, line ratios is shown in the bottom panel of Fig. \ref{fig:palomar_bpt}. In the top panel of Fig. \ref{fig:palomar_bpt}, we show the same BPT diagram with the line ratio measurements of \citet{1997ApJS..112..315H}. One may see that the distributions of the galaxies in both BPT diagrams are very similar, which indicates that the differences in the analysis procedures do not affect the main statistical results obtained for the Palomar Survey.

Table \ref{tab:nuclei_classification} shows a comparison of the detection rates and classifications of the galactic nuclei between both samples. The errors of the detection rates were calculated using a binomial distribution, taking into account both sample sizes. All results presented for the Palomar Survey are based on our measurements, except for the Type 1 AGNs, whose detection number (11) is based on the results found by \citet{1997ApJS..112..391H}, including those detected with confidence (seven) and those detected with less certainty (four). The detection rates are similar, within the errors, for LINERs, Seyferts and \hii\, regions. However, Transitions Objects are seen in galaxies from the Palomar Survey, but not in the \divingTD sample. If Transition Objects are related to a composite LINER/\hii\, spectrum \citep{1993ApJ...417...63H, 2006MNRAS.372..961K}, then an explanation for this result is that the observed spectra for the Palomar Survey have an effective aperture that was either 1$\times$4 arcsec$^2$ or 2$\times$4 arcsec$^2$, while the aperture of the nuclear spectra extracted from the galaxies of the \divingTD sample is equal to the FWHM of the PSF of the data cubes, which is $<$ 1 arcsec for most objects. In other words, the larger aperture of the Palomar Survey observations increases the probability of contamination by a circum-nuclear \hii\, region. This hypothesis was tested using the mini-\divingTD sample in paper II, where the number of Transitions Objects increased by a factor of two when the spectra of the objects from the mini-\divingTD sample were extracted from the data cubes with the same aperture as the galaxies from the Palomar Survey. 

\begin{table*}
    \centering
    \caption{Classification of the nuclear spectra for the ETGs from the \divingTD Project and the Palomar Survey. }
    \begin{tabular}{ccccc}
        \hline
        Classification & All ETGs \divingTD & All ETGs \divingTD & All ETGs Palomar & All ETGs Palomar\\
        & Number & \% & Number & \% \\
        \hline
        LINERs& 27 & 48\pms7 & 29 & 37\pms5 \\
        Seyferts& 2 &4\pms2  & 4 & 5\pms2 \\
        Transition objects & 0 & 0 & 7 & 9\pms3 \\
        \hii regions & 0 & 0 & 0 & 0  \\
        Weak emission line objects & 19 &34\pms6 & 7 & 9\pms3 \\
        Emission line nuclei & 48 &86\pms5 & 47 & 60\pms5  \\
        Pure absorption line nuclei & 8 &14\pms5  & 31 & 40\pms5 \\
        Type 1 AGNs & 16 & 29\pms6 & 11 & 14\pms4 \\
        All & 56 & 100 & 78 & 100 \\
        \hline
    \end{tabular}
    \label{tab:nuclei_classification}
\end{table*}

Other observed differences between both samples are the detection rate of emission lines and the fractions of pure absorption line nuclei and type 1 AGNs. In all cases, this is related to the fact that the data from the \divingTD sample were obtained using a larger telescope and a more modern instrument when compared to the Palomar Survey observations. In addition, we were able to isolate the nuclear region of the galaxies in a more efficient way, which increases the S/N of the nebular emission with respect to the starlight. Even the higher fraction of WELOs is associated with this reason. Since the detection limit of nebular emission in the Palomar Survey is higher, when the \nii+\halpha\, lines are observed, usually one may also see the \hbeta\, and the \oiii$\lambda$5007 lines.

\subsection{Ionization sources} \label{sec:ionization_sources}

Since a LINER-like spectrum may be produced in different ways, such as LLAGNs \citep{1983ApJ...269L..37H, 1983ApJ...264..105F}, shocks \citep{1980A&A....87..152H, 2008ApJS..178...20A} or HOLMES \citep{1994A&A...292...13B, 2008MNRAS.391L..29S}, a careful investigation is needed to indicate the most likely ionization source. Table \ref{tab:nuc_classification} shows all sample ETGs with detected broad H$\alpha$ emission (Section \ref{sec:blr}), X-ray cores (from \citealt{2017ApJ...835..223S} and \citealt{2020ApJ...900..124B}), and detected [Ne {\sc v}] lines in MIR spectra (from \citealt{2013MNRAS.432..374R}). Of all galaxies with LINER-like nuclei (including the WELOs), only 11 do not have a BLR, X-ray cores or coronal lines. If EW(\halpha) is also used as a proxy for nuclear activity \citep{2011MNRAS.413.1687C}, then the galaxies NGC 1533 and NGC 4546 also have AGNs, since both have EW(\halpha) $>$ 6 \AA. In addition, NGC 1326 and NGC 1947 also have EW(\halpha) values that may indicate the presence of a weak AGN. Thus, only seven emission line nuclei do not have a clear sign of a non-stellar ionizing source. 

It is worth mentioning that LINERs with low EW(\halpha) values do not exclude the existence of a LLAGN. At least eight galaxies with EW(\halpha) $<$ 2 \AA\, possess a broad H$\alpha$ component. Even three pure absorption line nuclei have X-rays cores (NGC 1407, NGC 1427 and NGC 3923). A possible interpretation is that not all ionizing photons emitted by the LLAGNs are being absorbed in the neighbourhood of the SMBH. It may happen if the central region of these galaxies has a high gas porosity \citep{2013A&A...555L...1P,2016A&A...588A..68G}. However, one must bear in mind that HOLMES may provide a number of ionizing photons that is enough to produce EW(\halpha) values between 1.5 and 2.5 \AA\, \citep{2011MNRAS.413.1687C}. Thus, there is a possibility that only a small fraction of the narrow \halpha\, emission in nuclei with low EW(\halpha) values is indeed caused by nuclear activity. 

\begin{figure*}
    \centering
    \includegraphics[scale=1.2]{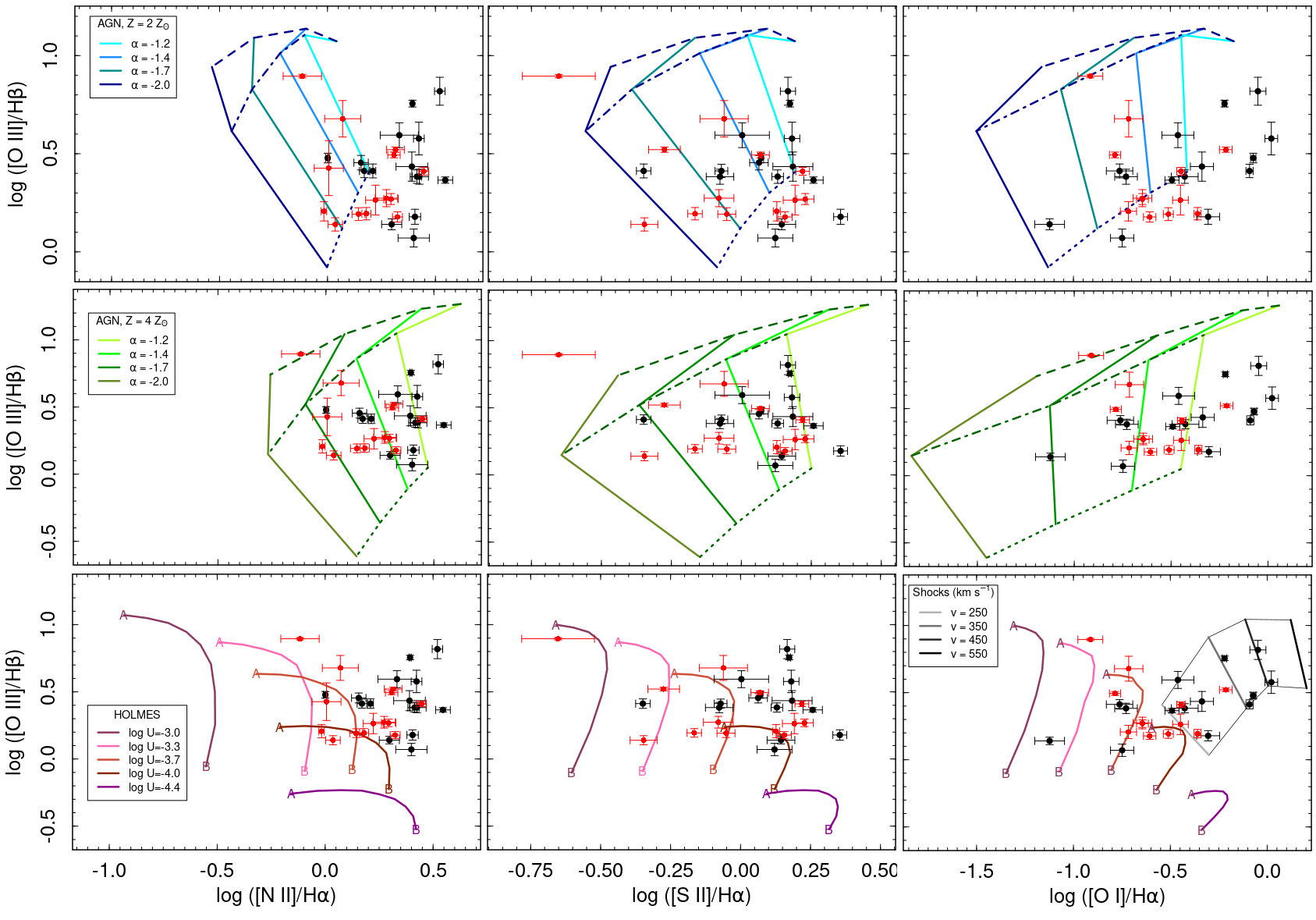}
    \caption{Same diagnostic diagrams as in Fig. \ref{fig:diagnostic_diagrams_d3d}, but separated by type 1 (black circles) and type 2 (red circles) objects. Grids with dusty AGN models \citep{2004ApJS..153....9G} are shown for a metallicity Z = 2$\mathrm{Z_\odot}$ (top panels) and for Z = 4$\mathrm{Z_\odot}$ (central panels). The power-law indexes ($\mathrm{F_\nu \propto \nu^\alpha)}$ are shown in the plots, and the dotted lines are for log U = -4.0, the dot-dashed lines are for log U = -3.0 and the dashed lines are for log U = -2.0. In the bottom panels, we present the grids with models of photoionization by hot old stellar populations (HOLMES, \citealt{2008MNRAS.391L..29S}), and fast shock models without a precursor with solar abundances and $\mathrm{n_e}$ = 1000 cm$^{-3}$ \citep{2008ApJS..178...20A}. For the HOLMES models, log U are shown in the plots, and each curve contains the predicted line ratios for metallicities between 1.0 (marked as ``A'') and 6.3 Z$_\odot$ (marked as ``B''). For the fast shock models, the velocities are shown in the plots, and the thin lines are related to magnetic fields of 100 (bottom-right line) and 316 $\mu$G (top-left line).}
    
    \label{fig:bpt_photoionization_models}
\end{figure*}

In Fig. \ref{fig:bpt_photoionization_models}, we show the same BPT diagrams as in Fig. \ref{fig:diagnostic_diagrams_d3d}, but with type 1 AGNs marked as black circles, while the galaxies without a BLR are marked as red circles. We also inserted predicted line ratios from three different models: dusty AGNs with metallicities of 2 and 4 $\mathrm{Z_\odot}$, a hydrogen density of $\sim$ 1000 cm$^{-3}$, ionization parameters -4.0 $\leq$ log U $\leq$ -2.0, and power-law index -2.0 $\leq$ $\alpha$ $\leq$ -1.2 \citep{2004ApJS..153....9G}; photoionization models by HOLMES with metallicities 1.0 $\leq$ Z/$\mathrm{Z_\odot}$ $\leq$ 6.3, a gas density of 500 cm$^{-3}$ and ionization parameters -4.4 $\leq$ log U $\leq$ -3.0 \citep{2008MNRAS.391L..29S}; fast shock models without a precursor with solar abundances, $\mathrm{n_e}$ = 1000 cm$^{-3}$, velocities between 250 and 550 \kms and magnetic fields between 100 and 316 $\mu$G \citep{2008ApJS..178...20A}. In the \oiii/\hbeta\, versus \nii/\halpha\, diagram, one may see that most of the type 1 objects and three nuclei without a broad component (NGC 1316, NGC 4546 and NGC 6868) have line ratios that may be only explained by the dusty AGN models. A similar result is seen in the \oiii/\hbeta\, versus \sii/\halpha\, diagram.  However, the \oiii/\hbeta\, versus \oi/\halpha\, diagram reveals that six objects (NGC 1411, NGC 1553, NGC 1052, NGC 4958, NGC 4546 and NGC 2217) have \oi/\halpha\, ratios that require fast shock models. It is worth noticing that both $\mathrm{n_e}$ and line widths measurements for the NLR of the sample galaxies are in accordance with the shock model limits presented in Fig. \ref{fig:bpt_photoionization_models}. 

\subsection{Comparison between elliptical and lenticular galaxies} \label{sec:properties_E_S0}

Of the 56 ETGs in the \divingTD sample, 30 are ellipticals and 26 are lenticular galaxies. Table \ref{tab:comparison_classification_E_S0} shows the nuclear classification and detection rates for both morphological types. We used the \nii/\halpha\, based BPT in Table \ref{tab:comparison_classification_E_S0}, since most of the galaxies have LINER-like nuclei within the errors even in the other two BPT diagrams. The exceptions are NGC 4984 and NGC 5128; both objects have bona fide Seyfert spectra. The errors of the detection rates were calculated using a binomial distribution, thus it reflects the sample sizes for both morphological types. The most significant difference lies in the LINER+Seyfert classification. Lenticular galaxies have (65\pms9)\% of their nuclei with such types. It is a 2$\sigma$ difference when compared to the fraction of the same nuclear types in ellipticals [(40\pms9)\%]. The difference between the detection rates of emission lines is also slightly higher in lenticulars: (92\pms5)\% and (80\pms7)\% for S0 and E galaxies, respectively. It corresponds to a 1.4$\sigma$ difference. Other works in the literature have also reported that lenticular galaxies are more likely to have emission lines than ellipticals \citep{2006MNRAS.366.1151S,1996A&AS..120..463M}, although \citet{1986AJ.....91.1062P} and \citet{2010A&A...519A..40A} did not find any difference in the detection rate of nebular emission in both types. These results may be related to the fact that galaxies with larger bulge-to-disc ratios are more likely to have jet-dominated AGNs, with a weak optical emission \citep{2005MNRAS.362...25B,2014ARA&A..52..589H}. 

\begin{table}
    \centering
    \caption{Classification of the nuclear spectra separated by elliptical and lenticular galaxies. }
    \begin{tabular}{ccccc}
        \hline
        Classification  & E & E & S0 & S0 \\
        & Number & \% & Number & \% \\
        \hline
        LINERs  & 12 & 40\pms9 & 15 & 57\pms10\\
        Seyferts  & 0 & 0 & 2  & 8\pms5\\
        Weak emission line objects & 12 & 40\pms9  & 7 &27\pms9  \\
        Emission line nuclei & 24 &80\pms7  & 24 &92\pms5  \\
        Pure absorption line nuclei  & 6 &20\pms7  & 2 &8\pms5 \\
        Type 1 AGNs & 9 & 30\pms8 & 7 &27\pms9 \\
        All &  30 &100  & 26 &100  \\
        \hline
    \end{tabular}
    \label{tab:comparison_classification_E_S0}
\end{table}

We also compared the nebular properties of the nuclei from both morphological types, including their line ratios. We used the Mann-Whitney (MW) test to check if the distributions of the parameters are indeed different for E and S0 galaxies. Table \ref{tab:comparison_E_S0_properties} shows the medians of the parameters for the whole sample and is also separated by ellipticals and lenticular objects. It also shows the statistical p-value calculated with the MW test, which corresponds to the probability of rejecting the null hypothesis that the two distributions are drawn from the same parent population. Fig. \ref{fig:histograms_comparison} shows the distributions of EW(\halpha), E(B-V) and L(\halpha), separated by morphological types. 

\begin{table}
    \centering
    \caption{Median of the distributions of the emission line parameters, except for M$_B$, which is related to the median of the distribution of the absolute magnitudes (taken from Table \ref{tab:sample_galaxies}). We show the medians for all sample galaxies and also separated by elliptical and lenticular objects. The p-values were calculated using the Mann-Whitney (MW) test and they correspond to the probability of rejecting the null hypothesis that the distributions of elliptical and lenticular galaxies are drawn from the same parent population. }
    \begin{tabular}{ccccc}
    \hline
    Parameter & Median & Median & Median & p-value \\
              & All & E & S0 & MW test\\
    \hline
    $\mathrm{M_B}$ & -20.66 & -21.22 & -20.10 & 0.00002 \\
    \oiii/\hbeta & 2.57 & 2.42 & 2.58 & 0.370 \\
    \nii/\halpha & 1.88 & 2.15 & 1.53 & 0.033 \\
    \sii/\halpha & 1.01 & 1.18 & 0.87 & 0.136 \\
    \oi/\halpha & 0.33 & 0.34 & 0.31 & 0.760 \\
    log L(\halpha)  (erg s$^{-1}$) &38.94 & 38.96 & 38.85 & 0.813 \\ 
    E(B-V) & 0.45 & 0.41 & 0.48 & 0.310 \\
    EW(H$\alpha$) (\AA) & 1.29 & 0.77 & 1.69 & 0.273 \\
    
    \hline
    \end{tabular}
    
    \label{tab:comparison_E_S0_properties}
\end{table}

\begin{figure}
    \centering
    \includegraphics[scale=0.5]{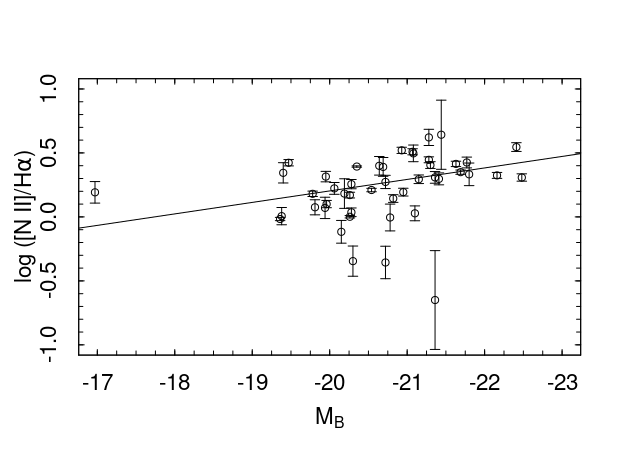}
    \caption{Plot of the \nii/\halpha\, line ratios versus the absolute magnitude of the galaxies. The solid line corresponds to the best-fit curve obtained using the Tukey's bisquare method.}
    \label{fig:correlation_nii_Mb}
\end{figure}

The only parameter that presents a significant difference between E and S0s is the \nii/\halpha\, ratio. However, the MW test also revealed that the distributions of absolute magnitudes of both morphological types are also different, with a p-value of 0.00002 (see also Fig. \ref{fig:hist_sample_parameters}). Thus, this difference in the \nii/\halpha\, ratio between both populations may be associated with the fact that, on average, the elliptical galaxies from the sample are more luminous (and, consequently, more massive) than the lenticular objects. A Kendall's correlation test shows that \nii/\halpha\, is indeed correlated with $\mathrm{M_B}$, with a p-value of 0.002 (here, the p-value corresponds to the probability of rejecting the null hypothesis that the two parameters are not correlated). Fig. \ref{fig:correlation_nii_Mb} compares the \nii/\halpha\, ratios with $\mathrm{M_B}$. A linear regression using the Tukey's bisquare method indicates that

\begin{equation}
    \mathrm{log([N\, {\sc II}]/H\alpha) = (-0.09\pm0.03)M_B - (1.6\pm0.6).} 
    \label{eq:nii_mb_relation}
\end{equation}
The correlation between \nii/\halpha\, and $\mathrm{M_B}$ was also seen by \citet{1986AJ.....91.1062P} using long-slit spectra. Since \nii/\halpha\, is sensitive to metallicity changes in the ionized gas \citep{1994ApJ...429..572S,2002MNRAS.330...69D,2004ApJS..153....9G,2019ARA&A..57..511K}, one possibility is that this correlation may arise because of the so-called mass-metallicity relation in galaxies \citep{1979A&A....80..155L, 2004ApJ...613..898T, 2009MNRAS.396L..71V}. An alternative scenario is that more luminous galaxies may have AGNs with harder power-law continua. Fig. \ref{fig:bpt_photoionization_models} indicates that the \nii/\halpha\, ratio is correlated with the power-law indexes. This difference in excitation conditions was also discussed as a possibility to explain the \nii/\halpha\, versus $\mathrm{M_B}$ relation by \citet{1986AJ.....91.1062P}.

\section{Conclusions} \label{sec:conclusion} 

In this work, we analysed the nebular emission from the nuclear region of all early-type galaxies brighter than 12.0 mag (B band) in the Southern Hemisphere and with galactic latitude |b| $< $ 15\degree. All 56 ETGs of this sample belong to the \divingTD project and they were observed using the GMOS-IFU, installed at both Gemini telescopes, which provided data cubes with seeing-limited angular resolutions, with a median value of 0.70 arcsec. Below, we summarise our main conclusions:

\begin{itemize}
    \item We detected emission lines in the nucleus of 48 galaxies of the sample. If the \divingTD survey is representative of the local Universe, since it is a statistically complete sample, then we propose that (86$\pm$5)\% of all nearby galaxies have nebular emission in their nuclei, while only (14\pms5)\% of the objects have pure absorption line nuclei. 
    \item Using BPT diagnostic diagrams, we found that (52$\pm$7)\% of the objects may be classified as LINERs or Seyfert galaxies. We did not detect any \hii regions or Transition Objects in this sample. The remaining (34$\pm$6)\% of the galaxies do not show H$\beta$ or \oiii\, lines in their spectra. Since these objects have low values of EW(\halpha), we classified them as weak emission line objects. 
    \item A broad line region, which is a typical signature of type 1 AGNs, is present in (29\pms6)\% of all ETGs. Their detection is based on a broad component seen in the H$\alpha$ line and also on the fact that they emerge from a spatially unresolved source located in the central region of the galaxies, as viewed in images taken from the red wing of the broad H$\alpha$ component.  
    \item A comparison between the elliptical and lenticular galaxies of the sample (respectively 30 and 26 objects) indicates that S0s have a higher incidence of emission lines in their nuclei [(92$\pm$5)\% in S0s against (80$\pm$7)\% in Es]. In addition, the fraction of galaxies classified as LINERs+Seyfert among the S0s is also higher than in Es [(65$\pm$9)\% in S0s against (40$\pm$9)\% in Es]. When it comes to nebular properties, we found that the \nii /H$\alpha$ ratio is slightly higher on elliptical galaxies (median values for S0 and Es are 2.15 and 1.53, respectively). Given that the elliptical galaxies are more luminous than the lenticular objects in the \divingTD sample, this result may be associated with the mass-metalicity relation of galaxies or to differences in the excitation conditions of the gas. A correlation between \nii /H$\alpha$ and the absolute magnitude of the galaxies also gives support to both scenarios.
    \item When we compare the results of the ETGs from the \divingTD project with the ETGs from the Palomar Survey \citep{1997ApJ...487..568H}, we found that the detection rate of emission line nuclei is higher in the \divingTD sample. Most of this difference is due to the higher incidence of weak emission line objects in the ETGs from the \divingTD project. In addition, the detection rate of type-1 AGNs is also higher among the ETGs from the \divingTD project. These results are a direct consequence of the fact that the observations of the \divingTD project are more sensitive to weak components, since they were performed with a more modern instrument with a better spatial resolution than the Palomar Survey observations. 
    
    \item Another important difference between both \divingTD and Palomar surveys is the number of Transition Objects. At least (9$\pm$3)\% of the Palomar ETGs belong to this class. Assuming that most Transition Objects are associated with a LINER emission with contamination from \hii regions, the use of IFU data with seeing-limited spatial resolution allowed us to avoid the light from possible star-forming regions in the circumnuclear environment. This result is in accordance with the findings presented in the mini \divingTD sample in paper II. 
    
    \item Of the 48 galaxies with emission line nuclei, 41 have signs of LLAGNs, as certified by the presence of a BLR, X-ray cores, the presence of coronal lines in the MIR or by high values of EW(\halpha). We note that even nuclei with low values of EW(\halpha) have signatures of AGNs, although it is not possible to assure what fraction of the emission from the narrow components is indeed caused by nuclear activity, since hot-old stellar populations, such as p-AGB stars, may provide enough ionizing photons to explain the nebular emission seen in these cases. In addition, some objects have also indications of shocks in their nuclear region, as revealed by high \oi/\halpha\, line ratios or by broad components seen in the \oi$\lambda\lambda$6300, 6363 doublet. 
     
    \end{itemize}

\section*{Acknowledgements}

Based on observations obtained at the Gemini Observatory (processed using the Gemini IRAF package), which is operated by the Association of Universities for Research in Astronomy, Inc., under a cooperative agreement with the NSF on behalf of the Gemini partnership: the National Science Foundation (United States), the National Research Council (Canada), CONICYT (Chile), the Australian Research Council (Australia), Minist\'{e}rio da Ci\^{e}ncia, Tecnologia e Inova\c{c}\~{a}o (Brazil) and Ministerio de Ciencia, Tecnolog\'{i}a e Innovaci\'{o}n Productiva (Argentina). This research has made use of the NASA/IPAC Extragalactic Database (NED), which is operated by the Jet Propulsion Laboratory, California Institute of Technology, under contract with the National Aeronautics and Space Administration. We also acknowledge the usage of the HyperLeda database (http://leda.univ-lyon1.fr). We thank Conselho Nacional de Desenvolvimento Cient\'ifico e Tecnol\'ogico (CNPq) for support under grants 306790/2019-0, 304584/2022-3 (TVR) and 306063/2019-0 (RBM). KSC thanks Coordena\c{c}\~{a}o de Aperfeiçoamento de Pessoal de N\'{i}vel Superior - Brasil (CAPES) for the financial support under the grant 88887.629089/2021-00. MDS thanks FAPERGS (PROBIC) and Programa Institucional de Inicia\c{c}\~{a}o
Cient\'ifica da UFFS (PRO-ICT). Finally, we thank Natalia Vale Asari for providing us with a table with the results of the photoionization models of \citet{2008MNRAS.391L..29S} and also the referee for hers/his valuable comments that improved the quality of the paper.  

\section*{Data Availability}

Further detail about the DIVING$^\mathrm{3D}$ survey can be found at https://diving3d.maua.br. The raw GMOS/IFU data are available at the Gemini Science Archive (https://archive.gemini.edu/searchform). The treated data cubes can be requested at diving3d@gmail.com.



\bibliographystyle{mnras}
\bibliography{bibliography} 




\appendix

\section{Nuclear spectra} \label{appendix:nuclear_spectra}

All spectra from the spaxel located at the peak of the stellar continuum of the sample galaxies are shown in Fig. \ref{fig:nuclear_spectra}. The top panels for each galaxy show the observed spectra in black and the spectral synthesis results in red. The bottom panels present the residual spectra. 

\begin{figure*}

\begin{center}

\includegraphics[scale=0.35]{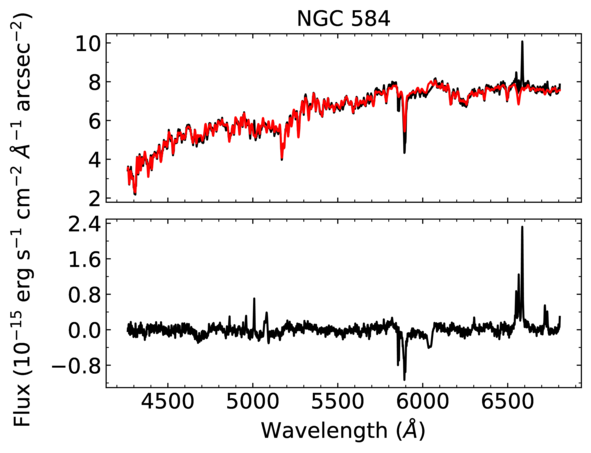}
\includegraphics[scale=0.35]{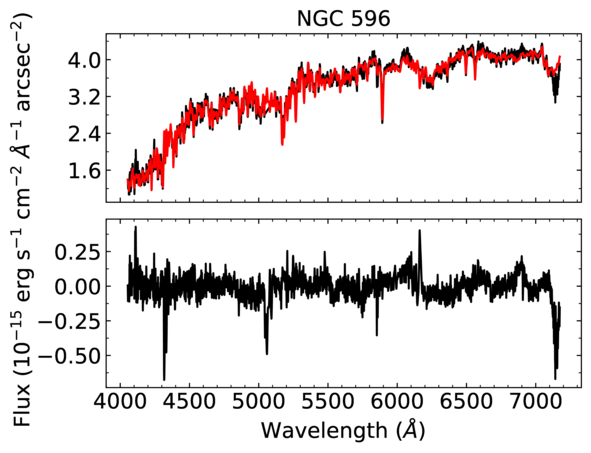}
\includegraphics[scale=0.35]{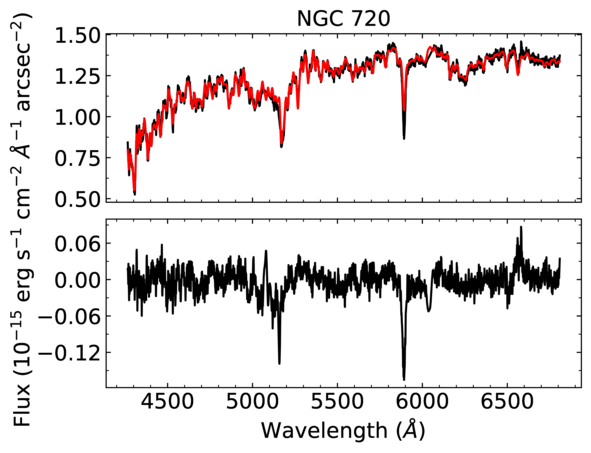}
\includegraphics[scale=0.35]{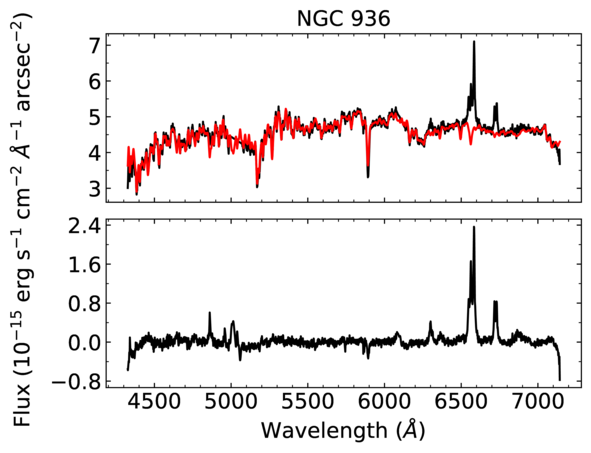}
\includegraphics[scale=0.35]{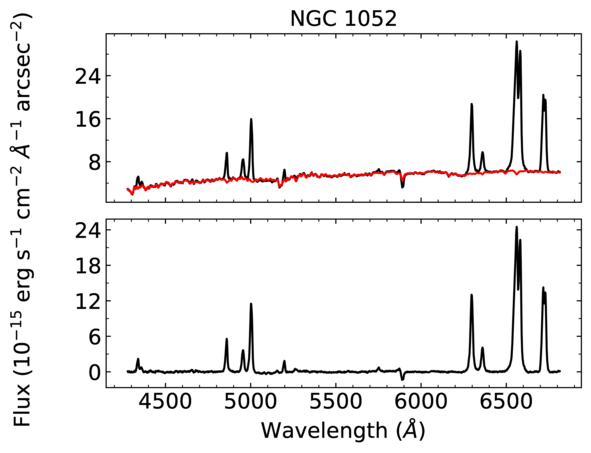}
\includegraphics[scale=0.35]{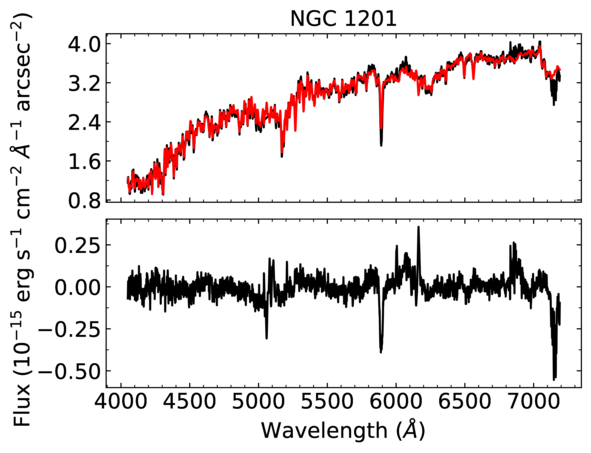}
\includegraphics[scale=0.35]{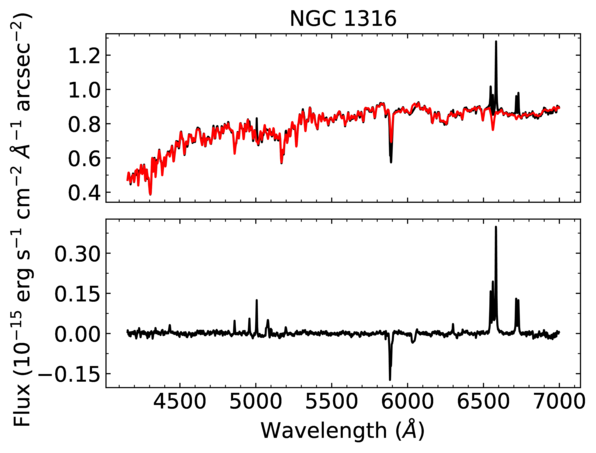}
\includegraphics[scale=0.35]{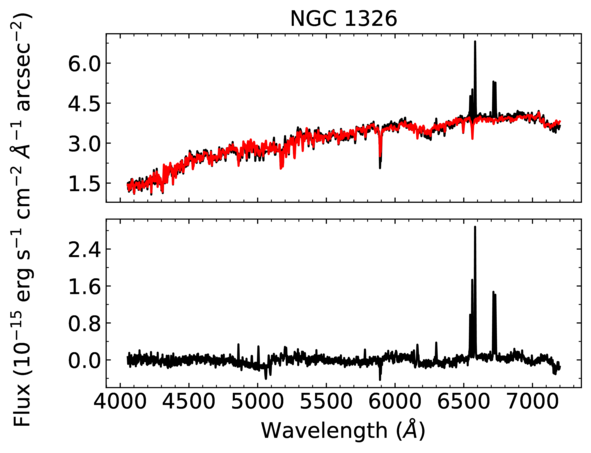}

\end{center}

\caption{Nuclear spectra of all sample galaxies, taken from the spaxels of the GMOS data cubes that are located at the peak of the stellar continuum of the objects. For each object, the top panel presents the observed spectrum (black) together with the spectral synthesis results (red), obtained with the {\sc ppxf} software. The bottom panels show the residual spectra. \label{fig:nuclear_spectra}}

\end{figure*}

\addtocounter{figure}{-1}

\begin{figure*}

\begin{center}

\includegraphics[scale=0.35]{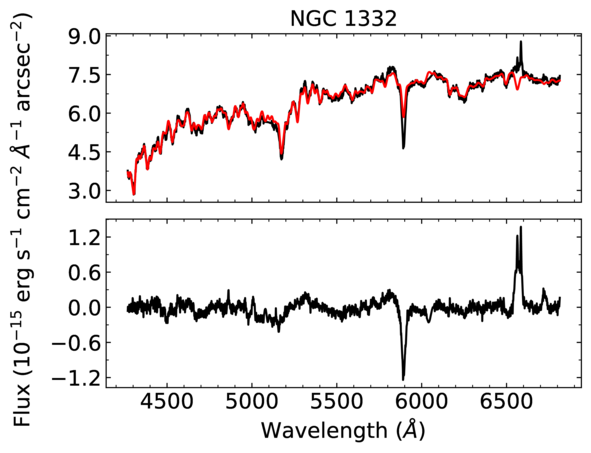}
\includegraphics[scale=0.35]{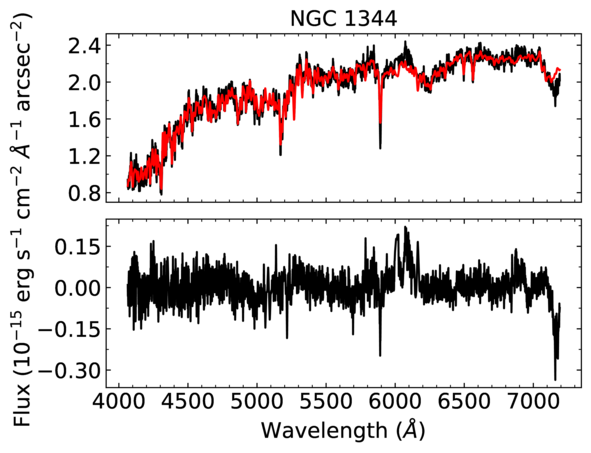}
\includegraphics[scale=0.35]{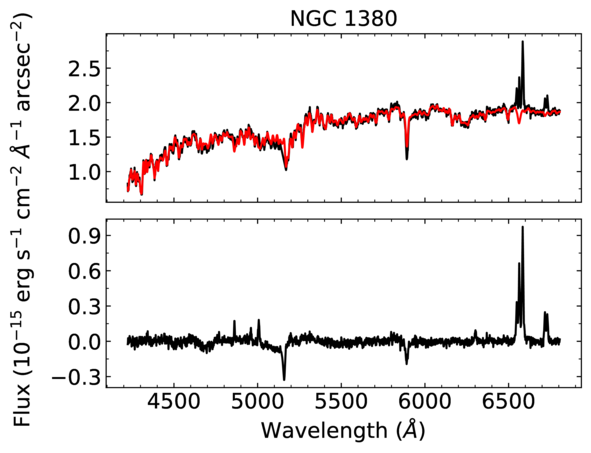}
\includegraphics[scale=0.35]{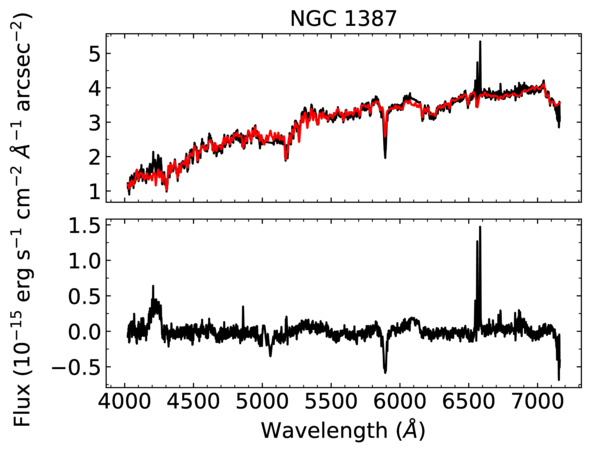}
\includegraphics[scale=0.35]{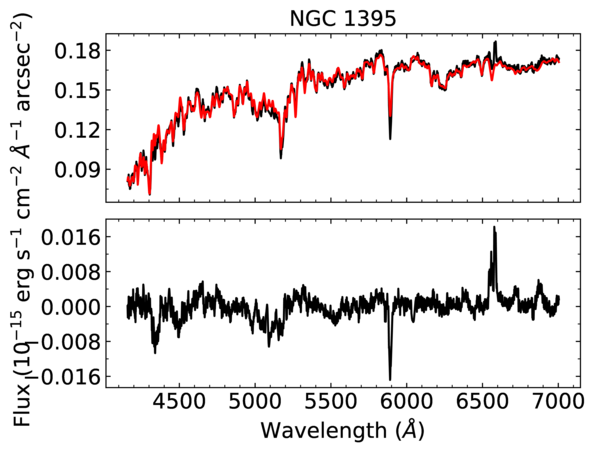}
\includegraphics[scale=0.35]{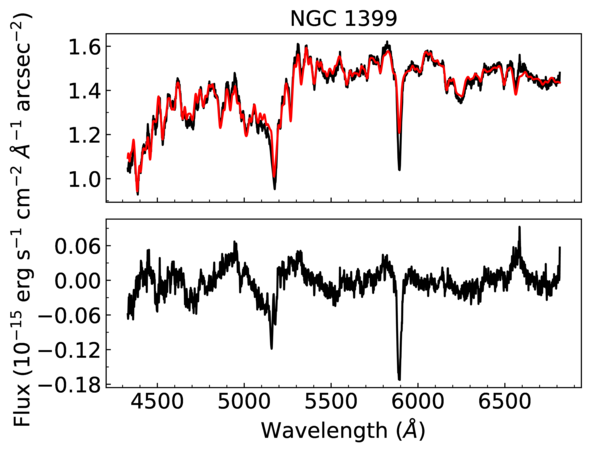}
\includegraphics[scale=0.35]{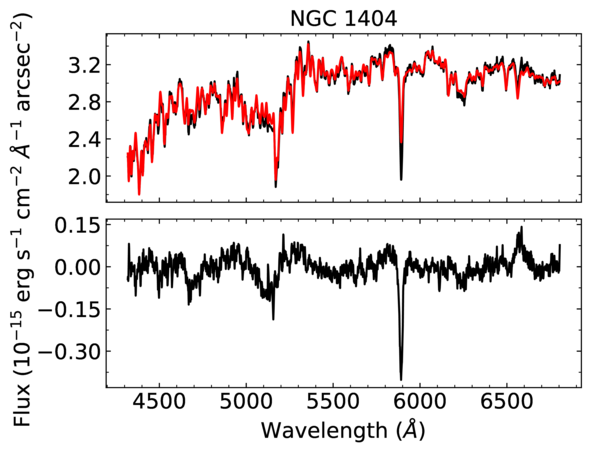}
\includegraphics[scale=0.35]{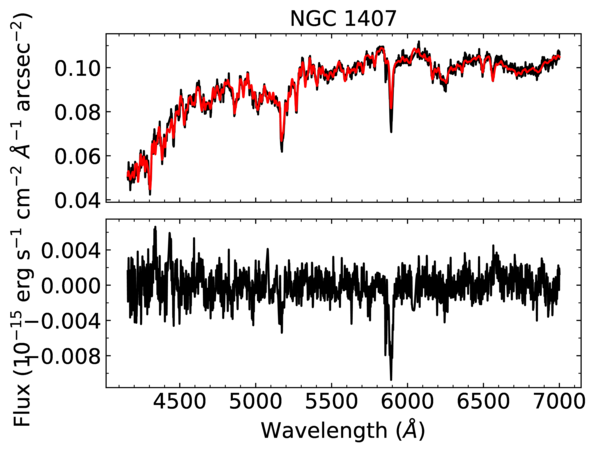}

\end{center}
\caption{--continued}

\end{figure*}

\addtocounter{figure}{-1}

\begin{figure*}

\begin{center}

\includegraphics[scale=0.35]{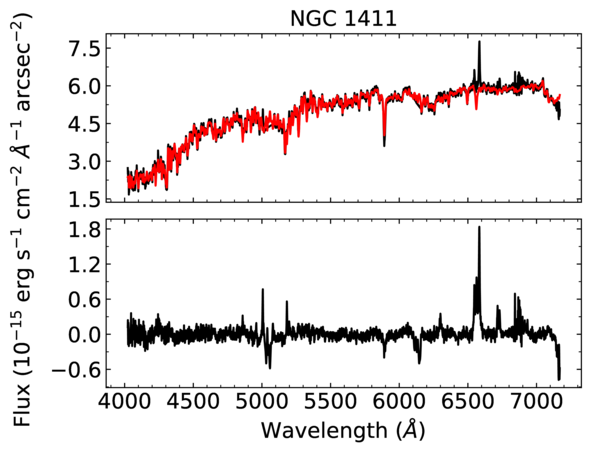}
\includegraphics[scale=0.35]{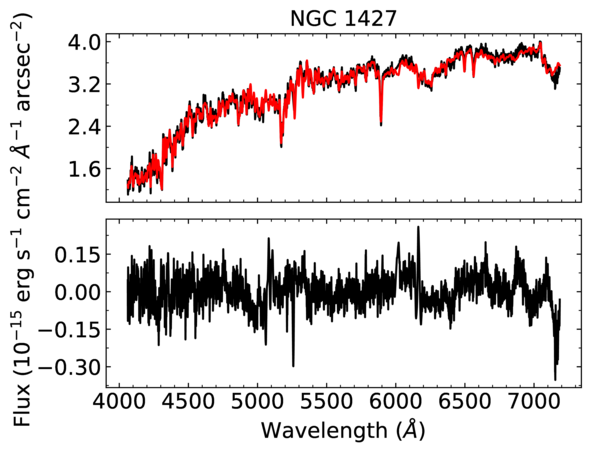}
\includegraphics[scale=0.35]{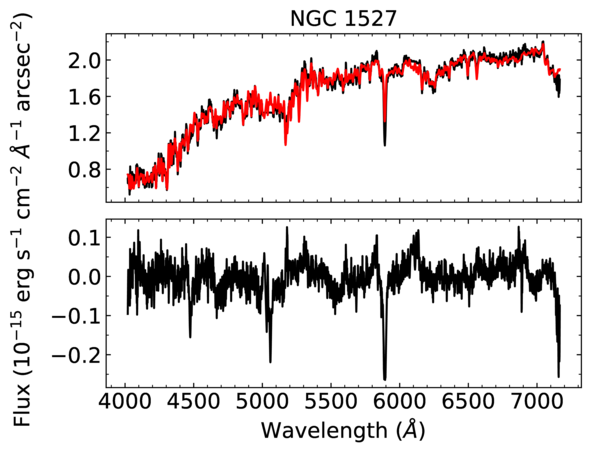}
\includegraphics[scale=0.35]{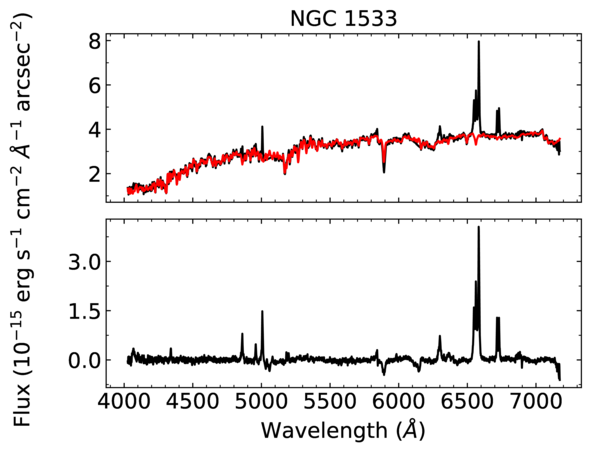}
\includegraphics[scale=0.35]{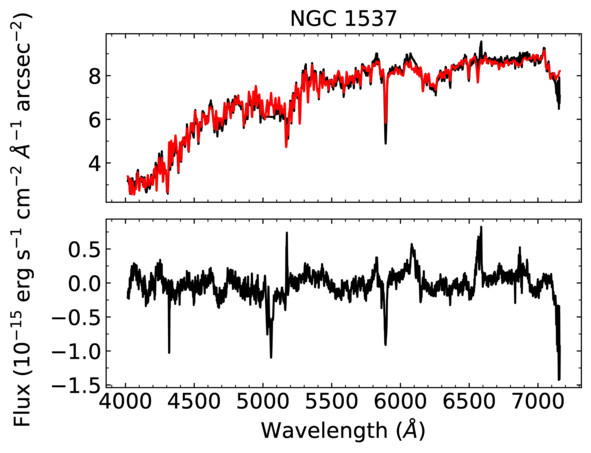}
\includegraphics[scale=0.35]{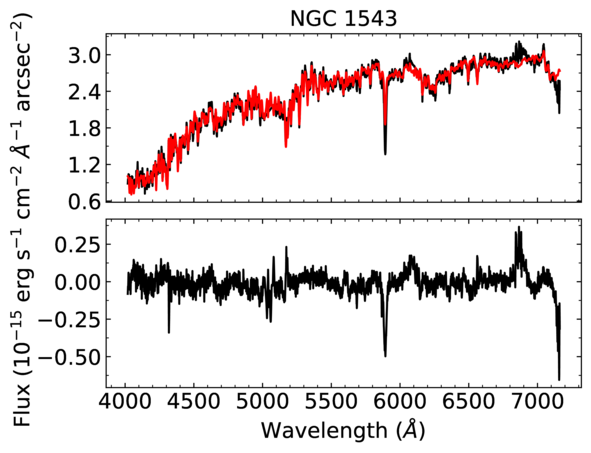}
\includegraphics[scale=0.35]{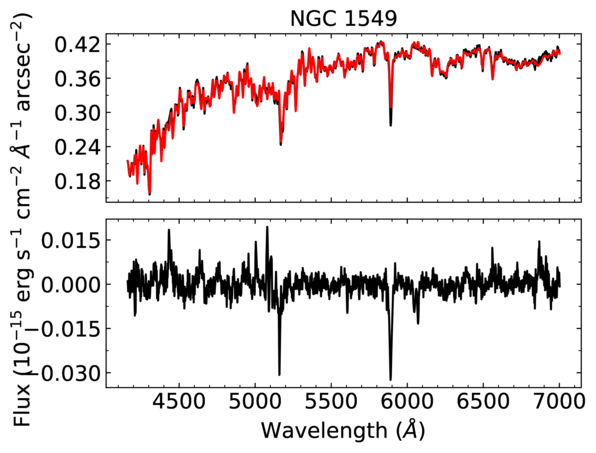}
\includegraphics[scale=0.35]{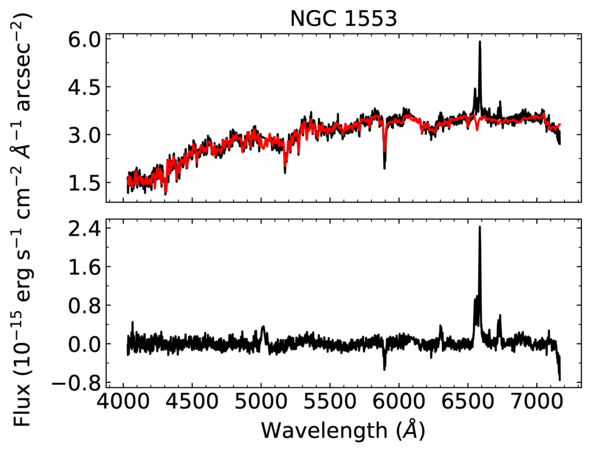}

\end{center}
\caption{--continued}

\end{figure*}

\addtocounter{figure}{-1}

\begin{figure*}

\begin{center}

\includegraphics[scale=0.35]{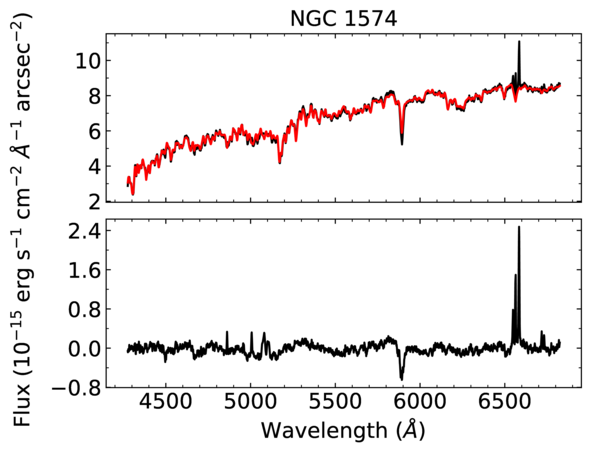}
\includegraphics[scale=0.35]{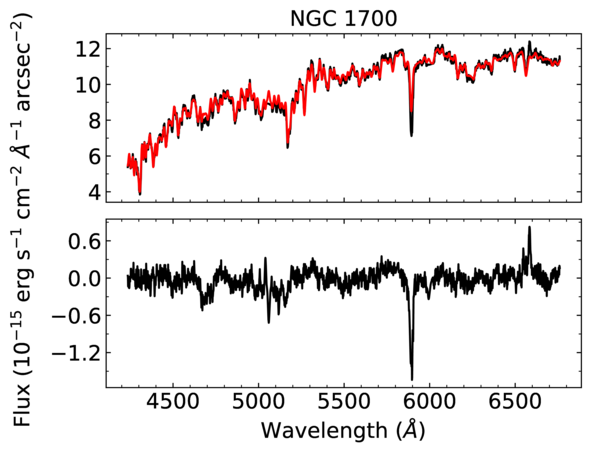}
\includegraphics[scale=0.35]{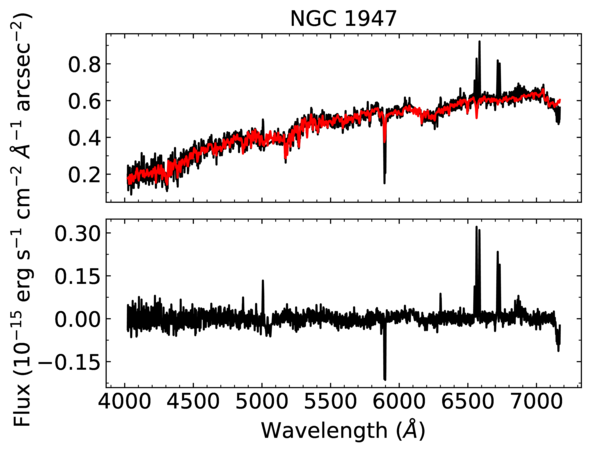}
\includegraphics[scale=0.35]{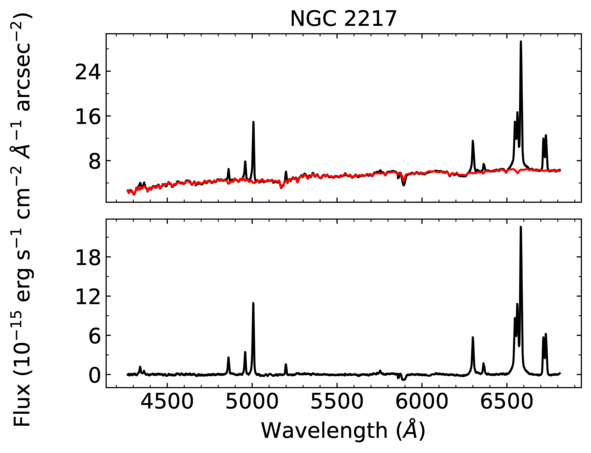}
\includegraphics[scale=0.35]{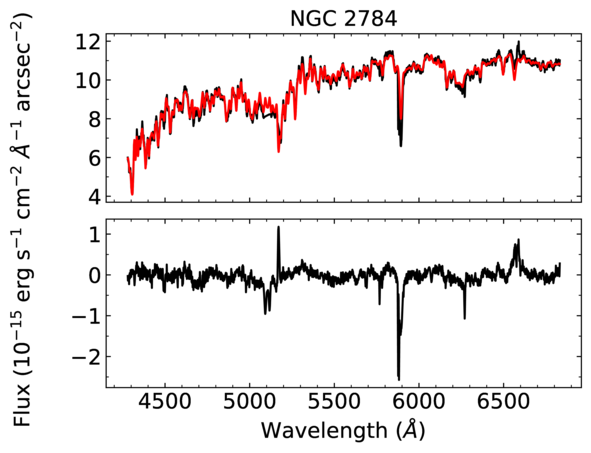}
\includegraphics[scale=0.35]{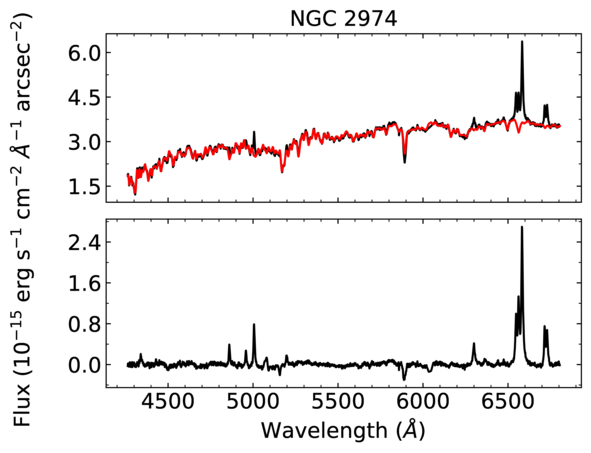}
\includegraphics[scale=0.35]{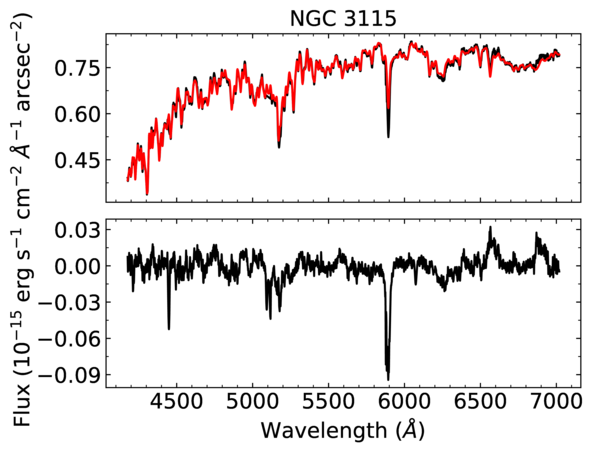}
\includegraphics[scale=0.35]{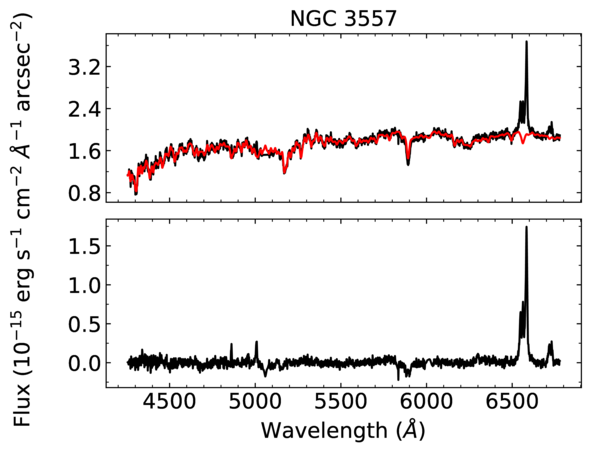}

\end{center}
\caption{--continued}

\end{figure*}

\addtocounter{figure}{-1}

\begin{figure*}

\begin{center}

\includegraphics[scale=0.35]{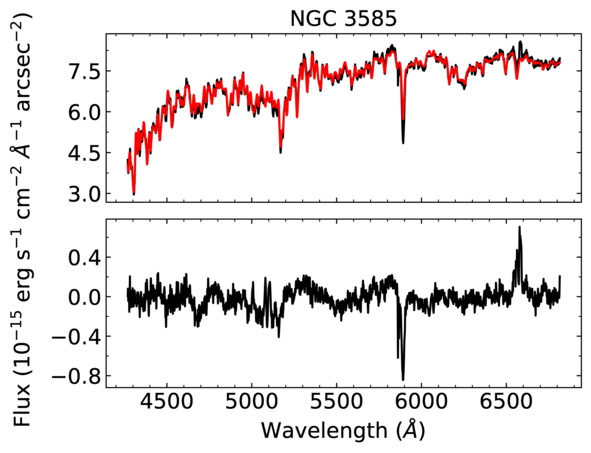}
\includegraphics[scale=0.35]{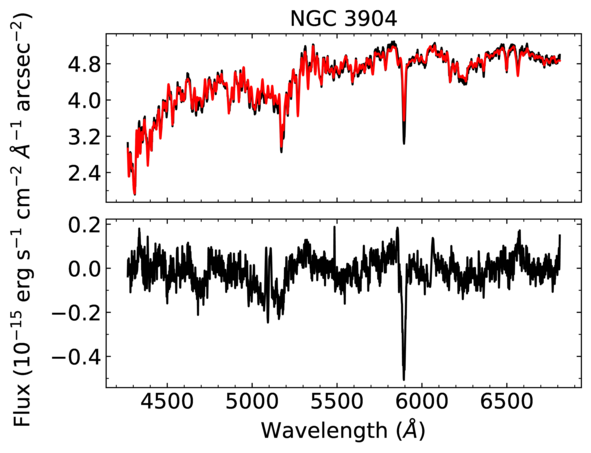}
\includegraphics[scale=0.35]{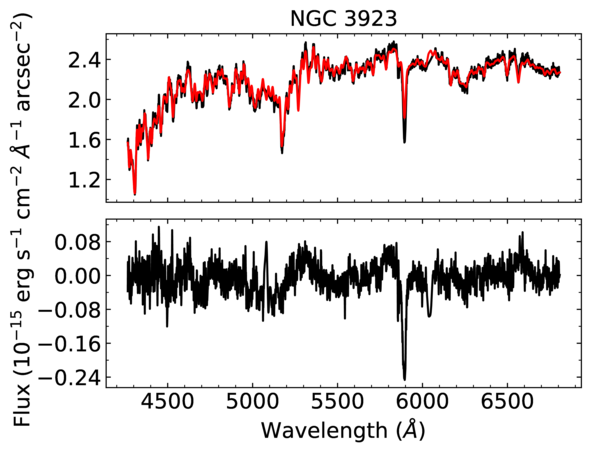}
\includegraphics[scale=0.35]{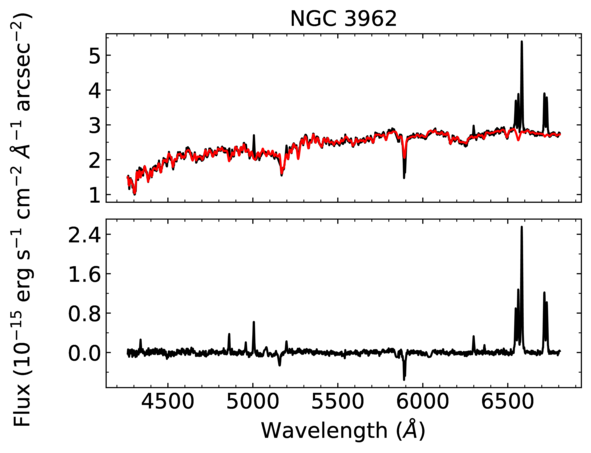}
\includegraphics[scale=0.35]{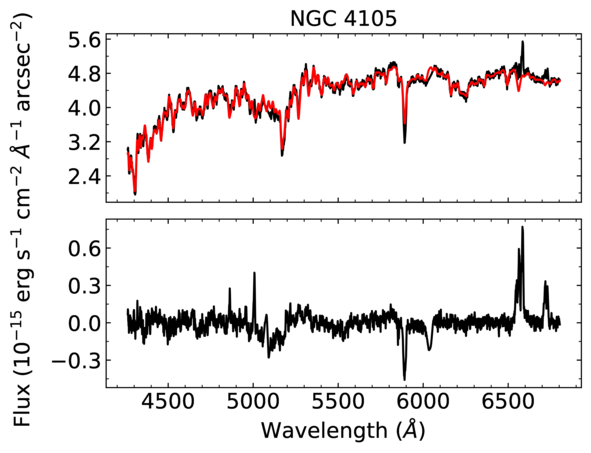}
\includegraphics[scale=0.35]{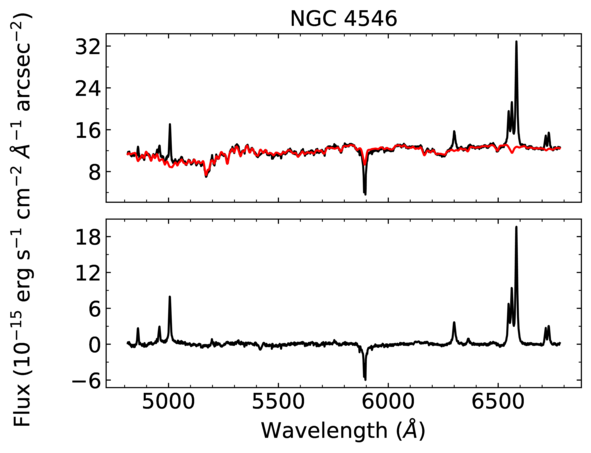}
\includegraphics[scale=0.35]{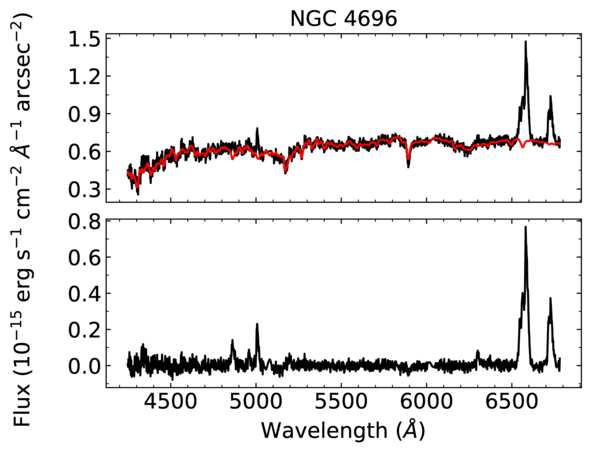}
\includegraphics[scale=0.35]{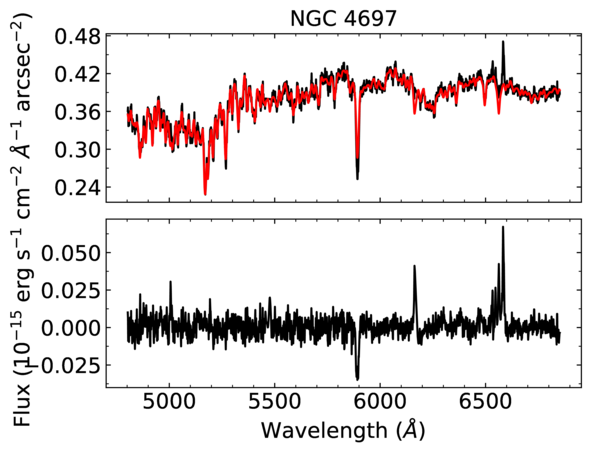}

\end{center}
\caption{--continued}

\end{figure*}

\addtocounter{figure}{-1}

\begin{figure*}

\begin{center}

\includegraphics[scale=0.35]{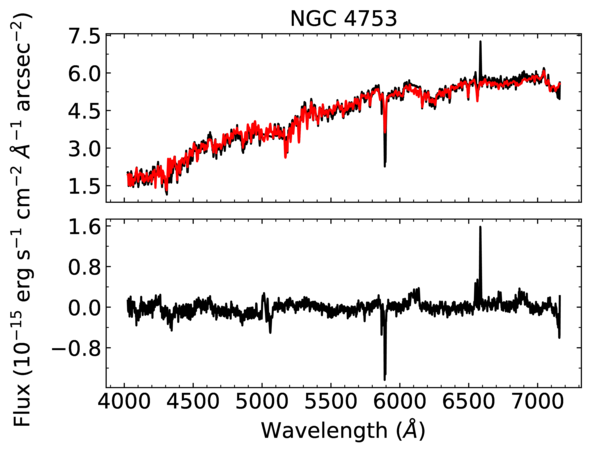}
\includegraphics[scale=0.35]{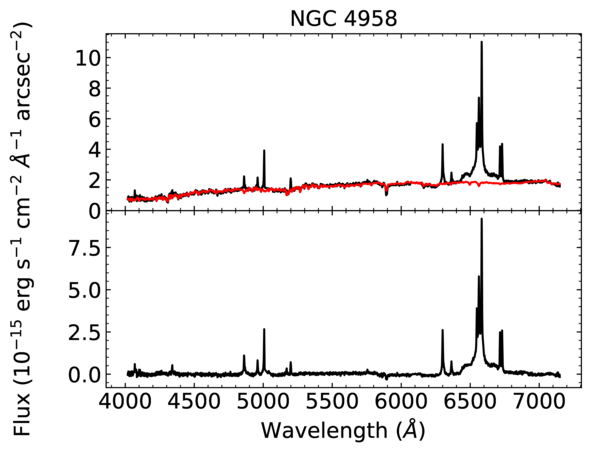}
\includegraphics[scale=0.35]{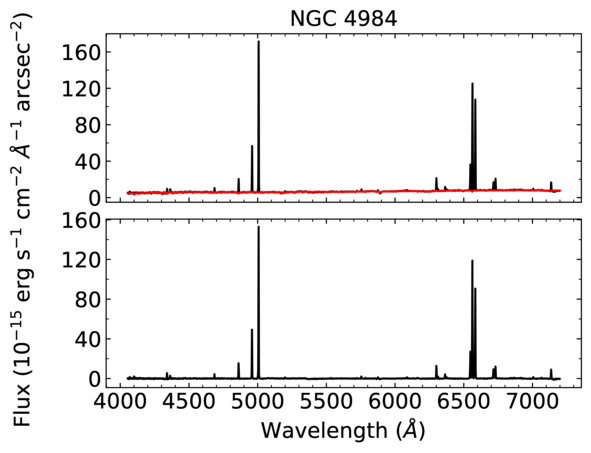}
\includegraphics[scale=0.35]{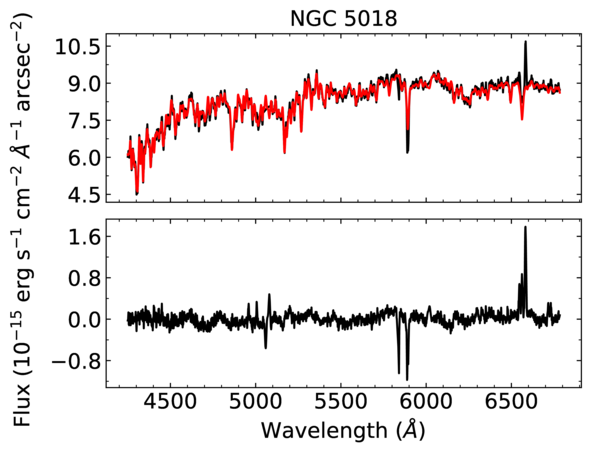}
\includegraphics[scale=0.35]{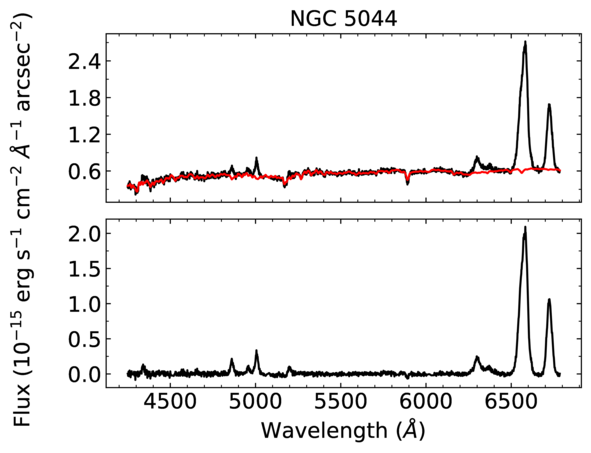}
\includegraphics[scale=0.35]{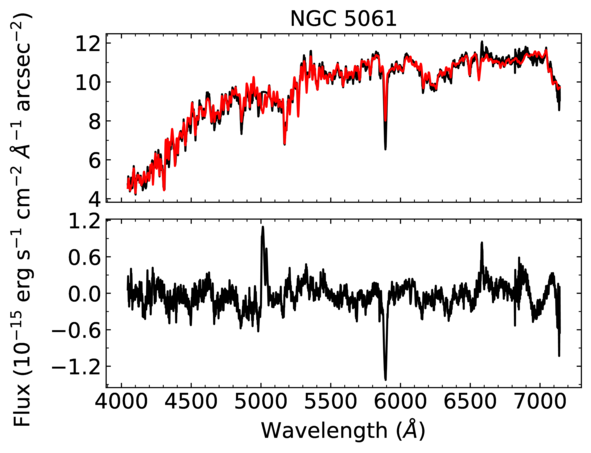}
\includegraphics[scale=0.35]{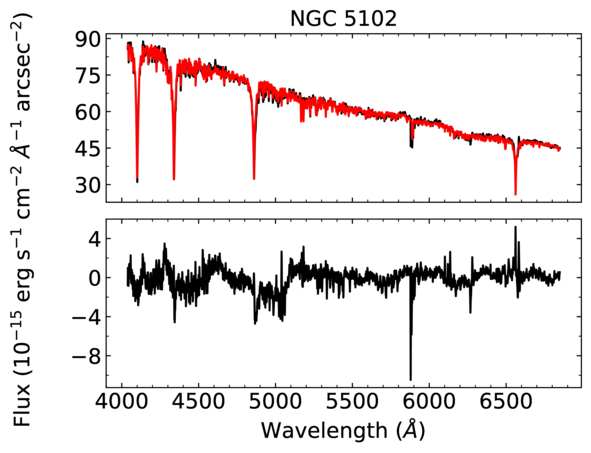}
\includegraphics[scale=0.35]{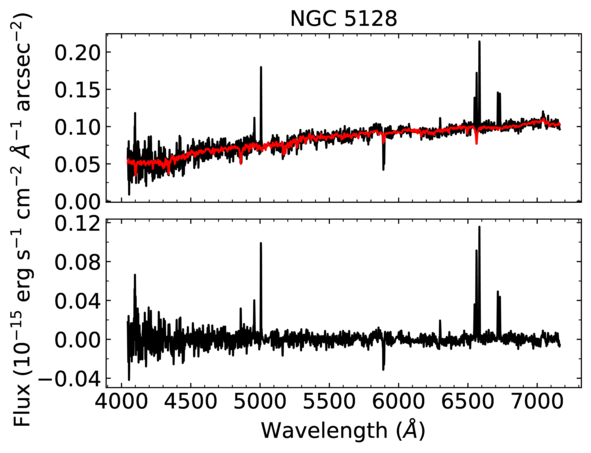}

\end{center}
\caption{--continued}

\end{figure*}

\addtocounter{figure}{-1}

\begin{figure*}

\begin{center}

\includegraphics[scale=0.35]{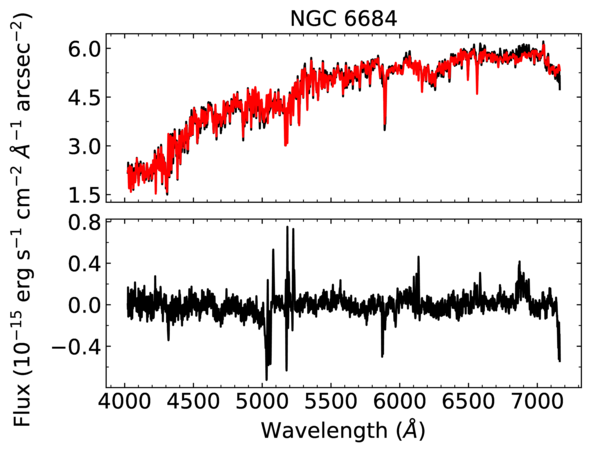}
\includegraphics[scale=0.35]{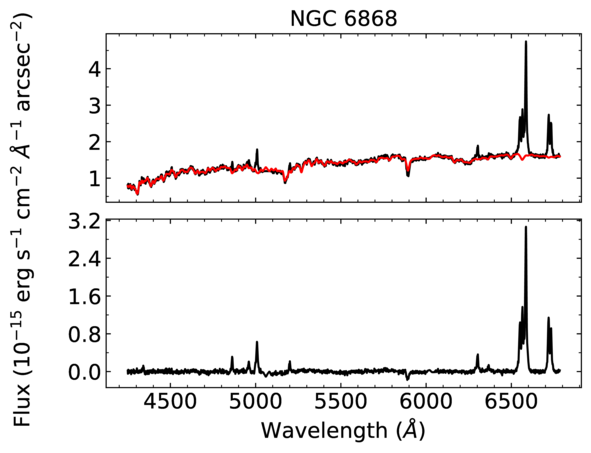}
\includegraphics[scale=0.35]{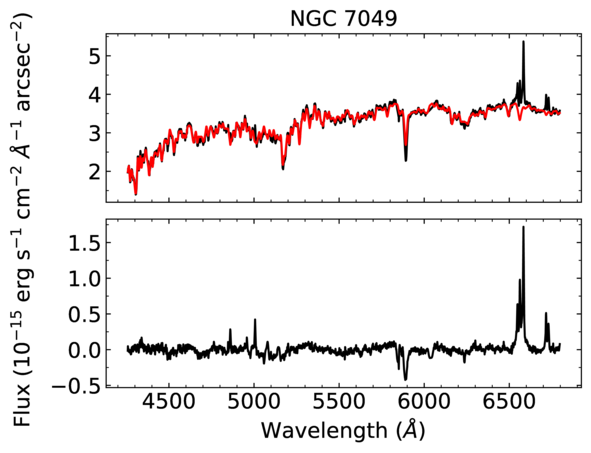}
\includegraphics[scale=0.35]{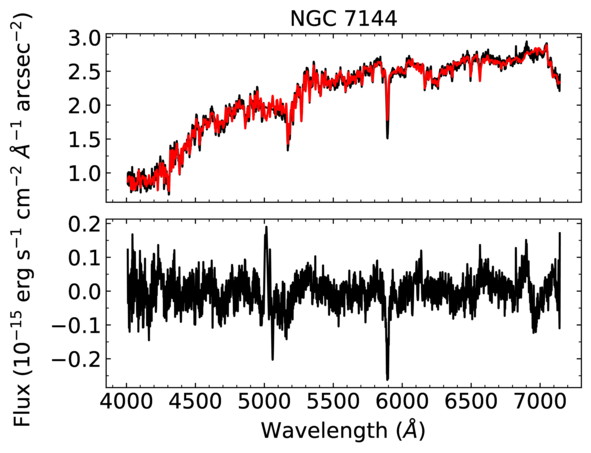}
\includegraphics[scale=0.35]{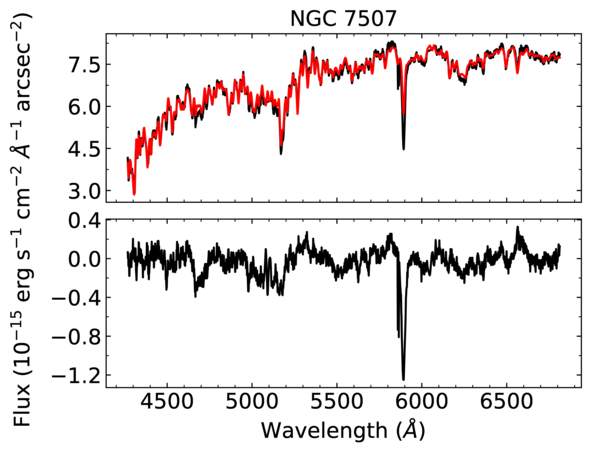}
\includegraphics[scale=0.35]{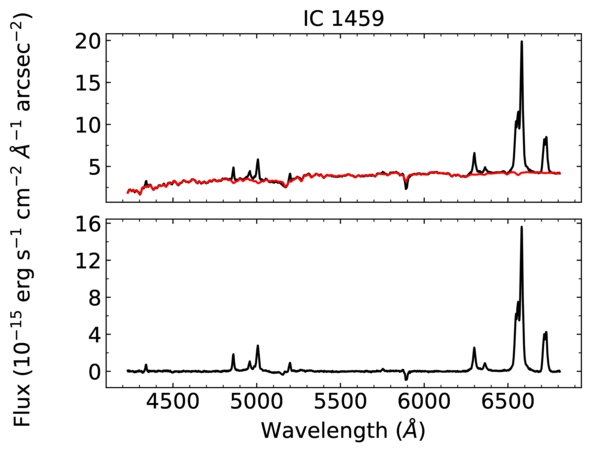}
\includegraphics[scale=0.35]{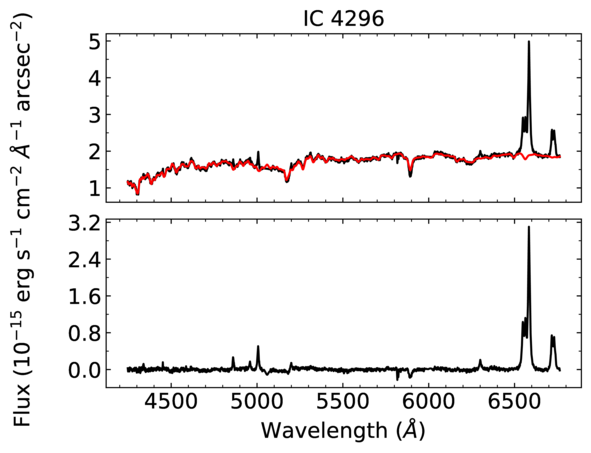}
\includegraphics[scale=0.35]{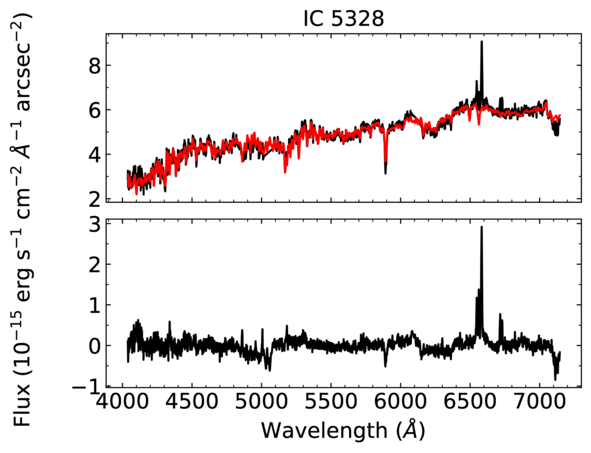}

\end{center}
\caption{--continued}

\end{figure*}

\section{Emission line profiles} \label{appendix:line_profiles}

In Fig. \ref{fig:red_line_profiles}, we show the H$\alpha$, [N {\sc ii}]$\lambda \lambda$6548, 6583 and [S {\sc ii}]$\lambda$6716, 6731 lines that were detected in the nuclear spectra of 48 galaxies of the sample. We also present the fitted Gaussian profiles to these emission lines. For the galaxies where a broad component is also seen in H$\beta$, we show the observations of this line together with the fitted profiles in Fig. \ref{fig:hbeta_line_profiles}. Finally, we present the [O {\sc i}]$\lambda\lambda$6300, 6363 doublet together with the fitted profiles of NGC 2217, NGC 5044 and IC 1459 in Fig. \ref{fig:oi_line_profiles}, since a broad component is also observed in these lines for these objects.

\begin{figure*}

\begin{center}

\includegraphics[scale=0.35]{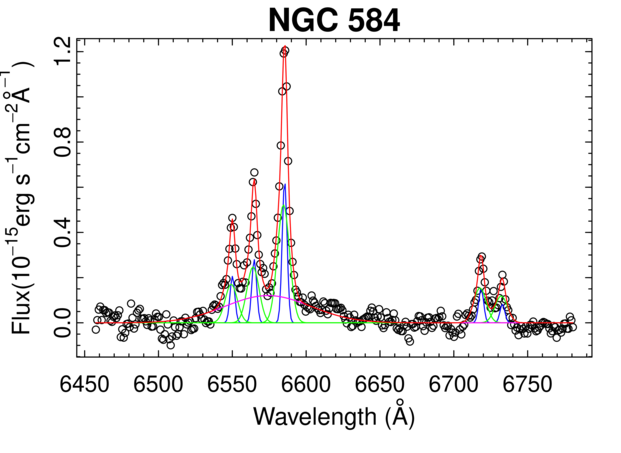}
\includegraphics[scale=0.35]{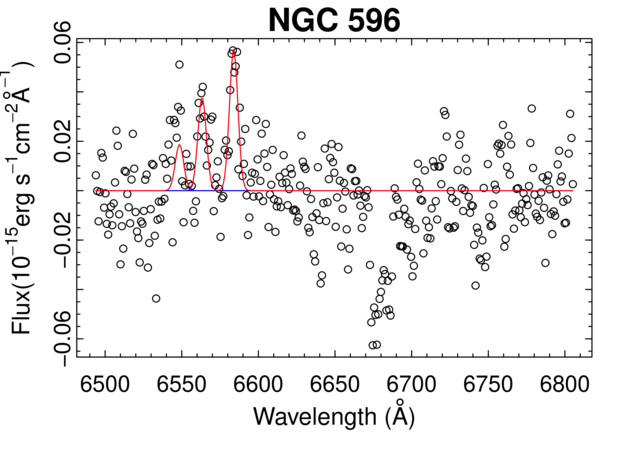}
\includegraphics[scale=0.35]{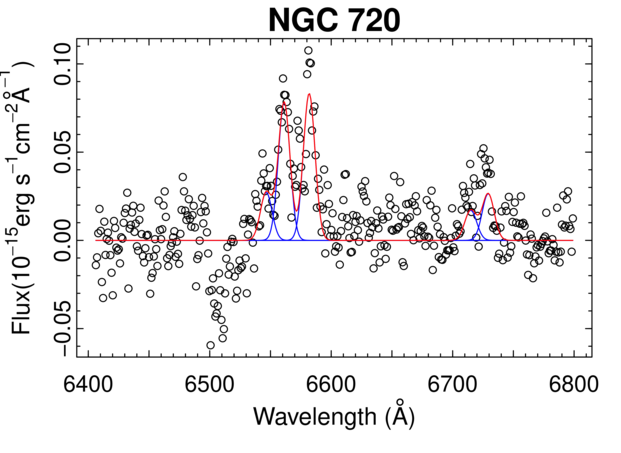}
\includegraphics[scale=0.35]{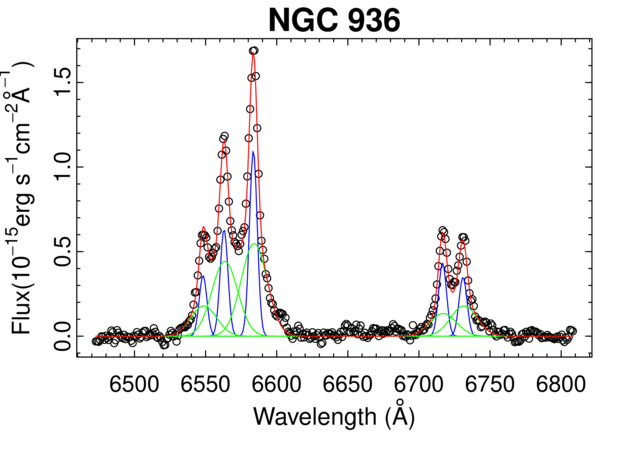}
\includegraphics[scale=0.35]{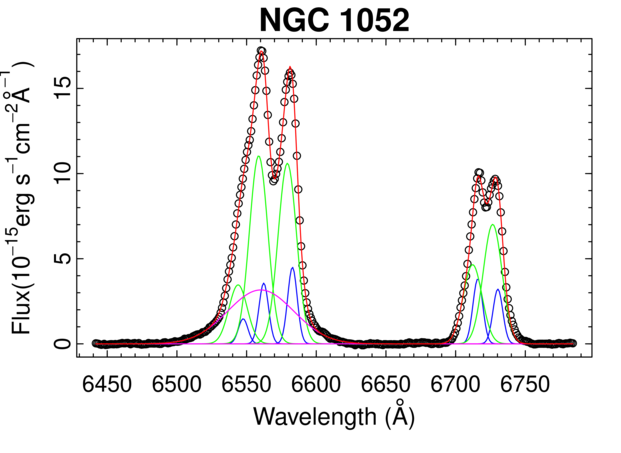}
\includegraphics[scale=0.35]{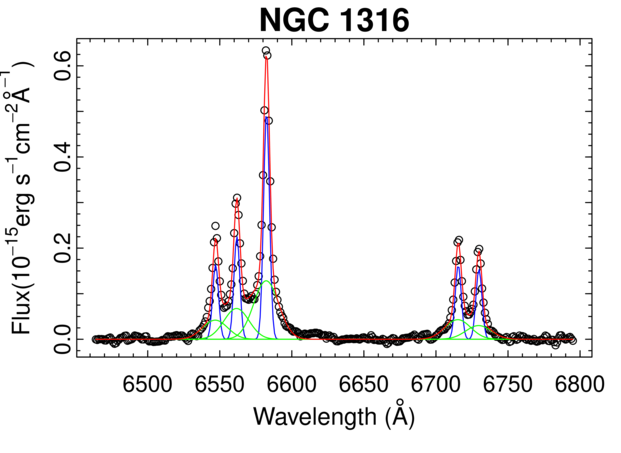}
\includegraphics[scale=0.35]{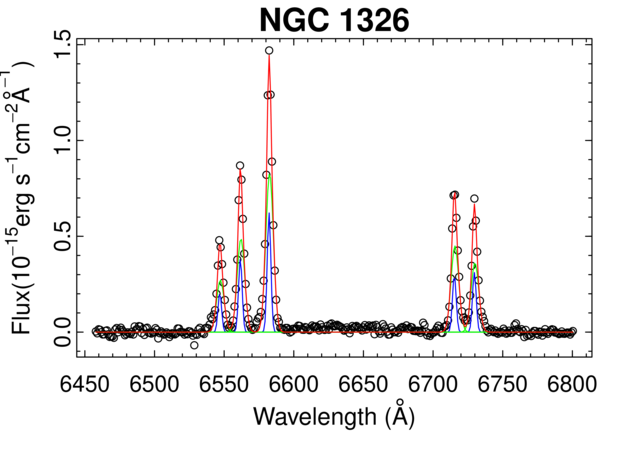}
\includegraphics[scale=0.35]{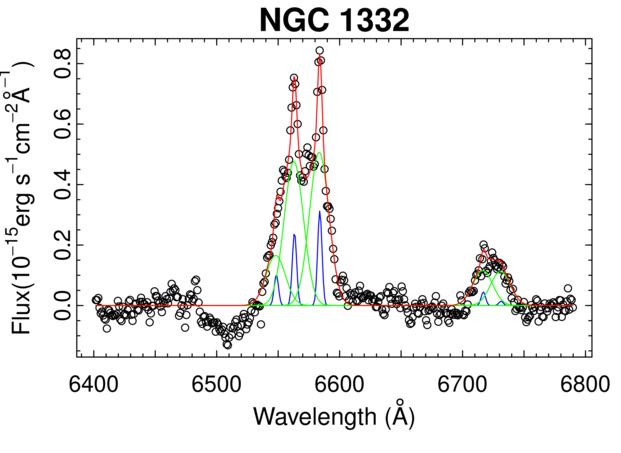}

\end{center}

\caption{H$\alpha$, [N {\sc ii}]$\lambda \lambda$6548, 6583 and [S {\sc ii}]$\lambda$6716, 6731 emission lines detected in 48 galaxies of the sample. The Gaussians correspond to the line decompositions. Gaussians from set 1 are shown in blue, those from set 2 are in green. The magenta Gaussian is associated with the broad component of H$\alpha$. The sum of all Gaussians is shown in red. The orange Gaussians are related with additional corrections that had to be performed in a few objects of the sample, probably due to a bad starlight subtraction. The purple Gaussian in NGC 4958 is associated with a circular relativistic Keplerian disc (Ricci \& Steiner 2020). \label{fig:red_line_profiles}}

\end{figure*}

\addtocounter{figure}{-1}

\begin{figure*}

\begin{center}

\includegraphics[scale=0.35]{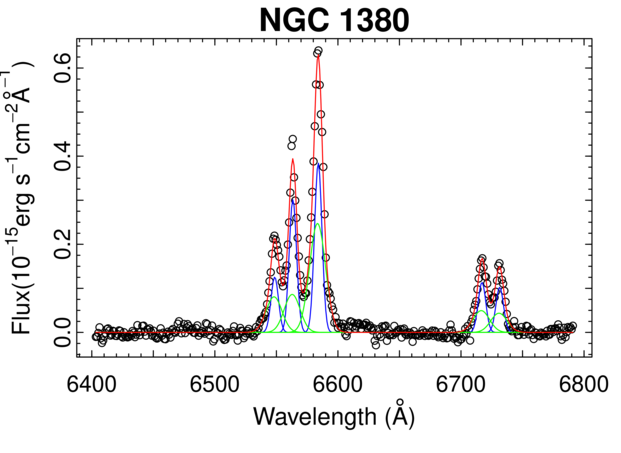}
\includegraphics[scale=0.35]{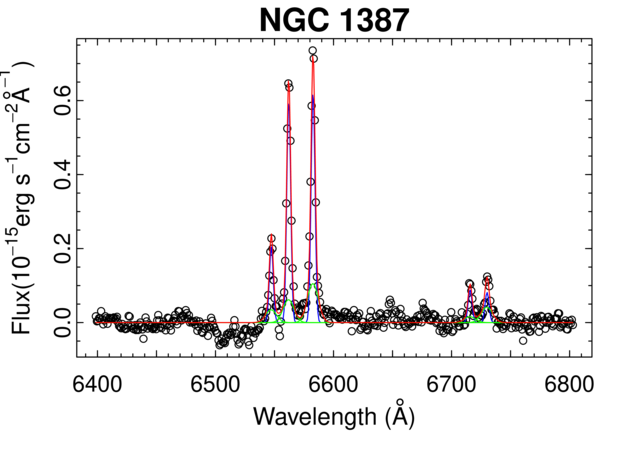}
\includegraphics[scale=0.35]{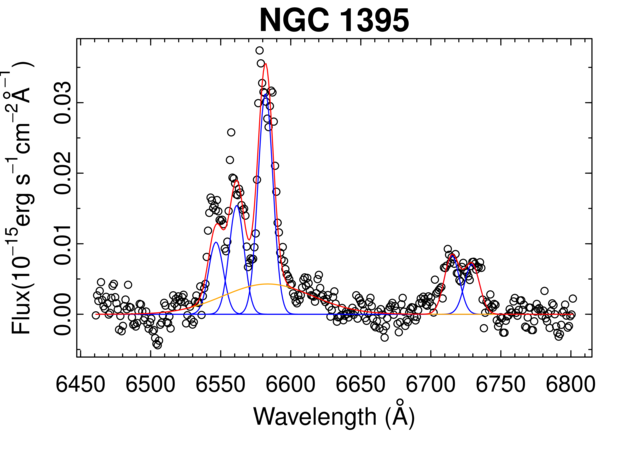}
\includegraphics[scale=0.35]{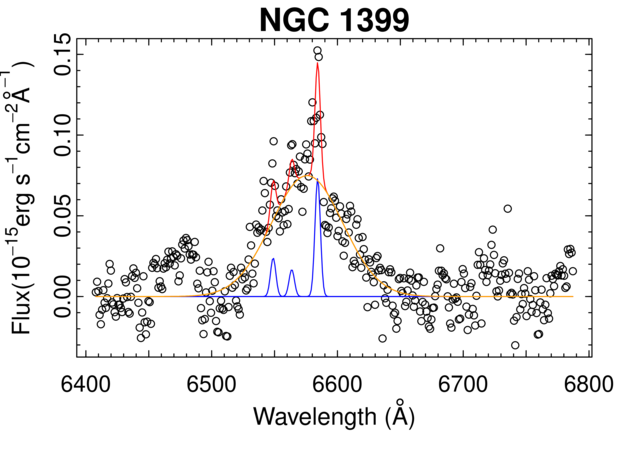}
\includegraphics[scale=0.35]{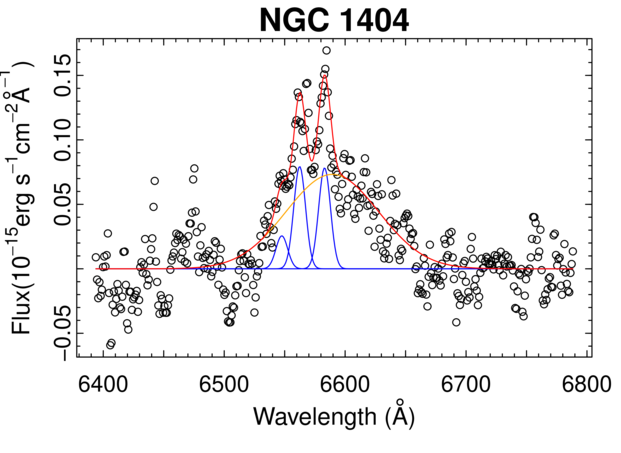}
\includegraphics[scale=0.35]{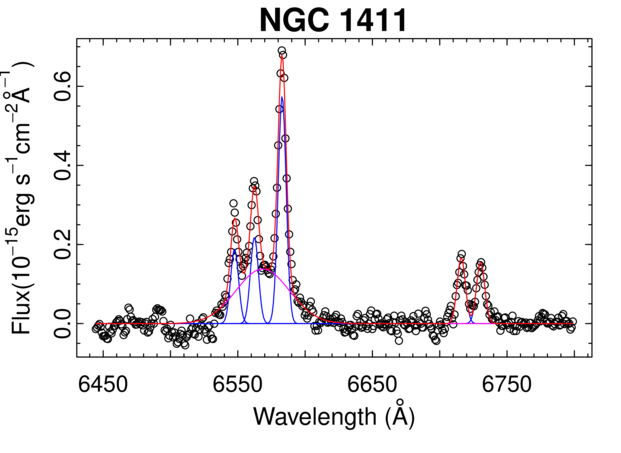}
\includegraphics[scale=0.35]{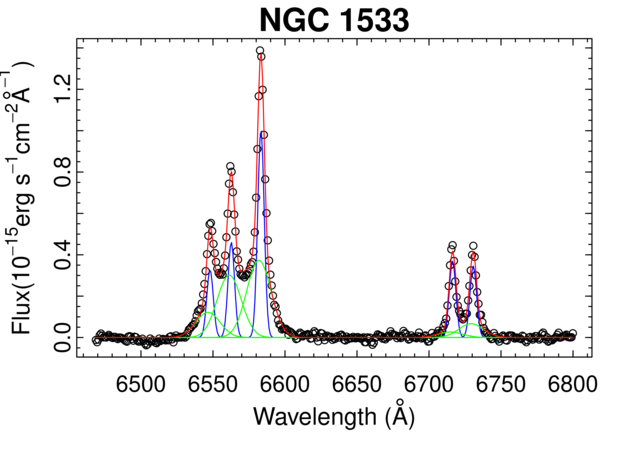}
\includegraphics[scale=0.35]{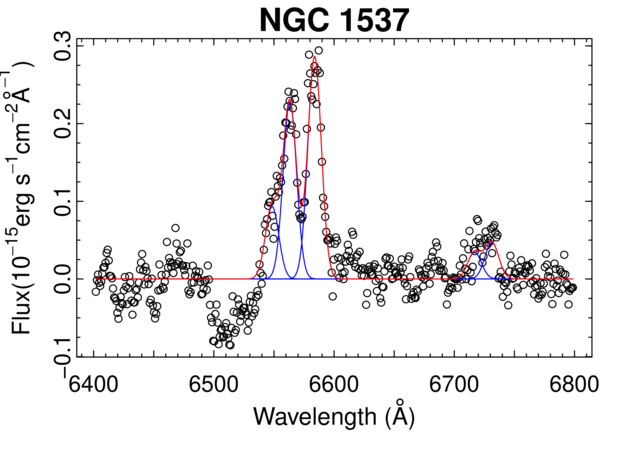}

\end{center}
\caption{-- continued}

\end{figure*}

\addtocounter{figure}{-1}

\begin{figure*}

\begin{center}

\includegraphics[scale=0.35]{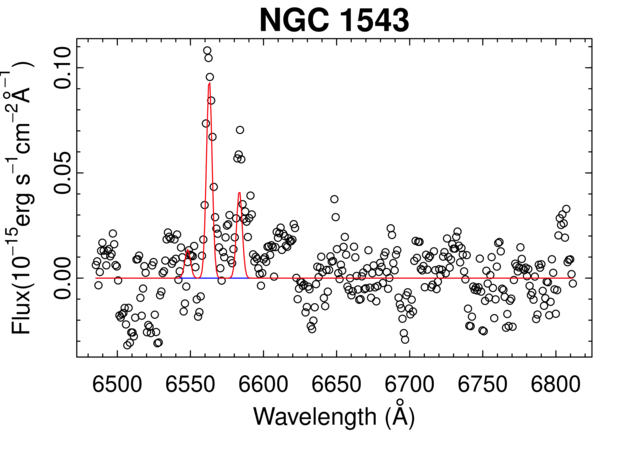}
\includegraphics[scale=0.35]{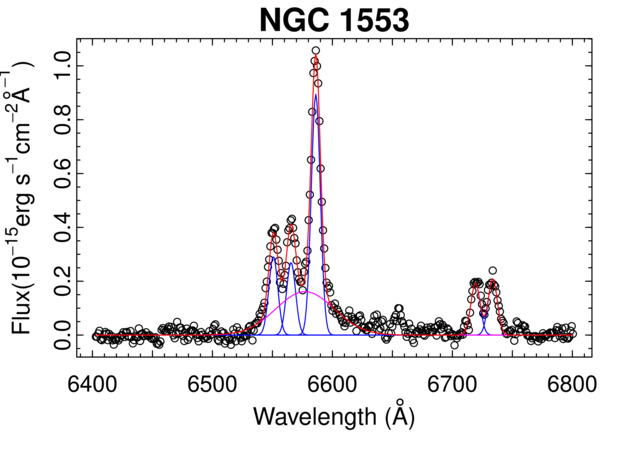}
\includegraphics[scale=0.35]{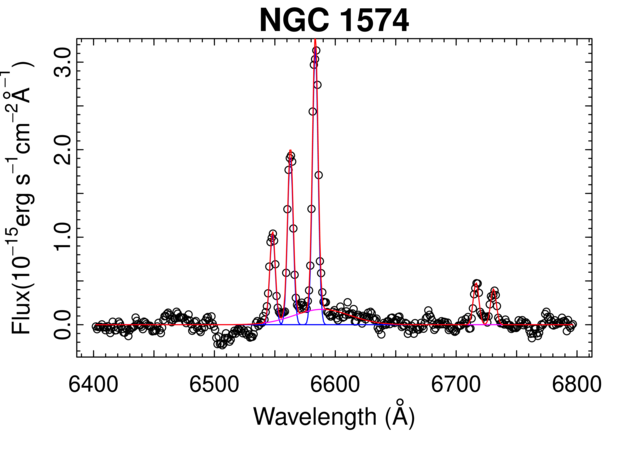}
\includegraphics[scale=0.35]{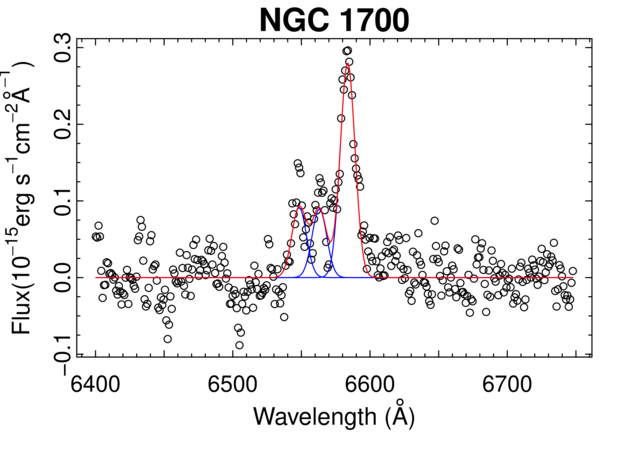}
\includegraphics[scale=0.35]{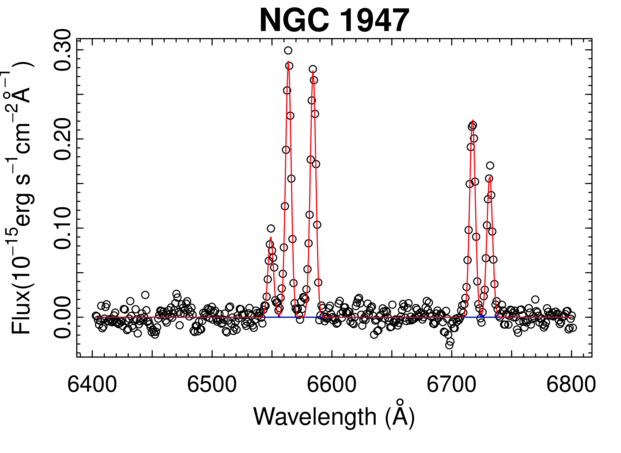}
\includegraphics[scale=0.35]{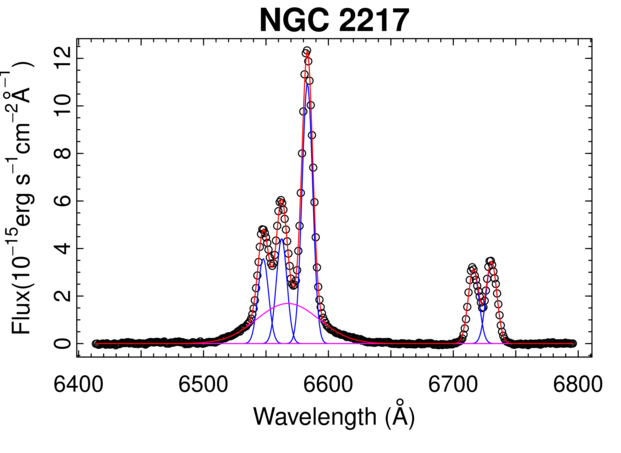}
\includegraphics[scale=0.35]{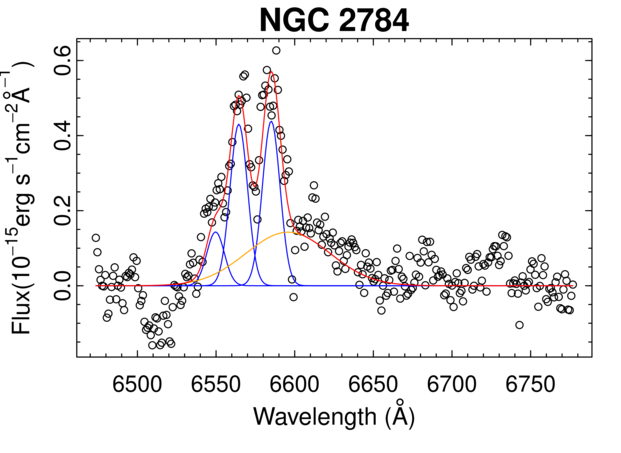}
\includegraphics[scale=0.35]{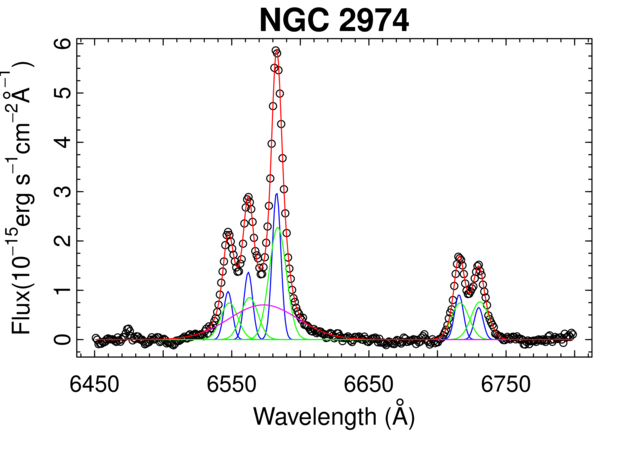}

\end{center}
\caption{-- continued}

\end{figure*}

\addtocounter{figure}{-1}

\begin{figure*}

\begin{center}

\includegraphics[scale=0.35]{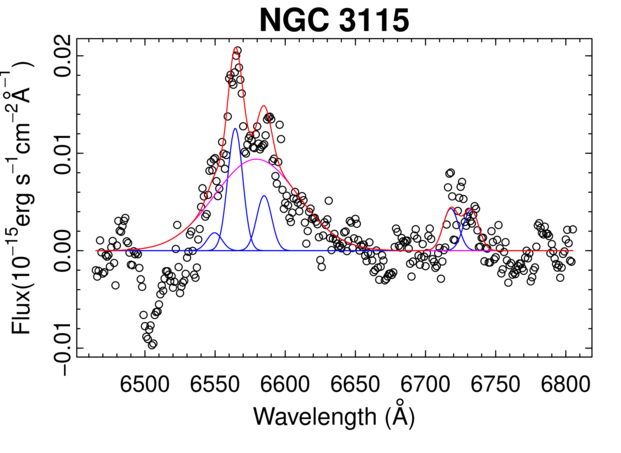}
\includegraphics[scale=0.35]{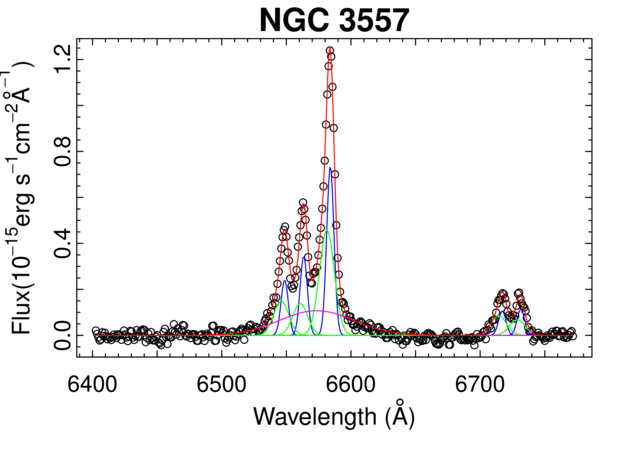}
\includegraphics[scale=0.35]{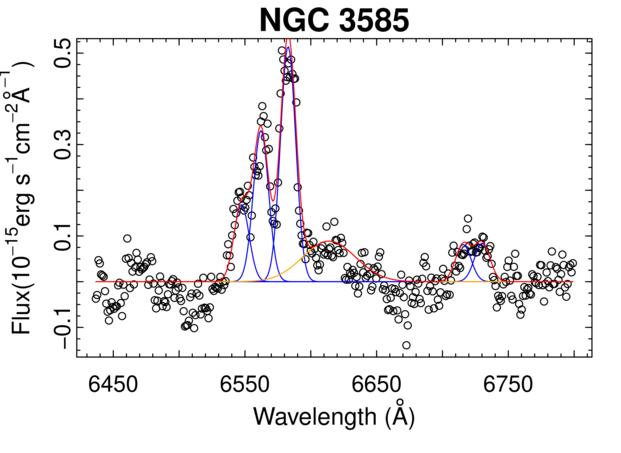}
\includegraphics[scale=0.35]{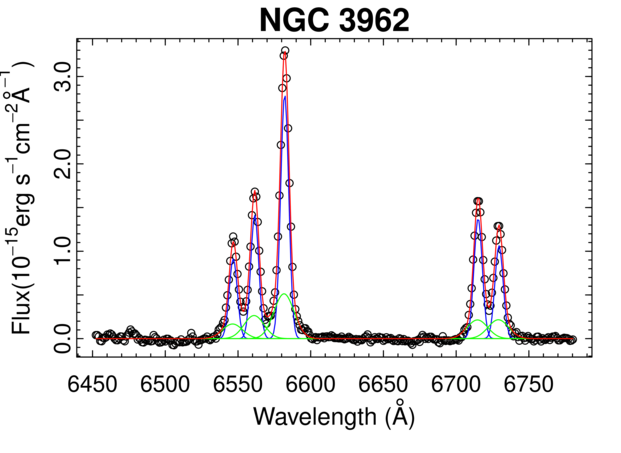}
\includegraphics[scale=0.35]{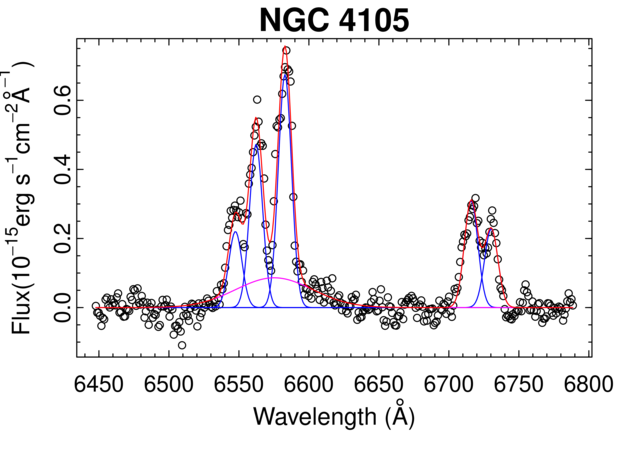}
\includegraphics[scale=0.35]{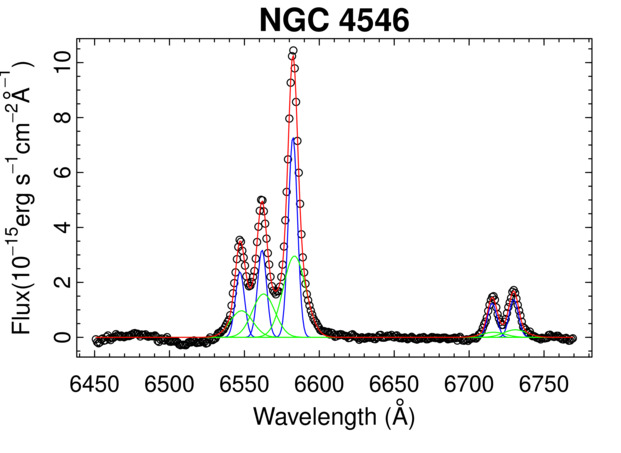}
\includegraphics[scale=0.35]{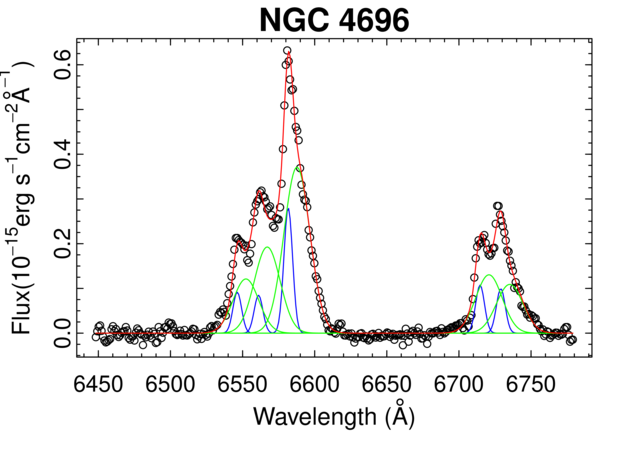}
\includegraphics[scale=0.35]{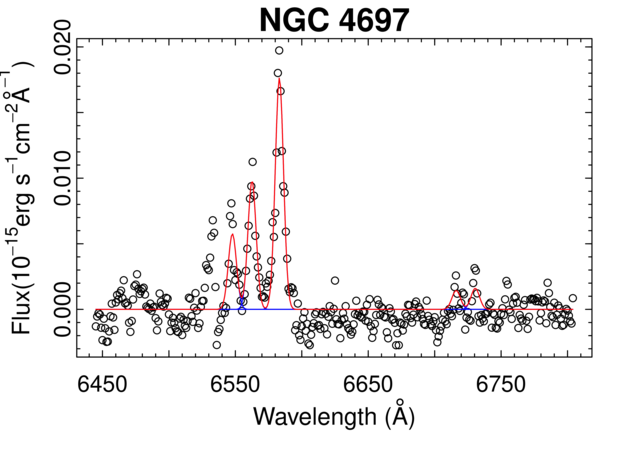}

\end{center}
\caption{-- continued}

\end{figure*}

\addtocounter{figure}{-1}

\begin{figure*}

\begin{center}

\includegraphics[scale=0.35]{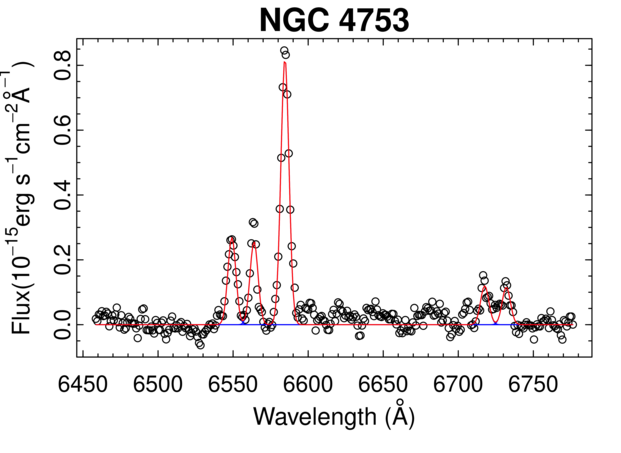}
\includegraphics[scale=0.35]{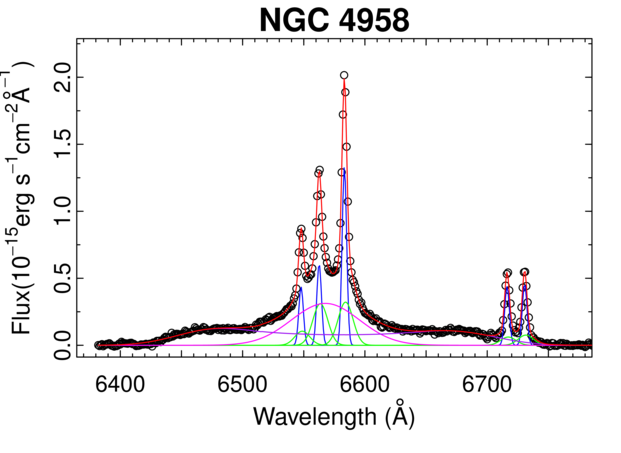}
\includegraphics[scale=0.35]{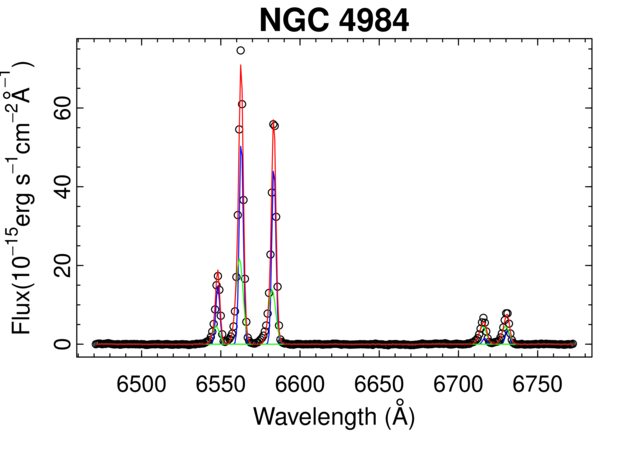}
\includegraphics[scale=0.35]{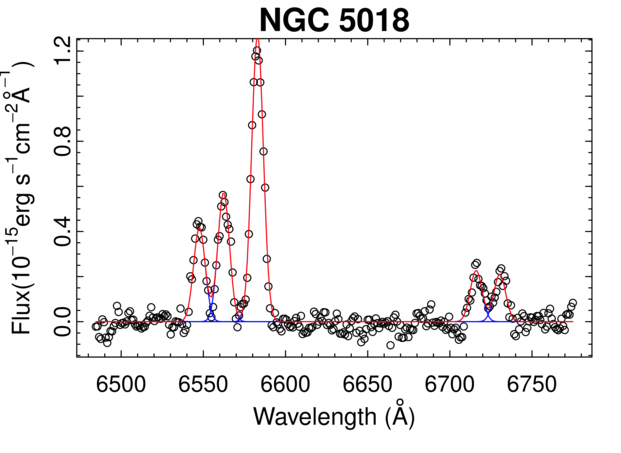}
\includegraphics[scale=0.35]{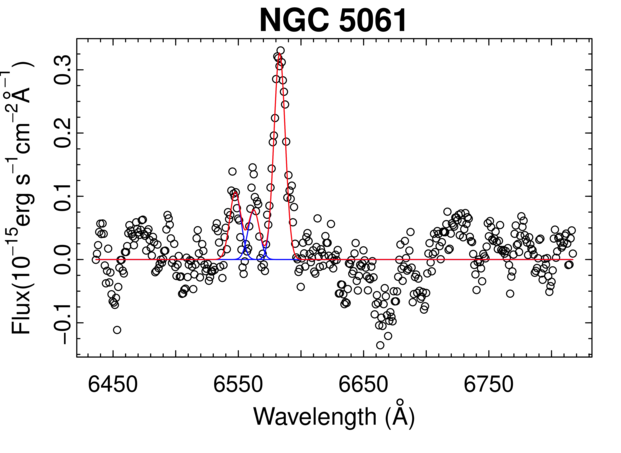}
\includegraphics[scale=0.35]{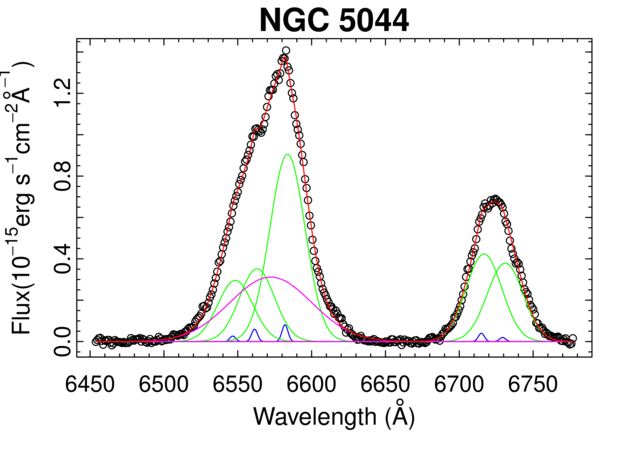}
\includegraphics[scale=0.35]{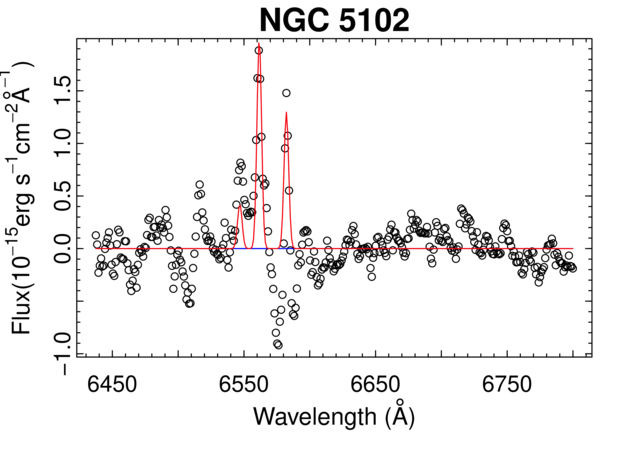}
\includegraphics[scale=0.35]{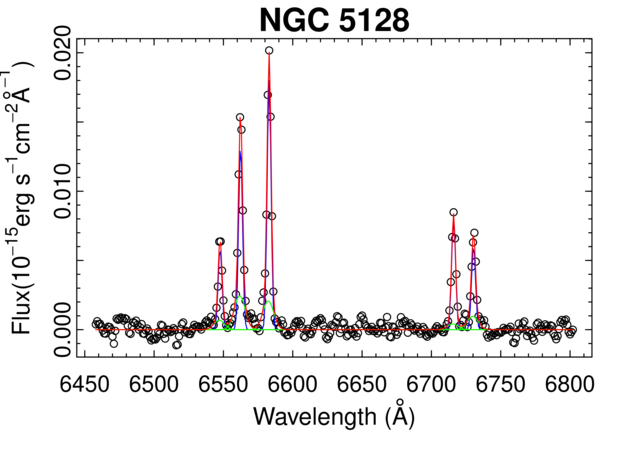}

\end{center}
\caption{-- continued}

\end{figure*}

\addtocounter{figure}{-1}

\begin{figure*}

\begin{center}

\includegraphics[scale=0.35]{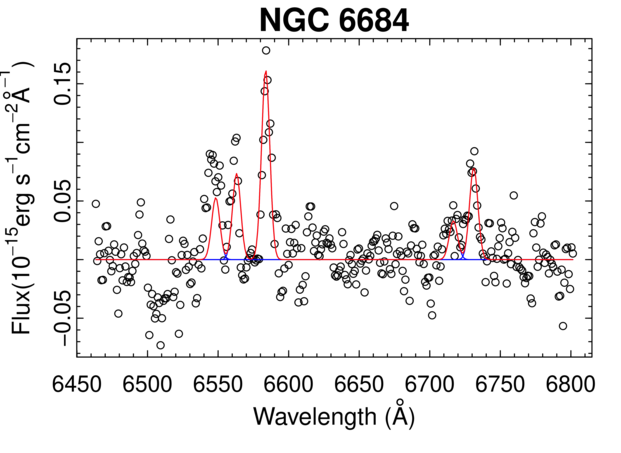}
\includegraphics[scale=0.35]{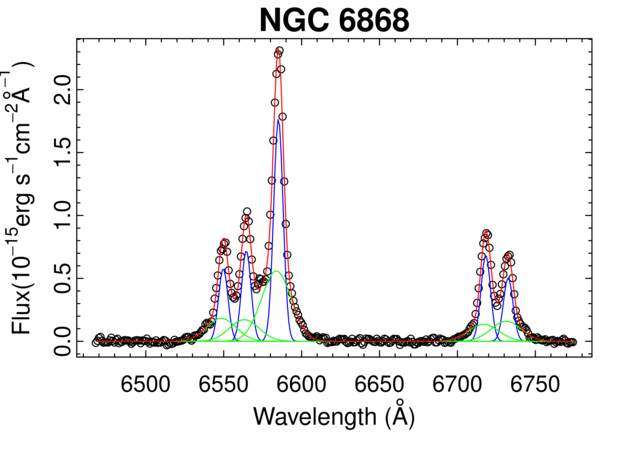}
\includegraphics[scale=0.35]{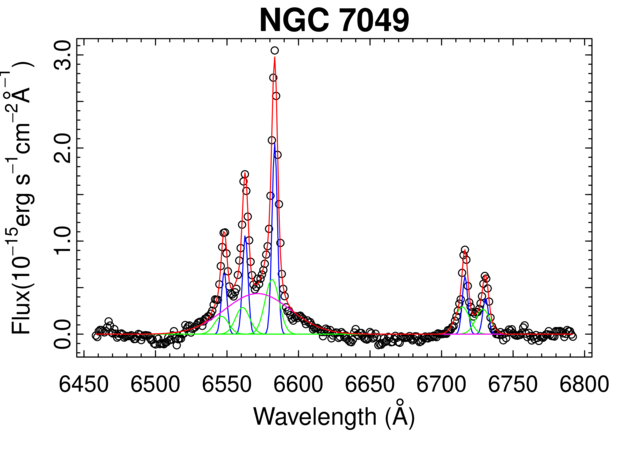}
\includegraphics[scale=0.35]{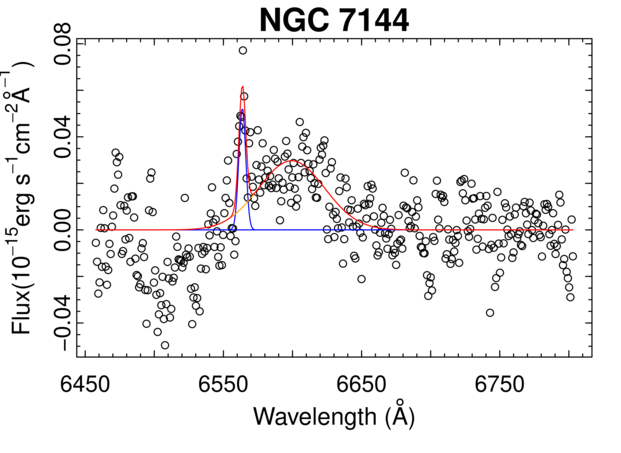}
\includegraphics[scale=0.35]{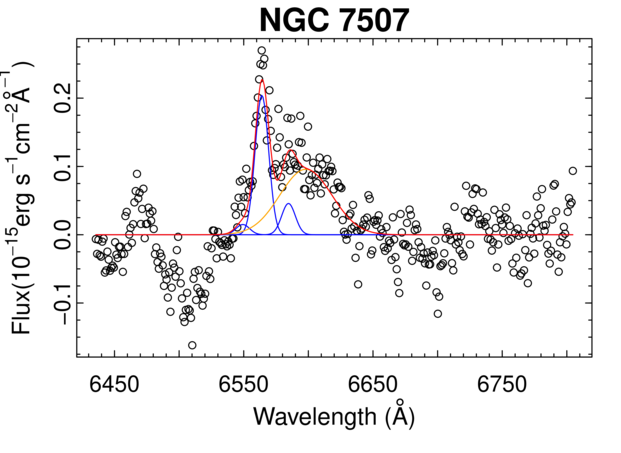}
\includegraphics[scale=0.35]{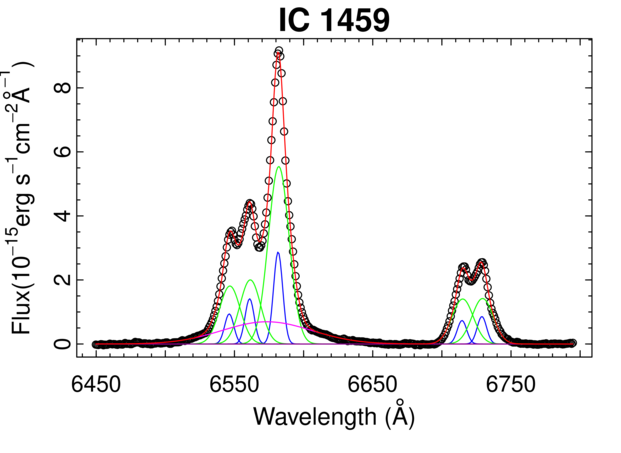}
\includegraphics[scale=0.35]{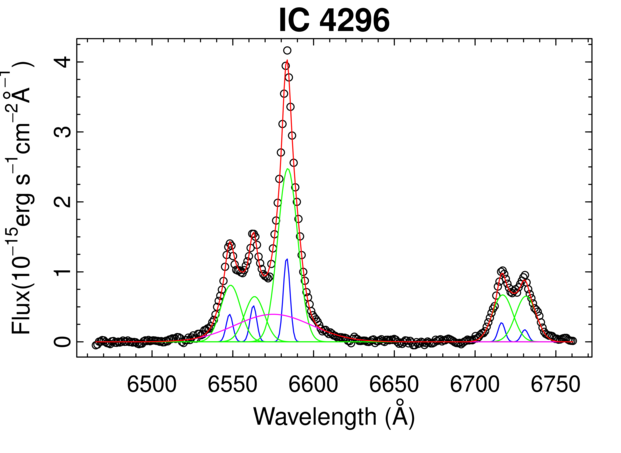}
\includegraphics[scale=0.35]{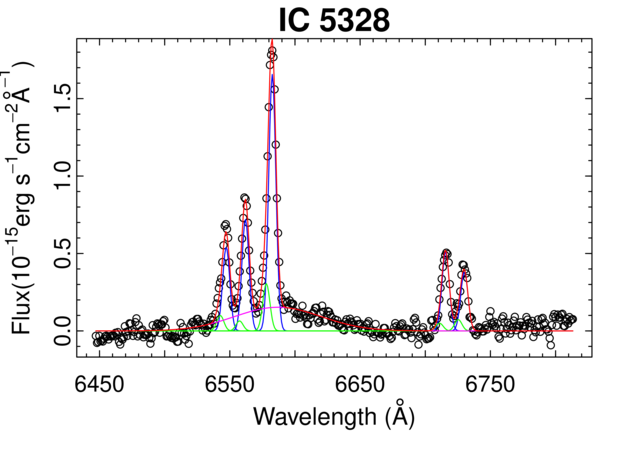}

\end{center}
\caption{-- continued}

\end{figure*}

\begin{figure*}

\begin{center}

\includegraphics[scale=0.35]{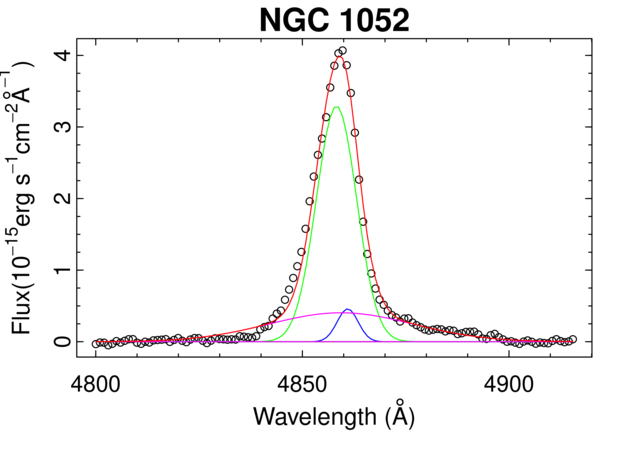}
\includegraphics[scale=0.35]{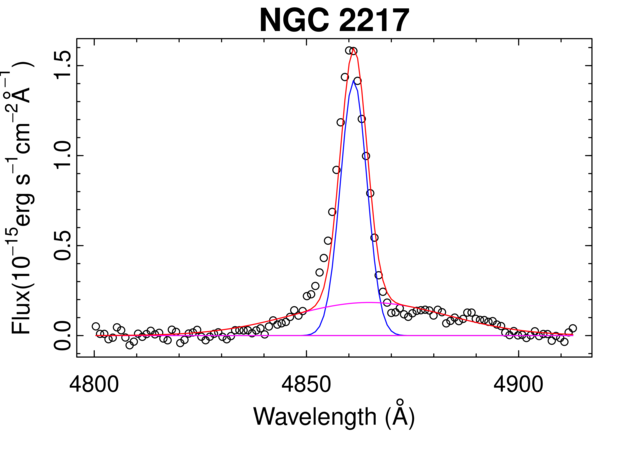}
\includegraphics[scale=0.35]{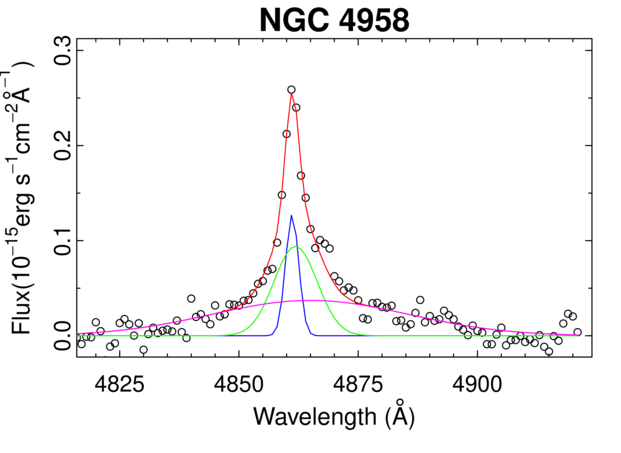}
\includegraphics[scale=0.35]{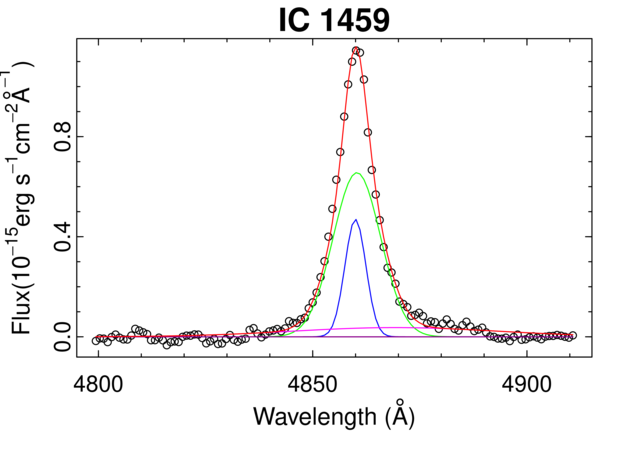}

\end{center}
\caption{Line profiles of H$\beta$ for nuclei with a broad component in this emission line. Since these profiles were fitted assuming that the H$\beta$ line has the same kinematics as the H$\alpha$ line, the colours of the Gaussian functions are the same as in Fig. \ref{fig:red_line_profiles}. \label{fig:hbeta_line_profiles}}

\end{figure*}

\begin{figure*}

\begin{center}

\includegraphics[scale=0.35]{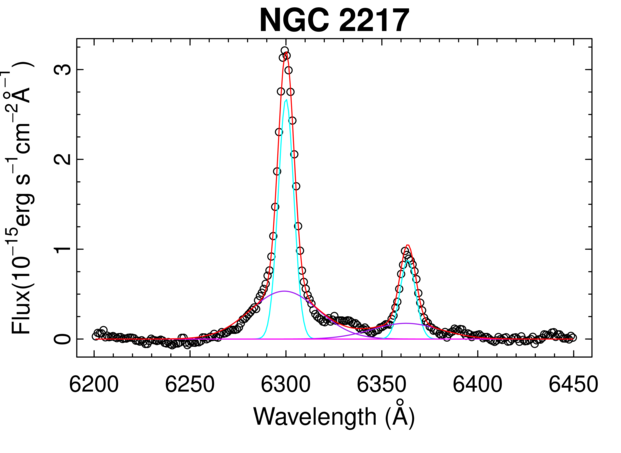}
\includegraphics[scale=0.35]{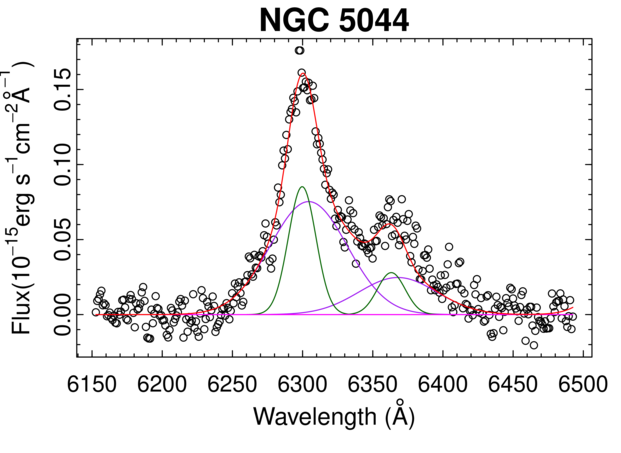}
\includegraphics[scale=0.35]{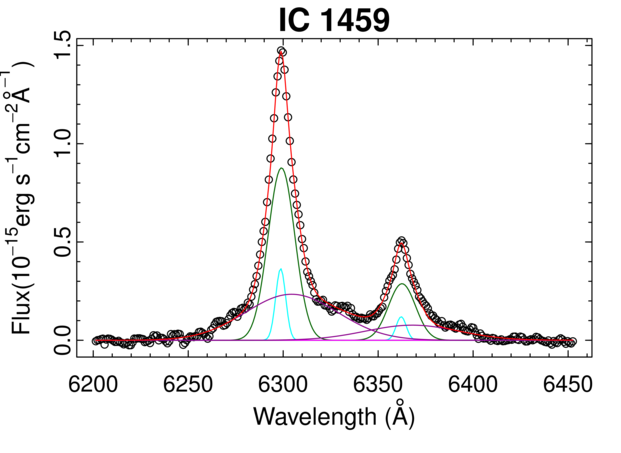}

\end{center}
\caption{Line profiles of the [O {\sc i}]$\lambda\lambda$6300, 6363 for nuclei with a broad component in this feature. Here, the dark green and the cyan Gaussians are related to the narrow component of the doublet, while the dark magenta Gaussians correspond to the broad component of the [O {\sc i}] lines. \label{fig:oi_line_profiles}}

\end{figure*}

\section{Spatial profiles of the broad line regions}

In Fig. \ref{fig:blr_spatial_profiles}, we present the radial profiles of the images taken from the red wing of the broad H$\alpha$ component of the galaxies with a BLR. The black points correspond to the observed profiles, while the red lines are related to the fitted Gaussian functions that are associated with the PSF of the data cubes. The profiles of the stellar component are shown as blue crosses for a comparison with the profile from the spatially unresolved BLRs. 

\begin{figure*}

\begin{center}

\includegraphics[scale=0.35]{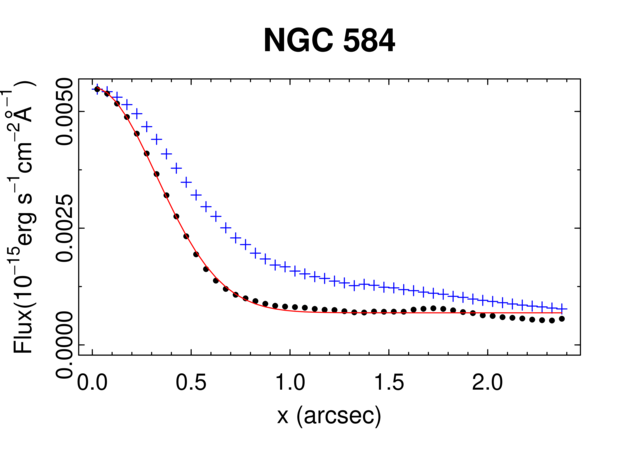}
\includegraphics[scale=0.35]{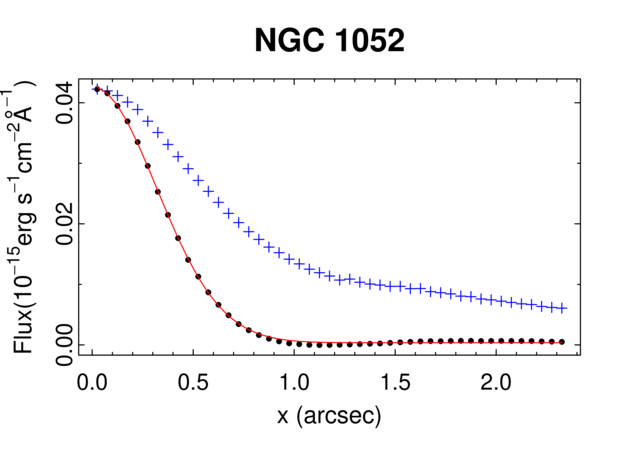}

\includegraphics[scale=0.35]{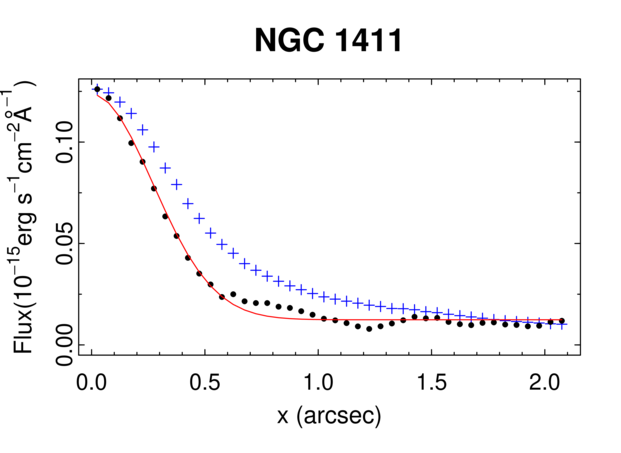}
\includegraphics[scale=0.35]{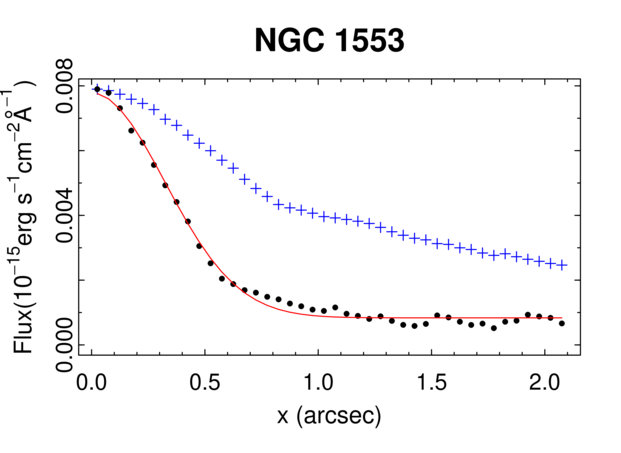}

\includegraphics[scale=0.35]{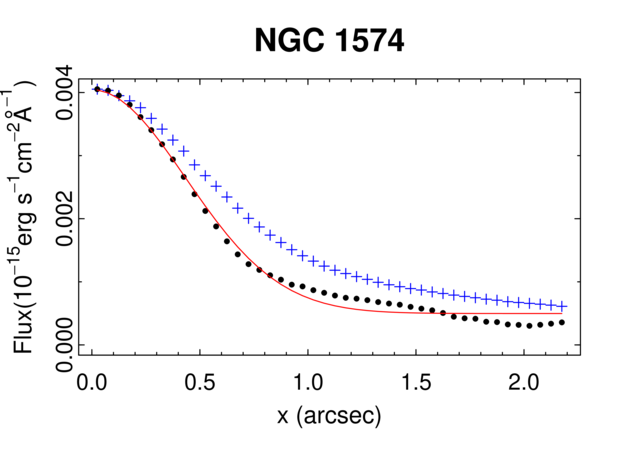}
\includegraphics[scale=0.35]{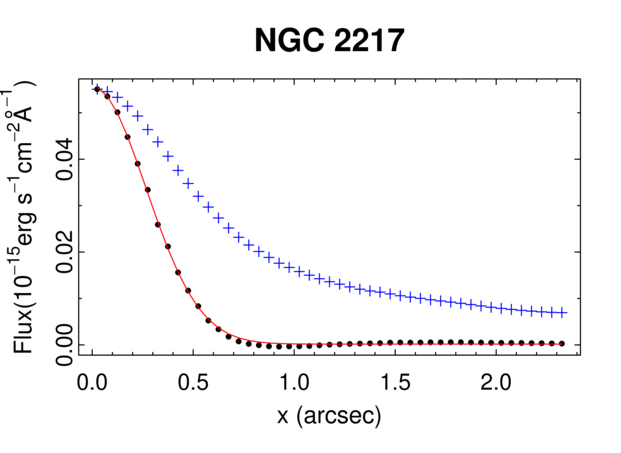}

\includegraphics[scale=0.35]{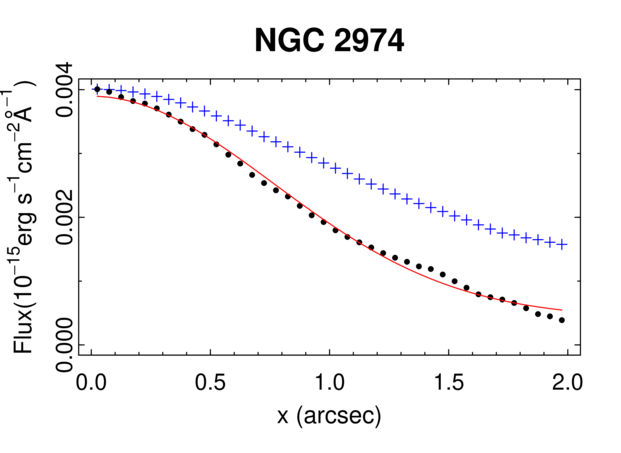}
\includegraphics[scale=0.35]{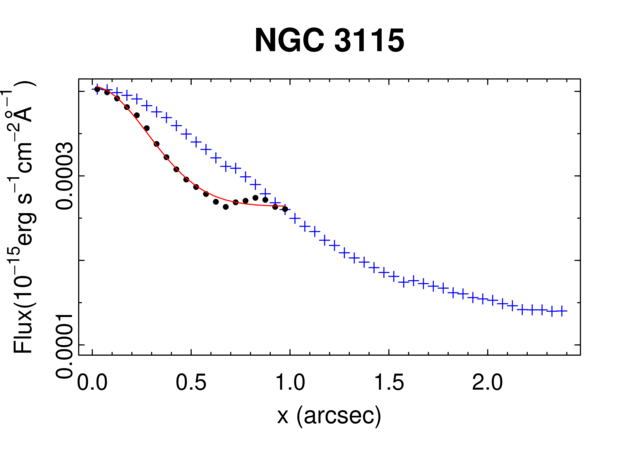}

\end{center}

\caption{Radial profiles of images extracted from the red wings of the broad component of H$\alpha$ of 16 galaxies of the sample. The red line corresponds to the PSFs of the gas cubes, whose FWHMs are shown in Table 1, and the blue crosses are related to the radial profile of the stellar component, extracted from an image of the stellar continuum. The maximum stellar flux was renormalised to the maximum emission of the BLR, so a convenient comparison may be done. The comparison between the stellar and the broad component profiles in these galaxies suggests that the BLR is real and not an effect of a bad starlight subtraction. \label{fig:blr_spatial_profiles}}

\end{figure*}

\addtocounter{figure}{-1}

\begin{figure*}

\begin{center}

\includegraphics[scale=0.35]{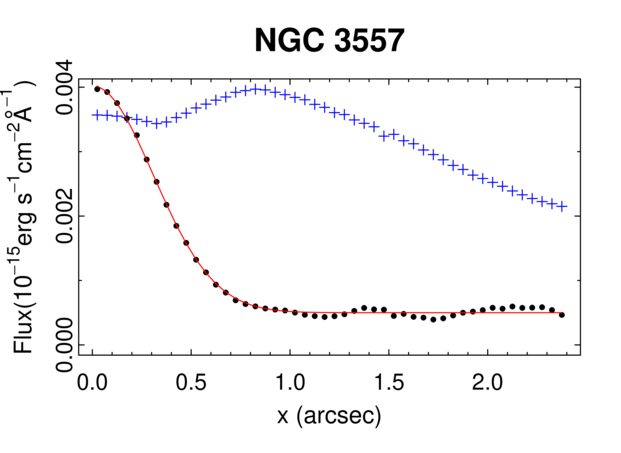}
\includegraphics[scale=0.35]{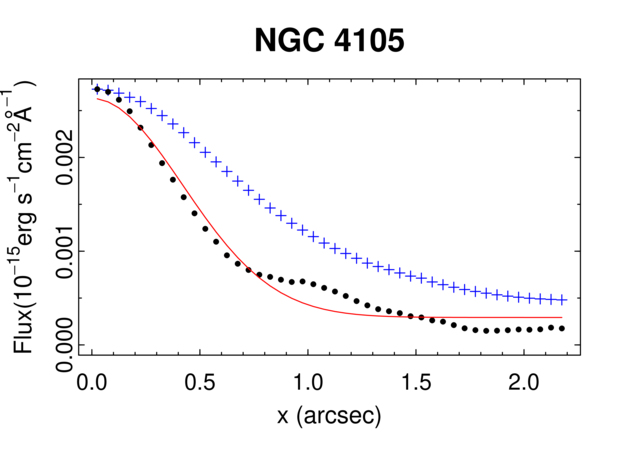}

\includegraphics[scale=0.35]{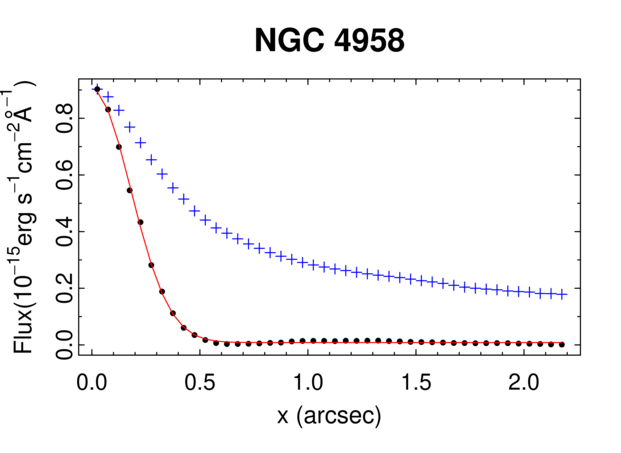}
\includegraphics[scale=0.35]{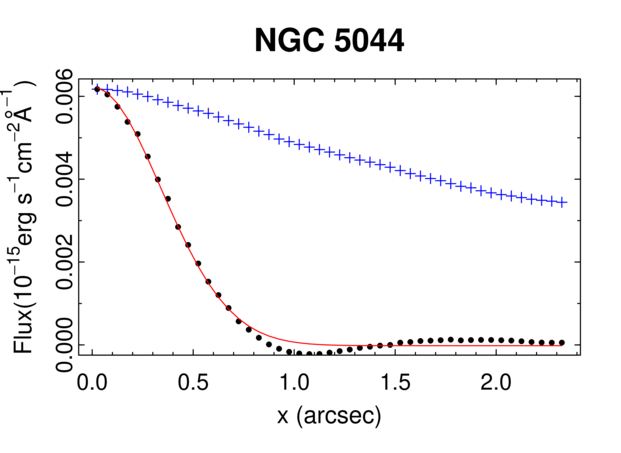}

\includegraphics[scale=0.35]{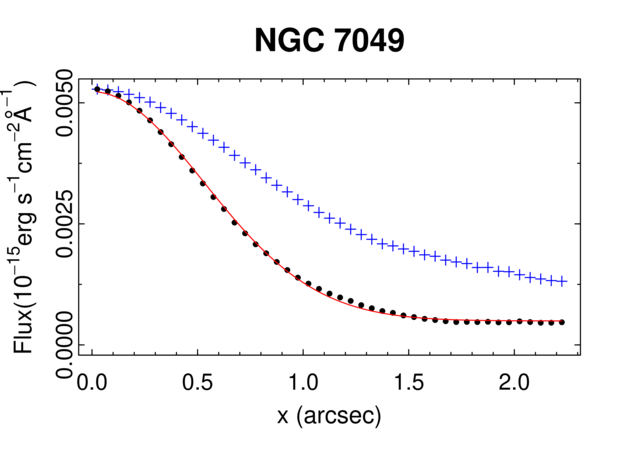}
\includegraphics[scale=0.35]{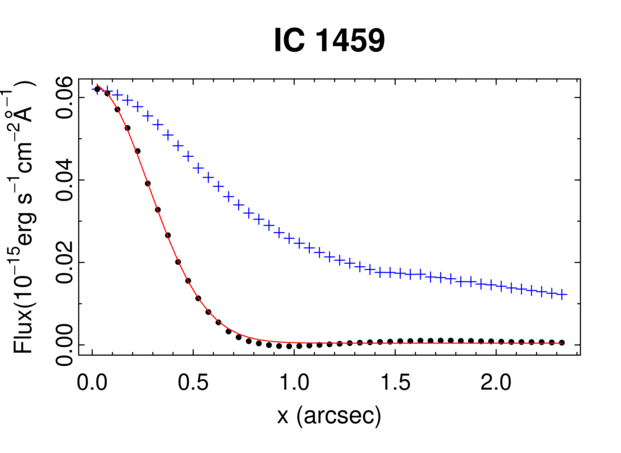}

\includegraphics[scale=0.35]{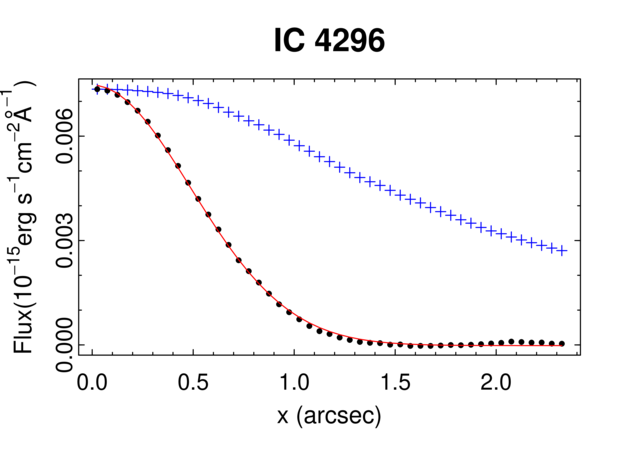}
\includegraphics[scale=0.35]{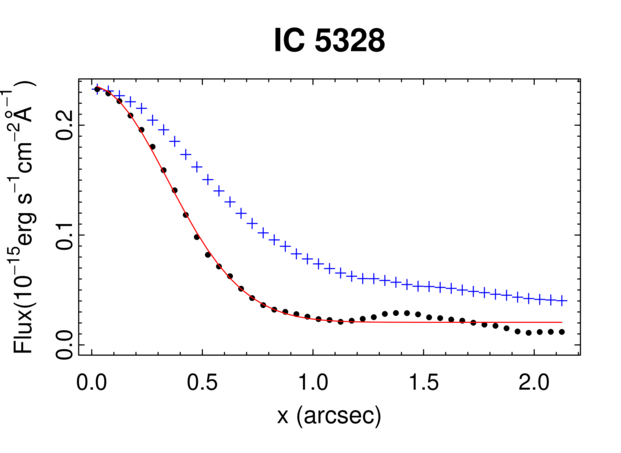}
\end{center}
\caption{-- continued}

\end{figure*}


\bsp	
\label{lastpage}
\end{document}